\documentclass[11pt,british]{article}
\PassOptionsToPackage{linecolor=blue,backgroundcolor=blue!25,bordercolor=blue,textsize=scriptsize}{todonotes}
\usepackage[utf8]{inputenc}
\usepackage{geometry}
\usepackage[dvipsnames]{xcolor}
\usepackage{babel}
\usepackage{booktabs}
\usepackage{mathtools}
\usepackage{amsmath}
\usepackage{amsthm}
\usepackage{amssymb}
\usepackage{setspace}
\usepackage{microtype}
\usepackage[unicode=true,pdfusetitle,
 bookmarks=true,bookmarksnumbered=false,bookmarksopen=false,
 breaklinks=false,pdfborder={0 0 0},pdfborderstyle={},backref=false,colorlinks=true]
 {hyperref}
\hypersetup{colorlinks=true,urlcolor=darkblue,anchorcolor=darkblue,citecolor=darkblue,filecolor=darkblue,linkcolor=darkblue,menucolor=darkblue,linktocpage=true,pdftitle=Numerical spectra of the Laplacian for line bundles on Calabi-Yau hypersurfaces}

\makeatletter

\providecommand{\tabularnewline}{\\}

\usepackage{tikz}
\usetikzlibrary{backgrounds}
\usepackage[framemethod=tikz]{mdframed}
\AtBeginEnvironment{mdframed}{%
\tikzset{every picture/.style={}}%
}
\mdfsetup{roundcorner=.5ex,font=\small}
{\begin{mdframed}[backgroundcolor=white]}%
{\end{mdframed}}
{\begin{mdframed}[backgroundcolor=white]}%
{\end{mdframed}}

\usepackage{slashed}	 	
\usepackage{amsfonts} 		
\usepackage{cite}	
\SetTracking{encoding={*}, shape=sc}{0} 	
\allowdisplaybreaks		
\date{\today} 		
\numberwithin{equation}{section}	
\usepackage[title,header]{appendix}

\makeatletter
\g@addto@macro\bfseries{\boldmath}
\makeatother

\usepackage{xcolor}
\definecolor{darkblue}{rgb}{0.1,0.0,0.5}

\makeatletter
\g@addto@macro\@floatboxreset\centering
\makeatother

\let\originalleft\left
\let\originalright\right
\renewcommand{\left}{\mathopen{}\mathclose\bgroup\originalleft}
\renewcommand{\right}{\aftergroup\egroup\originalright}

\usetikzlibrary{cd}

\makeatother

\begin{document}

\global\long\def\rep#1{\boldsymbol{#1}}%
\global\long\def\repb#1{\overline{\boldsymbol{#1}}}%
\global\long\def\dd{\text{d}}%
\global\long\def\ii{\text{i}}%
\global\long\def\ee{\text{e}}%
\global\long\def\Dorf{L}%
\global\long\def\tDorf{\hat{L}}%
\global\long\def\GL#1{\text{GL}(#1)}%
\global\long\def\Orth#1{\text{O}(#1)}%
\global\long\def\SO#1{\text{SO}(#1)}%
\global\long\def\Spin#1{\text{Spin}(#1)}%
\global\long\def\Symp#1{\text{Sp}(#1)}%
\global\long\def\Uni#1{\text{U}(#1)}%
\global\long\def\SU#1{\text{SU}(#1)}%
\global\long\def\Gx#1{\text{G}_{#1}}%
\global\long\def\Fx#1{\text{F}_{#1}}%
\global\long\def\Ex#1{\text{E}_{#1}}%
\global\long\def\ExR#1{\text{E}_{#1}\times\mathbb{R}^{+}}%
\global\long\def\ex#1{\mathfrak{e}_{#1}}%
\global\long\def\gl#1{\mathfrak{gl}_{#1}}%
\global\long\def\SL#1{\text{SL}(#1)}%
\global\long\def\Stab{\operatorname{Stab}}%
\global\long\def\vol{\operatorname{vol}}%
\global\long\def\Vol{\operatorname{Vol}}%
\global\long\def\tr{\operatorname{tr}}%
\global\long\def\ad{\operatorname{ad}}%
\global\long\def\ext{\mbox{\large{\ensuremath{\wedge}}}}%
\global\long\def\AdS#1{\text{AdS}_{#1}}%
\global\long\def\op#1{\operatorname{#1}}%
\global\long\def\im{\operatorname{im}}%
\global\long\def\re{\operatorname{re}}%
\global\long\def\eqspace{\mathrel{\phantom{{=}}{}}}%
\global\long\def\bZ{\mathbb{Z}}%
\global\long\def\bC{\mathbb{C}}%
\global\long\def\bP{\mathbb{P}}%
\global\long\def\bR{\mathbb{R}}%
\global\long\def\feyn#1{\slashed{#1}}%
\global\long\def\id{\operatorname{id}}%
\global\long\def\ap{\alpha'}%
\global\long\def\ct#1{\mathtt{#1}}%
\global\long\def\cc#1{\text{\ensuremath{\underbar{\ensuremath{\mathtt{#1}}}}}}%
\global\long\def\id{\boldsymbol{1}}%
\global\long\def\transpose{{\scriptscriptstyle \mathsf{T}}}%

\begin{titlepage}
\begin{flushright} 
\end{flushright}
\vfill
\begin{center} 
{\setstretch{1.4}\LARGE\bf Numerical spectra of the Laplacian for line bundles\\ on Calabi--Yau hypersurfaces\par} 
\vskip 1cm 
Anthony Ashmore,\textsuperscript{a} Yang-Hui He,\textsuperscript{b,c} Elli Heyes\textsuperscript{c} and Burt A.\,Ovrut\textsuperscript{d} 
\vskip 1cm
\textit{\small{}\textsuperscript{a}Enrico Fermi Institute \& Kadanoff Center for Theoretical Physics,\\ University of Chicago, Chicago, IL 60637, USA}
\\[.2cm]
\textit{\small\textsuperscript{b}London Institute for Mathematical Sciences, Royal Institution,\\ London, W1S 4BS, UK}
\\[.2cm]
\textit{\small\textsuperscript{b}Merton College, University of Oxford, OX1 4JD, UK}
\\[.2cm]
\textit{\small\textsuperscript{b}School of Physics, NanKai University, Tianjin, 300071, P.R. China}
\\[.2cm]
\textit{\small\textsuperscript{c}Department of Mathematics, City, University of London,\\ EC1V 0HB, UK}
\\[.2cm]
\textit{\small\textsuperscript{d}Department of Physics, University of Pennsylvania,\\ Philadelphia, PA 19104, USA}
\end{center}
\vfill
\begin{center} \textbf{Abstract} \end{center}
\begin{quote} 
We give the first numerical calculation of the spectrum of the Laplacian acting on bundle-valued forms on a Calabi--Yau three-fold. Specifically, we show how to compute the approximate eigenvalues and eigenmodes of the Dolbeault Laplacian acting on bundle-valued $(p,q)$-forms on K\"ahler manifolds. We restrict our attention to line bundles over complex projective space and Calabi--Yau hypersurfaces therein. We give three examples. For two of these, $\mathbb{P}^3$ and a Calabi--Yau one-fold (a torus), we compare our numerics with exact results available in the literature and find complete agreement. For the third example, the Fermat quintic three-fold, there are no known analytic results, so our numerical calculations are the first of their kind. The resulting spectra pass a number of non-trivial checks that arise from Serre duality and the Hodge decomposition. The outputs of our algorithm include all the ingredients one needs to compute physical Yukawa couplings in string compactifications.
\end{quote}
\vfill
{\begin{NoHyper}\let\thefootnote\relax\footnotetext{\tt \!\!\!\!\!\!\!\!\!\!\! ashmore@uchicago.edu, hey@maths.ox.ac.uk, elli.heyes@city.ac.uk, ovrut@elcapitan.hep.upenn.edu}\end{NoHyper}}
\end{titlepage}

\microtypesetup{protrusion=false} 
\tableofcontents 
\microtypesetup{protrusion=true} 

\section{Introduction and summary}

Heterotic string theory has provided a plethora of string models with realistic low-energy physics~\cite{Braun:2005ux,Lukas:1998yy,Donagi:1999ez,Bouchard:2005ag,Blumenhagen:2006ux,Lebedev:2006kn,Candelas:2007ac,Lebedev:2008un,Ambroso:2010pe,MayorgaPena:2012ifg,Anderson:2009mh,Anderson:2011ns,Anderson:2013xka}. The standard ingredients in these models are a Calabi--Yau three-fold $X$, equipped with a Ricci-flat metric, and a vector bundle $V$ whose connection solves the hermitian Yang--Mills equation~\cite{Candelas:1985en}. Upon compactifying on $X$, one finds a four-dimensional effective theory with $\mathcal{N}=1$ supersymmetry governed by a Kähler potential and a superpotential. By judicious choices of the three-fold and the vector bundle, one can find MSSM-like models incorporating a variety of desirable features. In principle, the masses and couplings in these models can be computed directly from the geometry of $X$ and $V$. However, even after decades of work, we are still unable to compute these numbers for all but the simplest examples. A substantial part of the difficulty can be attributed to the lack of explicit expressions for non-trivial Calabi--Yau metrics and hermitian Yang--Mills connections.

Some general features, such as the number of generations or the vanishing of certain couplings, can be inferred from topological or algebraic data of the three-fold $X$ and bundle $V$~\cite{Greene:1986ar,Greene:1986bm,Greene:1986jb,Matsuoka:1986vg,Greene:1987xh,Donagi:2000zs,Braun:2006me,Anderson:2010tc}. However, the details of the resulting four-dimensional physics -- Yukawa couplings and so on -- are determined by a Kähler potential and a superpotential, both of which depend on the metric on $X$ and connection on $V$. Without this data, it is generally not possible to compute masses or couplings, thereby preventing us from making precise particle physics predictions using string theory.

With little to no chance of ever discovering analytic expressions for the relevant metrics or connections, there has been considerable focus on using numerical methods to compute these objects. Numerous algorithms have been devised for numerically determining Ricci-flat metrics and hermitian Yang--Mills connections on Calabi--Yau manifolds, including position-space techniques~\cite{Headrick:2005ch}, spectral methods~\cite{math/0512625,Douglas:2006rr,Douglas:2006hz,Braun:2007sn,Headrick:2009jz,Cui:2019uhy,Anderson:2010ke,Anderson:2011ed,Cui:2023eqr}, and more recent advances employing machine learning and neural networks~\cite{Anderson:2020hux,Ashmore:2019wzb,Douglas:2020hpv,Jejjala:2020wcc,Ashmore:2021ohf,Ashmore:2021rlc,Larfors:2021pbb,Larfors:2022nep,Gerdes:2022nzr,Berglund:2022gvm}. Building on these, there are now works which take the first steps in using these numerical metrics for computations, such as finding the spectrum of the Laplacian on scalars and $(p,q)$-forms~\cite{Braun:2008jp,Ashmore:2020ujw}, checking the swampland distance conjecture as a function of complex structure moduli~\cite{Ashmore:2021qdf}, discovering chaos in two-dimensional sigma models~\cite{Afkhami-Jeddi:2021qkf}, and relating level crossing in the spectrum to the presence of attractor points~\cite{Ahmed:2023cnw}. 

The focus of the present paper is to give all the ingredients necessary for computing the superpotential and Kähler potential in simple examples. In more detail, as is well known to experts, in order to derive the matter sector of the four-dimensional effective theory that descends from the heterotic string on a Calabi--Yau three-fold $X$ admitting a bundle $V$, one has to carry out the following steps:
\begin{enumerate}
\item Calculate the Calabi--Yau metric on $X$ for a particular point in both complex and Kähler moduli space.
\item Calculate the hermitian Yang--Mills connection on $V$.
\item Calculate the zero modes of a certain twisted Dirac operator. Since $X$ is Kähler, this is equivalent to finding bundle-valued differential forms which are harmonic with respect to the Dolbeault Laplacian $\Delta_{\bar{\partial}_{V}}$ associated to the twisted Dolbeault differential $\bar{\partial}_{V}$~\cite{Candelas:1985en,Strominger:1985it,Strominger:1985ks}.
\item Find an orthonormal basis for the harmonic modes (or compute the matter-field Kähler metric from inner products of these modes).
\item Calculate the physical superpotential from integrals of wedge products of the normalised harmonic modes. 
\end{enumerate}
The focus of this paper is step three. (In fact, our numerical approach means that step four comes for ``free'', as we will see.) We give the first numerical calculation of the spectrum and eigenmodes of the Laplacian acting on bundle-valued forms on a Calabi--Yau three-fold. Specifically, we compute the approximate spectrum and eigenmodes of the Dolbeault Laplacian acting on bundle-valued scalars ($(0,0)$-forms) and $(0,1)$-forms.\footnote{Calculations were carried out on a laptop computer using custom-written code in Mathematica~\cite{Mathematica}. The authors hope to release a package in the near future.} We restrict our attention to line bundles over Calabi--Yau $n$-folds constructed as hypersurfaces in a single ambient projective space. With the eigenmodes in hand, we are able to compute an orthonormal basis of harmonic modes and thus the correctly normalised superpotential which determines physical Yukawa couplings. Unfortunately, the examples we consider are too simple to admit non-vanishing cubic superpotential couplings, and so there are no matter-field Yukawa couplings to calculate. However, this proof-of-concept calculation still represents a significant step towards calculating a Yukawa coupling in a physically relevant compactification.

It should be emphasized that even decoupled from the physical context which motivated our study, our calculations are of wider interest. The methods we use are applicable to line bundles over any K\"ahler manifold, such as complex projective space. More generally, explicit and numerical eigenmodes of the bundle-valued Laplacian should be useful to geometers and mathematical physicists alike.

The organisation of the paper is as follows. In Section \ref{sec:Phenomenology-and-the} we review how four-dimensional physics is determined by the geometry of a Calabi--Yau metric and a hermitian Yang--Mills connection. We also outline the concepts needed to define the Dolbeault Laplacian on a bundle and specify the precise eigenvalue problem that we will solve. In Section \ref{sec:The-spectrum-of-P3}, we apply our numerical method to the toy example of line bundles over complex projective space. We present the known analytic results for the spectrum, describe how to convert the eigenvalue problem into one of finite-dimensional linear algebra, and then compare the exact and numerical results. In Section \ref{sec:The-torus-as-CY1}, we consider the simplest Calabi--Yau hypersurface, namely the flat torus described by a cubic equation in $\bP^{2}$. Again, we present the known exact results for the spectrum and then compare these to results for the bundle-valued scalar and $(0,1)$-form spectra computed numerically. In Section \ref{sec:Quintic-Calabi=002013Yau-three-folds}, we apply our numerical method to compute the spectrum for a line bundle over the Fermat quintic three-fold. Though there are no analytic results to compare with, we carry out a number of consistency checks that the spectrum should satisfy.

\subsection*{Summary and future directions}

To summarise, the results of this paper are:
\begin{itemize}
    \item The construction in Section \ref{subsec:An-approximate-basis} of a finite basis of bundle-valued differential forms on complex projective space, which can be used to approximate the space of eigenmodes of the Dolbeault Laplacian on both $\bP^N$ and hypersurfaces therein.
    \item Numerical calculations of the eigenmodes and eigenvalues of the Dolbeault Laplacian on $\mathcal{O}(m)$-valued scalars and $(0,1)$-forms on $\bP^3$. These are presented in Figures \ref{fig:P3_scalar_graph} and \ref{fig:P3_01_graph} and Tables \ref{tab:P3_scalar_table} and \ref{tab:P3_01_table}. We find perfect agreement with known exact results for the scalar spectrum and perform a number of consistency checks for the $(0,1)$ spectrum. We have not found exact results for the bundle-valued $(0,1)$-form spectrum in the literature, so, to the best of the authors' knowledge, our numerical calculation is the first of its kind.
    \item Numerical calculations of the eigenmodes and eigenvalues of the Dolbeault Laplacian on $\mathcal{O}(m)$-valued scalars and $(0,1)$-forms on a torus. The torus is a Calabi--Yau one-fold, allowing us to compare our numerical results with exact predictions. These results are presented in Figures \ref{fig:T2_scalar_graph} and \ref{fig:T2_01_graph} and Tables \ref{tab:T2_scalar_table} and \ref{tab:T2_01_table}. Again, we find perfect agreement with the known exact results.
    \item The first numerical calculations of the eigenmodes and eigenvalues of the Dolbeault Laplacian on $\mathcal{O}(m)$-valued scalars and $(0,1)$-forms on a Calabi--Yau three-fold. We focus on the Fermat quintic three-fold and give the results in Figures \ref{fig:quintic_scalar_graph} and \ref{fig:quintic_01_graph} and Tables \ref{tab:quintic_scalar_table} and \ref{tab:quintic_01_table}. Here, there are no analytic results to compare with; instead, we perform a number of non-trivial checks on the spectrum which come from Serre duality and the Hodge decomposition.
\end{itemize}

This paper focuses on the examples of line bundles over Calabi--Yau manifolds in a single ambient projective space. In future work, we plan to extend this in two ways. First, we will move from hypersurfaces in a single projective space to complete intersection Calabi--Yau (CICY) manifolds, defined by multiple equations in products of projective space. Even without moving to non-abelian bundles, these examples are rich enough to admit Standard Model-like theories with non-vanishing Yukawa couplings~\cite{Anderson:2009mh,Anderson:2011ns,Anderson:2012yf,Anderson:2013xka,GrootNibbelink:2015lme,GrootNibbelink:2015dvi,GrootNibbelin:2016ovb,Braun:2017feb}. Second, we plan to generalise our algorithm to non-abelian bundles, starting with the examples considered by Douglas et al.~\cite{Douglas:2006hz}. This will allow us to compute the spectrum and corresponding Yukawa couplings for compactifications which give rise to realistic physics, such as the so-called heterotic Standard Models~\cite{hep-th/0512177,hep-th/0502155,Braun:2005ux,Bouchard:2005ag,Anderson:2009mh,Ambroso:2010pe,1112.1097,Anderson:2011ns,Anderson:2012yf,Anderson:2013xka,GrootNibbelink:2015dvi,GrootNibbelink:2015lme,1007.0203,hep-th/9903052}. Of particular interest for the authors is a certain $\SU 4$ bundle over the $\bZ_{3}\times\bZ_{3}$ symmetric Schoen three-fold discussed in \cite{hep-th/0502155,Braun:2005zv,hep-th/0512177, Braun:2006ae,Ambroso:2010pe,Marshall:2014kea,Marshall:2014cwa,Ovrut:2014rba,Ovrut:2015uea,Deen:2016vyh,Ovrut:2018qog,Dumitru:2018jyb,Dumitru:2018nct,Dumitru:2019cgf,Ashmore:2020ocb,Ashmore:2020wwv,Ashmore:2021xdm,Dumitru:2021jlh,Dumitru:2022apw,Dumitru:2022eri} and analysed numerically by Braun et al.~\cite{Braun:2007sn}. These advances, together with progress on non-perturbative superpotentials~\cite{Witten:1999eg,Buchbinder:2002ic,Beasley:2003fx,Basu:2003bq,Braun:2007xh,Braun:2007tp,Braun:2007vy,Bertolini:2014dna,Buchbinder:2016rmw,Buchbinder:2019eal,Buchbinder:2019hyb,Buchbinder:2017azb,Buchbinder:2018hns,Buchbinder:2002pr,Anderson:2015yzz} and moduli stabilisation~\cite{Becker:2005nb,Becker:2006xp,Anderson:2010mh,Melnikov:2011ez,Anderson:2011cza,Anderson:2011ty, Anderson:2014xha,Anderson:2013qca,delaOssa:2014cia,delaOssa:2015maa,Garcia-Fernandez:2015hja,Candelas:2016usb,Ashmore:2018ybe,Blesneag:2018ygh,Ashmore:2019rkx,Ashmore:2020wwv}, should allow concrete computations of masses and couplings in top-down string models in the near future.

More generally, the numerical methods we have employed may be useful for studying the geometry of Kähler manifolds and bundles in their own right. For example, the scalar spectrum of $\Delta_{\bar{\partial}_{V}}$ can be interpreted in terms of the quantum mechanics of a charged particle moving on the curved manifold. Indeed, many of the known analytic results for the spectrum of this Laplacian were first found from considerations of particles moving on projective space or Riemann surfaces~\cite{10.2748/tmj/1178228026,2302.11691,TEJEROPRIETO2006288}. Moreover, the methods we have described can also be applied to the full $(p,q)$-form spectrum. Though general $(p,q)$-form eigenmodes of the Laplacian are not immediately relevant for string compactifications, they surely encode many interesting quantities which characterise manifolds and bundles, such as regularised heat kernels and analytic torsions. All of these are accessible using the techniques of this paper. We hope to return to these ideas in future works.

\section{Phenomenology and the Dolbeault Laplacian\label{sec:Phenomenology-and-the}}

Supersymmetric Minkowski compactifications of the $\Ex 8\times\Ex 8$ heterotic string without three-form flux are specified by a choice of Calabi--Yau $n$-fold $(X,g)$, where $g$ is a Ricci-flat Kähler metric, and a principal $G$-bundle with $G\subset\Ex 8\times\Ex 8$, whose curvature $F$ satisfies the hermitian Yang--Mills equation:
\begin{equation}
F_{ij}=F_{\bar{i}\bar{j}}=0,\qquad g^{i\bar{j}}F_{i\bar{j}}\propto\id,\label{eq:HYM}
\end{equation}
where $i,j$ label holomorphic coordinates on the Calabi--Yau, $\id$ is the identity element of $G$, and the constant of proportionality in the second equation is a real number known as the slope which is determined by the choice of $G$-bundle. A solution to \eqref{eq:HYM} is equivalent to the bundle being holomorphic and admitting a hermitian metric on its fibres which is ``Hermite--Einstein''. Unfortunately, there are no explicitly known Calabi--Yau metrics on three-folds, nor Hermite--Einstein metrics on bundles over Calabi--Yau manifolds. Instead, one must turn to numerical techniques. With these now available, the question becomes what physically interesting quantities we want to compute. Among many possible applications, the motivation of this work is the computation of physical Yukawa couplings in string models. To set the scene, we quickly review how a four-dimensional $\mathcal{N}=1$ effective theory is derived from a ten-dimensional string compactification.

\subsection{Supersymmetry, Yukawa couplings and the matter-field Kähler metric\label{subsec:Supersymmetry,-Yukawa-couplings}}

In addition to a metric, dilaton and $B$ field, the bosonic sector of the heterotic string has an $\Ex 8\times\Ex 8$ gauge field $A$. Matter fields in four dimensions come from a decomposition of this gauge field and its associated gaugino. Consider a background of the form $\bR^{1,3}\times X$, where $X$ is Calabi--Yau, and focus on a single $\Ex 8$ factor. Assuming that $X$ admits a principal $G$-bundle with $G\subset\Ex 8$, the gauge group $H$ in four dimensions is given by the commutant of $G$ in $\Ex 8$. For concreteness, consider an illustrative example\footnote{For simplicity, we ignore any discrete factors.} where $G=\SU 3$ and $H=\Ex 6$.\footnote{The special case where $V = TX$ is known as the ``standard embedding''. Many quantities of interest can be computed using the techniques of special geometry without needing an explicit metric on $X$~\cite{Strominger:1985ks,Candelas:1987se,Candelas:1990pi,Greene:1986bm,Greene:1986jb,Greene:1987xh}. Unfortunately, it is difficult to find acceptable MSSM-like physics in these simple models, so one is forced to consider more general vector bundles.}

The matter multiplets can then be read off from a decomposition of the $\rep{248}$ representation in which the ten-dimensional gaugino transforms under $\SU 3\times\Ex 6$ as
\begin{equation}
\rep{248}\to\bigoplus_{\rep r,\rep R}(\rep{r,}\rep R)=(\rep 8,\rep 1)\oplus(\rep 1,\rep{78})\oplus(\rep 3,\rep{27})\oplus(\repb 3,\repb{27}) ,
\label{eq:su3}
\end{equation}
where here and below, we use ${\rep r}$ to denote a representation of $\SU 3$ and ${\rep R}$, that of $\Ex 6$.
From this, we observe that the low-energy theory can contain matter transforming as the $\rep 1$, $\rep{27}$ or $\repb{27}$ of $\Ex 6$, corresponding to bundle moduli, families and anti-families. Using standard Kaluza--Klein analysis on Kähler manifolds~\cite{Green:1987sp,Green:1987mn}, one finds that these matter fields come from harmonic bundle-valued $(0,1)$-forms $\psi_{\rep r}$, where the relevant bundles are vector bundles $V_{\rep r}$ over $X$ associated to the principal $\SU 3$ bundle over $X$ and the $\SU 3$ representation $\rep r$. For example, from \eqref{eq:su3}, the number of $\rep{27}$ families present in four dimensions is counted by the number of harmonic $V_{\rep 3}$-valued $(0,1)$-forms on $X$, where $V_{\rep 3}$ is a rank-three vector bundle on $X$ whose fibres admit an action of $\SU 3$ in the $\rep 3$ representation. The number of these harmonic $(0,1)$-forms, and hence the number of families, is equal to the dimension\footnote{As is standard in the literature, $H^{q}(X,V)$ denotes the $V$-valued $(0,q)$-form sheaf cohomology, and not the de Rham cohomology.} of $H^{1}(X,V_{\rep 3})$. Similarly, the bundle moduli are counted by $H^{1}(X,V_{\rep 8})\simeq H^{1}(X,V_{\rep 3}\otimes V_{\rep 3}^{*})$.

Since the effective theory has $\mathcal{N}=1$ supersymmetry, it is determined by a superpotential and a Kähler potential~\cite{Wess:1992cp}. These objects are fixed by the geometry of the compactification in the following way. Let $\psi_{\rep r}^{I}$ be a basis for $H^{1}(X,V_{\rep r})$ which is not necessarily harmonic. If there is a singlet in the product $\rep r\times\rep{r'}\times\rep{r''}$, there can be a holomorphic Yukawa coupling of the form
\begin{equation}\label{eq:superpotential}
\lambda_{IJK}(\rep r,\rep{r'},\rep{r''})=\int_{X}\Omega\wedge\tr(\psi_{\rep r}^{I}\wedge\psi_{\rep{r'}}^{J}\wedge\psi_{\rep{r''}}^{K}),
\end{equation}
where $\Omega$ is the holomorphic $(3,0)$-form on $X$ and the trace indicates a projection to the $\SU 3$ singlet. Using the above decomposition \eqref{eq:su3}, we denote the four-dimensional chiral superfields associated via Kaluza--Klein reduction to each $\psi_{\rep r}^{I}$ by $C_{\rep R}^{I}$. The superpotential for these chiral superfields is then given by
\begin{equation}
W=\lambda_{IJK}(\rep R,\rep{R'},\rep{R''})C_{\rep R}^{I}C_{\rep{R'}}^{J}C_{\rep{R''}}^{K},
\end{equation}
where we have relabelled the Yukawa couplings by the four-dimensional gauge group, i.e.~their $\Ex 6$ representations. Given that a singlet appears in $\rep 8^{3}$, $\rep 8\cdot\rep 3\cdot\repb 3$, $\rep 3^{3}$ and their conjugates, the possible types of Yukawa couplings are $\rep 1^{3}$, $\rep 1\cdot\rep{27}\cdot\repb{27}$, $\rep{27}^{3}$ and $\repb{27}^{3}$.

The above are often known as holomorphic Yukawa couplings as they are quasi-topological in the sense that $\lambda_{IJK}$ can be computed using representatives of $H^{1}(X,V_{\rep r})$ which are not harmonic (this follows straightforwardly from $\dd\Omega=0$). However, the \emph{physical} Yukawa couplings depend on the normalisation of the kinetic terms for the chiral superfields. This normalisation is fixed by the matter-field Kähler metric, given by
\begin{equation}\label{eq:matter kahler}
\mathcal{G}_{IJ}=\int_{X}\bar{\star}_{V}\psi_{\rep r}^{I}\wedge\psi_{\rep r}^{J},
\end{equation}
which is simply the inner product between harmonic representatives of $H^{1}(X,V_{\rep r})$. Due to the need for harmonic forms and the presence of the Hodge star on bundle-valued forms $\star_{V}$, this depends on knowledge of the Calabi--Yau metric on $X$, the Hermite--Einstein metric on the fibres of $V_{\rep r}$ and the zero modes of the Dolbeault Laplacian on $(0,1)$-forms valued in $V_{\rep r}$. There is now much progress in computing Calabi--Yau and Hermite--Einstein metrics numerically, so this paper will focus on the bundle-valued harmonic modes. In particular, the aim of this paper is to understand how to compute these ingredients numerically for the simple case where $\rep r$ is a representation of $\Uni 1$, corresponding to $V_{\rep r}$ being a line bundle over $X$.

\subsection{The Dolbeault Laplacian on a vector bundle\label{subsec:The-Dolbeault-Laplacian}}

In the previous section, we saw that physical Yukawa couplings can be obtained by computing overlap integrals of harmonic representatives of certain bundle-valued cohomologies. Here, we will be a little more precise about the equations that these representatives should satisfy and the particular eigenvalue problem that we will solve. We first review the general formalism of bundles on Kähler manifolds and the Dolbeault Laplacian acting on bundle-valued $(p,q)$-forms. In practice, computing Yukawa couplings needs only the $(0,1)$-form sector, but we find it useful to keep our discussion more general. 

Let $X$ be a compact, complex manifold of complex dimension $n$ (real dimension $d=2n$) with Kähler metric $g_{i\bar{j}}$ and Kähler form $\omega$, and let $V$ be a rank-$r$ holomorphic vector bundle over $X$. The physically relevant case is when $V$ is any of the vector bundles $V_{\rep r}$, where the representation $\rep r$ appears in the decomposition of the $\rep{248}$, such as in \eqref{eq:su3}. However, for simplicity we will just write $V$. We denote by $\Omega^{p,q}(V)$ the space of $V$-valued $(p,q)$-forms. Unlike a real vector bundle, a holomorphic bundle comes with a canonical differential operator $\bar{\partial}_{V}$: 
\begin{equation}
\bar{\partial}_{V}\colon\Omega^{p,q}(V)\to\Omega^{p,q+1}(V).
\end{equation}
This generalises the usual Dolbeault operator $\bar{\partial}$ on $(p,q)$-forms, and likewise is nilpotent and obeys the Leibniz rule. Explicitly, let $\{E_{a}\}$ be a holomorphic frame for $V$, i.e.~on each patch of $X$, the $E_{a}$ form a basis for $\bC^{r}$ and have holomorphic transition functions valued in $\GL{r,\bC}$. A bundle-valued $(p,q)$-form $\alpha$ can then be written locally as
\begin{equation}
\alpha=\sum_{a=1}^{r}\alpha^{a}\otimes E_{a},
\end{equation}
where $\alpha^{a}\in\Omega^{p,q}(X)$ are standard $(p,q)$-forms on $X$.

A hermitian structure on $V$ is equivalent to a hermitian metric $G$ on the fibres of $V$ such that, for sections $s_{1},s_{2}\in\Omega^{0,0}(V)$,
\begin{equation}
G(s_{1},s_{2})=\overline{G(s_{2},s_{1})}=G_{\bar{a}b}\overline{s_{1}^{a}}s_{2}^{b},
\end{equation}
where $G_{\bar{a}b}$ is a positive-definite hermitian matrix. As with a conventional metric on a manifold, the hermitian structure gives an isomorphism between $V$ and the dual bundle $V^{*}$. Combining this with the Hodge star operator, we have a generalised Hodge star $\bar{\star}_{V}$ which maps $\Omega^{p,q}(V)$ to $\Omega^{n-p,n-q}(V^{*})$ and defines an inner product $\langle\cdot,\cdot\rangle$ as
\begin{equation}
\langle\alpha_{1},\alpha_{2}\rangle=\int_{X}\bar{\star}_{V}\alpha_{1}\wedge\alpha_{2}=\int_{X}(\alpha_{1},\alpha_{2})\vol,\label{eq:inner_product}
\end{equation}
where $\vol$ is the volume form defined by $\star1$ and $(\alpha_{1},\alpha_{2})$ is given by
\begin{equation}
(\alpha_{1},\alpha_{2})=\frac{1}{p!q!}\overline{(\alpha_{1})_{i_{1}\ldots i_{p}\bar{j}_{1}\ldots\bar{j}_{q}}^{a}}(\alpha_{2})_{k_{1}\ldots k_{p}\bar{l}_{1}\ldots\bar{l}_{q}}^{b}g^{\bar{i}_{1}k_{1}}\ldots g^{j_{1}\bar{l}_{1}}\ldots G_{\bar{a}b}.
\end{equation}
The adjoint of $\bar{\partial}_{V}$ is then defined relative to this inner product as $\langle\bar{\partial}_{V}\alpha_{1},\alpha_{2}\rangle=\langle\alpha_{1},\bar{\partial}_{V}^{\dagger}\alpha_{2}\rangle$ and is given explicitly by
\begin{equation}
\bar{\partial}_{V}^{\dagger}=(-1)^{np+1}\star_{V}\partial\star_{V}{},
\end{equation}
where $\partial$ is the standard Dolbeault differential.

Using these ingredients, we define the Dolbeault Laplacian as
\begin{equation}
\Delta_{\bar{\partial}_{V}}=\bar{\partial}_{V}^{\dagger}\bar{\partial}_{V}+\bar{\partial}_{V}\bar{\partial}_{V}^{\dagger},
\end{equation}
which is self-adjoint with respect to $\langle\cdot,\cdot\rangle$. A bundle-valued $(p,q)$-form $\alpha$ is then called \emph{harmonic} (or a \emph{zero mode}) if
\begin{equation}
\Delta_{\bar{\partial}_{V}}\alpha=0.\label{eq:laplacian}
\end{equation}
We emphasise that the statement that $\alpha$ is harmonic makes sense only with respect to a choice of Kähler metric $g$ on $X$ and hermitian structure $G$ on $V$. On a compact manifold, harmonic is equivalent to being both $\bar{\partial}_{V}$- and $\bar{\partial}_{V}^{\dagger}$-closed, with the harmonic forms giving the harmonic representatives of the Dolbeault cohomologies $H_{\bar{\partial}_{V}}^{p,q}(X,V)$. In what follows, it will be useful to define $h^{p,q}(V)=\dim H_{\bar{\partial}_{V}}^{p,q}(X,V)$ as the dimension of the $V$-valued $(p,q)$-form cohomology. There is also a Hodge decomposition which ensures that a bundle-valued $(p,q)$-form can be written uniquely as the sum of a harmonic, a $\bar{\partial}_{V}$-exact and a $\bar{\partial}_{V}^{\dagger}$-exact form. The bundle-valued sheaf cohomologies $H^{q}(X,V)$ are then defined as
\begin{equation}
H^{q}(X,V)\simeq H_{\bar{\partial}_{V}}^{0,q}(X,V).
\end{equation}

We now want to define a connection on $V$ and a corresponding differential operator $D$ that maps $\Omega^{p}(V)$ to $\Omega^{p+1}(V)$. Locally, $D=\dd+A$, where $A$ is a connection one-form (gauge field). Acting on a section $s\in\Omega^{0,0}(V)$, the curvature of $D$ is given by
\begin{equation}
D^{2}s=(\dd A+A\wedge A)\cdot s=F\cdot s,
\end{equation}
where $F\in\Omega^{2}(\op{End}V)$ acts on $s$ via the adjoint representation. A connection is compatible with the holomorphic structure of $V$ if the $(0,1)$-component of $D$ agrees with the Dolbeault differential, $D^{0,1}=\bar{\partial}_{V}$. Furthermore, the connection is hermitian if it is compatible with the hermitian structure on $V$ in the sense that
\begin{equation}
\dd(G(s_{1},s_{2}))=G(Ds_{1},s_{2})+G(s_{1},Ds_{2}),
\end{equation}
or equivalently $DG=0$. Note that compatibility with the holomorphic and hermitian structures uniquely determines the connection as the Chern connection of $G$. The Chern connection is characterised by a local one-form $A$ whose components in a holomorphic frame $\{E_{a}\}$ are given by
\begin{equation}\label{eq:A}
A^{a}{}_{b}=(G^{-1}\partial G)^{a}{}_{b}.
\end{equation}
Note that the Chern connection is type $(1,0$) by construction. The curvature $F$ of the Chern connection is then purely type $(1,1)$, and so defines a connection on a holomorphic bundle. 

The hermitian Yang--Mills equation \eqref{eq:HYM} can be expressed in terms of the hermitian structure $G$. First, as we mentioned above, the Chern connection of $G$ is automatically holomorphic as $F$ has no $(2,0)$ or $(0,2)$ components. The remaining condition is simply
\begin{equation}
g^{i\bar{j}}F_{i\bar{j}}=-g^{i\bar{j}}\partial_{\bar{j}}A_{i}=-g^{i\bar{j}}\partial_{\bar{j}}(G^{-1}(\partial_{i}G))\propto\mu(V)\,\id,\label{eq:hym2}
\end{equation}
where $\mu(V)$ is the slope of $V$, given by (more details on the slope can be found in Appendix \ref{subsec:The-slope}):
\begin{equation}
\mu(V)=\frac{1}{\op{rank}V}\int_{X}c_{1}(V)\wedge\omega^{n-1}.\label{eq:slope}
\end{equation}
A hermitian fibre metric $G$ which solves \eqref{eq:hym2} is known as Hermite--Einstein. Whether there exists a Hermite--Einstein metric on $V$ depends on the so-called stability of the bundle~\cite{Donaldson,UhlenbeckYau}. This can often be checked by somewhat laborious algebraic calculations, though the guarantee of existence is not constructive -- even if a given bundle is stable, it is often impossible to find an explicit expression for the corresponding Hermite--Einstein metric. This is especially true on manifolds without explicitly known metrics, such as for Ricci-flat metrics on Calabi--Yau manifolds.

For completeness, the covariant derivative of a section $s$ is
\begin{equation}
(D^{1,0}s)^{a}=\partial s^{a}+A^{a}{}_{b}s^{b},\qquad(D^{0,1}s)^{a}=\bar{\partial}s^{a}.
\end{equation}
For a holomorphic vector bundle, both $D^{1,0}$ and $D^{0,1}$ are nilpotent, while $D^{2}=F_{1,1}$. One then usually denotes these by $D^{1,0}=\partial_{V}=\partial+A$ and $D^{0,1}=\bar{\partial}_{V}=\bar{\partial}$, with the adjoint of $\bar{\partial}_{V}$ given by
\begin{equation}
\bar{\partial}_{V}^{\dagger}=(-1)^{np+1}\star\partial_{V}\star.\label{eq:adjoint}
\end{equation}
Following from this, one has the Bochner--Kodaira--Nakano identity~\cite{Bochner,Kodaira,Nakano,Demailly} which relates the $\bar{\partial}_{V}$-Laplacian to the $\partial_{V}$-Laplacian as
\begin{equation}
\Delta_{\bar{\partial}_{V}}=\Delta_{\partial_{V}}+[F,\Lambda],\label{eq:del_delb}
\end{equation}
where $\Lambda$ is contraction with the Kähler form $\omega$ on $X$. When $V$ is trivial, so that the curvature $F$ vanishes, this reduces to the usual relation between the $\partial$- and $\bar{\partial}$-Laplacians on a Kähler manifold, i.e., $\Delta_{\bar{\partial}}=\Delta_{\partial}$.

In order to link this back to the discussion of the previous section, we recall that certain $(0,1)$-form sheaf cohomologies count the number of four-dimensional matter fields in heterotic string compactifications. These cohomologies are spanned by harmonic modes which satisfy \eqref{eq:laplacian}, where $V$ should be replaced by the relevant bundles $V_{\rep r}$ associated to families, anti-families and so on. In addition, since the Dolbeault Laplacian depends on the metrics on both $X$ and the fibres of $V$, the connection one-form $A$ that appears in $\Delta_{\bar{\partial}_{V}}$ should be the appropriate hermitian Yang--Mills connection for $V_{\rep r}$, defined by a Hermite--Einstein metric on the fibres. The matter superfields relevant for Yukawa couplings then come from modes on $X$ that are $\Delta_{\bar{\partial}_{V}}$-harmonic representatives of 
\begin{equation}
H^{1}(X,V_{\rep r})\cong H_{\bar{\partial}_{V_{\rep r}}}^{0,1}(X,V_{\rep r}).
\end{equation}
With the necessary background on differential operators on holomorphic vector bundles now in place, we move on to consider the eigenvalue problem for $\Delta_{\bar{\partial}_{V}}$.

\subsection{The eigenvalue problem}

The general problem analysed in this work is finding the spectrum and eigenmodes of the Dolbeault Laplacian $\Delta_{\bar{\partial}_{V}}$ acting on $(p,q)$-forms valued in a vector bundle $V$. The particular examples we consider are those where the bundle $V$ is a line bundle over a compact Kähler manifold $X$. Furthermore, we will focus on computing the $(0,0)$- and $(0,1)$-form spectra. The spectrum of bundle-valued scalars will be useful for comparing with known results when $X$ is a projective space or a torus, while the $(0,1)$-form spectrum is what one needs to compute Yukawa couplings.

The eigenmodes $\phi\in\Omega^{p,q}(V)$ and eigenvalues $\lambda$ are defined by\footnote{Note that, in the case where the bundle is trivial, $V\simeq\mathcal{O}$, $\Delta_{\bar{\partial}_{V}}$ is equal to one-half of the de Rham Laplacian, so the spectrum of $\Delta_{\bar{\partial}_{V}}$ will be related to the usual spectrum by a factor of two.}
\begin{equation}
\Delta_{\bar{\partial}_{V}}\phi=\lambda\phi,\label{eq:eigenvalue}
\end{equation}
where the eigenvalues $\lambda$ are real and non-negative. The eigenmodes with zero eigenvalue, $\lambda=0$, are the ``harmonic'' or ``zero modes'' which span $H_{\bar{\partial}_{V}}^{p,q}(X,V)$. Since $X$ is assumed to be compact, the eigenvalues are discrete and have finite degeneracies. As we will see in examples, if the Kähler metric $g$ on $X$ admits either continuous or discrete symmetries, there may be multiple eigenmodes with the same eigenvalue. We will denote the $n$-th eigenvalue by $\lambda_{n}$ with multiplicity $\ell_{n}$ starting from $n=0$. Note that $\lambda_{0}$ always labels the smallest eigenvalue of $\Delta_{\bar{\partial}_{V}}$ even when $\lambda_{0}$ is not zero -- only when $\lambda_{0}=0$ do we refer to the corresponding eigenmodes as harmonic or zero modes. As usual, the eigenvalues scale with the volume of $X$ as $\lambda\sim\Vol(X)^{-2/d}$. We always normalise the volume of $X$ to one in the examples that follow.

Let us make a few comments on the expected structure of the spectrum of $\Delta_{\bar{\partial}_{V}}$. First, Serre duality implies
\begin{equation}
h^{p,q}(V)=h^{n-p,n-q}(V^{*}),
\end{equation}
so that the counting of zero modes of $\Delta_{\bar{\partial}_{V}}$ acting on $\Omega^{p,q}(V)$ and $\Omega^{n-p,n-q}(V^{*})$ should agree. In fact, since the Hodge star with conjugation $\bar{\star}_{V}$ commutes with the Laplacian, $\bar{\star}_{V}\Delta_{\bar{\partial}_{V}}=\Delta_{\bar{\partial}_{V^{*}}}\bar{\star}_{V}$, there is a relation between the entire tower of eigenmodes and eigenvalues. Denoting the set of $V$-valued $(p,q)$-form eigenmodes by $\{\phi\}_{V}^{p,q}$ and the corresponding eigenvalues as $\{\lambda\}_{V}^{p,q}$, one has
\begin{equation}
\begin{aligned}\bar{\star}_{V}\{\phi\}_{V}^{p,q} & =\{\phi\}_{V^{*}}^{n-p,n-q},\\
\{\lambda\}_{V}^{p,q} & =\{\lambda\}_{V^{*}}^{n-p,n-q}.
\end{aligned}
\label{eq:serre}
\end{equation}
Moreover, for $(0,q)$-forms, one can write this in terms of the canonical bundle $K_{X}$ of $X$ as $\{\lambda\}_{V}^{0,q}=\{\lambda\}_{K_{X}\otimes V^{*}}^{0,n-q}.$ We will use these relations as a non-trivial check on the numerical spectra in later parts of the paper.

Since the practicalities of solving for the eigenmodes and eigenvalues are covered thoroughly in the literature~\cite{Braun:2008jp,Ashmore:2020ujw,Ashmore:2021qdf,Ahmed:2023cnw}, we mention it only to fix some notation. For fixed $(p,q)$, let $\{\alpha_{A}\}$ be a basis for the vector space of complex-valued $(p,q)$-forms valued in $V$ on the manifold. This basis is infinite-dimensional, $A=1,\dots,\infty$, as we want to be able to express any element of $\Omega^{p,q}(V)$ as a linear combination of the basis with constant coefficients.\footnote{Recall that $\Omega^{p,q}(V)$ restricted to a point $x\in X$ is a finite-dimensional $\mathbb{C}$-vector space. If one does not restrict to a point but instead wants to describe the space of forms over the entire manifold, $\Omega^{p,q}(V)$ is an infinite-dimensional $\mathbb{C}$-vector space (or equivalently a finitely generated $C^\infty$-module).} 
The basis is not assumed to be orthonormal; the inner product \eqref{eq:inner_product} defines a matrix $O_{AB}$ as
\begin{equation}\label{O-AB}
O_{AB}\equiv\langle\alpha_{A},\alpha_{B}\rangle=\int_{X}\bar{\star}_{V}\alpha_{A}\wedge\alpha_{B},
\end{equation}
which captures the non-orthonormality. Similarly, the matrix elements of $\Delta_{\bar{\partial}_{V}}$ with respect to this basis are
\begin{equation}\label{del-AB}
\Delta_{AB}\equiv\langle\alpha_{A},\Delta_{\bar{\partial}_{V}}\alpha_{B}\rangle.
\end{equation}
The eigenvalue equation \eqref{eq:eigenvalue} can then be written in terms of the matrix elements as
\begin{equation}
\Delta_{AB}\phi_{B}=\lambda\,O_{AB}\phi_{B},\label{eq:gen_eigen}
\end{equation}
where $\phi=\phi_{C}\alpha_{C}$. This is then a generalised eigenvalue problem for $(\lambda,\phi_{A})$, albeit an infinite-dimensional one. Upon truncating $\{\alpha_{A}\}$ to finite-dimensional basis, one is left with a standard linear algebra problem to determine the eigenvalues $\lambda$ and the eigenvectors $\phi_{A}$, which in turn give the spectrum of $\Delta_{\bar{\partial}_{V}}$ and the expansion of the eigenmodes in terms of the truncated basis. Note that there is no reason to expect the basis modes $\alpha_{A}$ to themselves be eigenmodes of the Laplacian (in practice they are chosen to be numerically simple to compute). Thanks to this, truncating to a finite basis gives only an approximation of the spectrum and eigenmodes, with the dimension of the basis controlling the accuracy of the approximation. Our conventions for calculating the matrix elements in terms of the components of sections can be found in Appendix \ref{sec:Matrix-elements-of}.

Finally, we note that the matter-field Kähler metric \eqref{eq:matter kahler} is particularly straightforward to calculate once one has solved the eigenvalue problem. Explicitly, upon expanding the relevant harmonic representatives $\psi^{I}$ of $H^{1}(X,V)$ as $\psi^{I}=\psi_{A}^{I}\alpha_{A}$, the metric $\mathcal{G}_{IJ}$ can be written as
\begin{equation}
\begin{aligned}\mathcal{G}_{IJ} & =\int_{X}\bar{\star}_{V}\psi^{I}\wedge\psi^{J}=\overline{\psi_{A}^{I}}\psi_{B}^{J}\int_{X}\bar{\star}_{V}\alpha_{A}\wedge\alpha_{B}\\
 & \equiv\overline{\psi_{A}^{I}}O_{AB}\psi_{B}^{J}.
\end{aligned}
\end{equation}
In practice, when dealing with a generalised eigenvalue problem of the form \eqref{eq:gen_eigen}, Mathematica will return eigenvectors which are automatically $O_{AB}$-orthogonal~\cite{Mathematica}. They can then be made $O_{AB}$-orthonormal by simply rescaling the $\psi^{I}_{A}$ coefficients. In this basis, the field-space metric $\mathcal{G}_{IJ}$ is trivial.

\section{The spectrum of \texorpdfstring{$\Delta_{\bar{\partial}_{V}}$}{the Dolbeault Laplacian} on \texorpdfstring{$\protect\bP^{3}$}{P3}\label{sec:The-spectrum-of-P3}}

As in previous works~\cite{Braun:2008jp,Ashmore:2020ujw}, we begin with a study of three-dimensional complex projective space $\bP^{3}$ equipped with the Fubini--Study (FS) metric. Since much is known explicitly about projective space, this will provide an arena where we can check our numerical methods against exact results.

Recall that the Fubini--Study metric is the unique (up to scale) Kähler metric on $\bP^{3}$ with $\SU 4$ isometry, corresponding to the presentation of $\bP^3$ as a symmetric space:
\begin{equation}
\bP^{3}=\frac{\text{S}^{7}}{\Uni 1}=\frac{\SU 4}{\text{S}(\Uni 3\times\Uni 1)}.
\end{equation}
The Fubini--Study metric is defined by $g_{i\bar{j}}=\partial_{i}\bar{\partial}_{\bar{j}}K$, where $K$ is the Kähler potential
\begin{equation}
K=\frac{6^{1/3}}{2\pi}\log Z^{I}\bar{Z}_{I}.\label{eq:FS}
\end{equation}
Here $[Z^{0}:\dots:Z^{3}]$ are homogeneous coordinates on $\bP^{3}$ where, for example, on the patch $U_{0}=\{Z^{0}=1\}$, we have $Z^{I}=(1,z^{i})$ with $i=1,2,3$. The choice of prefactor in \eqref{eq:FS}  ensures $\Vol(\bP^{3})=1$. 

The bundles we consider are line bundles $V=\mathcal{O}(m)$ on $\bP^{3}$ for integer values of $m$. A hermitian metric on the fibres of $\mathcal{O}(m)$ is given by\footnote{See Appendix \ref{subsec:A-local-holomorphic} for a discussion of the components of $G$ relative to a choice of holomorphic frame for $\mathcal{O}(m)$.}
\begin{equation}
G=(Z^{I}\bar{Z}_{I})^{-m}.\label{eq:bundle_P3}
\end{equation}
Indeed, this is actually automatically Hermite--Einstein with respect to the Kähler metric defined by \eqref{eq:FS}. It is simple to see this by unwinding the various definitions in Section \ref{subsec:The-Dolbeault-Laplacian}, first by computing the connection as $A=\partial\log G$ and then the curvature as $F=\bar{\partial}A$. The slope $\mu$, defined in \eqref{eq:slope}, which appears as the constant of proportionality in the hermitian Yang--Mills equation, is then simply $\mu(\mathcal{O}(m))=m$ (see Appendix \ref{subsec:The-slope}).

\subsection{Analytic results\label{subsec:Analytic-results}}

The particular eigenvalue problem we want to solve is
\begin{equation}
\Delta_{\bar{\partial}_{V}}\phi=\lambda\phi,\label{eq:eigenvalue_problem}
\end{equation}
where $\phi$ is an $\mathcal{O}(m)$-valued $(0,0)$- or $(0,1)$-form. At this point, we consider the general problem of $\bP^N$ and specialise to $N=3$ when presenting our numerical results. First, we note that global holomorphic sections of $\mathcal{O}(m)$ are counted by $H_{\bar{\partial}_{V}}^{0,0}(\bP^N,\mathcal{O}(m))\simeq H^{0}(\bP^N,\mathcal{O}(m))$, which should match the number of harmonic/zero modes of the above Laplacian. On $\bP^N$, these cohomologies can be computed using the Bott formula:\footnote{More generally, the dimensions of $H_{\bar{\partial}_{V}}^{p,q}(X,\mathcal{O}(m))$ can be computed using Macaulay2~\cite{Macaulay2}. For example, defining $\bP^{3}$ using \texttt{C4 = QQ{[}x0,x1,x2,x3{]}} and \texttt{P3 = Proj C4}, the cohomologies can be computed using the command \texttt{HH\textasciicircum q(cotangentSheaf(p,P3){*}{*}OO\_P3(q)}.}
\begin{equation}
\begin{aligned}\dim H^{0}(\bP^{N},\mathcal{O}(m)) & =\binom{N+m}{m},\qquad m\geq0,\\
\dim H^{N}(\bP^{N},\mathcal{O}(m)) & =\binom{-m-1}{-N-m-1},\qquad m\leq-N-1,\\
\dim H^{p}(\bP^{N},\mathcal{O}(m)) & =0,\qquad\text{otherwise}.
\end{aligned}
\label{eq:Bott}
\end{equation}
In other words, the only non-vanishing cohomologies are $H^{0}(\bP^{N},\mathcal{O}(m))$ for $m\geq0$ and $H^{N}(\bP^{N},\mathcal{O}(m))$ for $m\leq-N-1$. On $\bP^3$,  we see that $\Omega^{0,0}(\mathcal{O}(m))$ will have zero modes, i.e.~$\lambda=0$, for $m\geq0$, while there are no zero modes at all for $\Omega^{0,1}(\mathcal{O}(m))$.

Away from zero modes, since both the Kähler metric on $\bP^{N}$ and Hermite--Einstein metric on $\mathcal{O}(m)$ are known, one might expect that one can solve for the full spectrum. Indeed, though this exact problem does not seem to have been considered in the literature before, there is a related problem from which we can extract the spectrum (at least for the scalar eigenmodes). Kuwabara~\cite{10.2748/tmj/1178228026} and Bykov and Smilga~\cite{2302.11691} analysed the spectrum of a Schrödinger operator on a line bundle $\mathcal{O}(m)$ over $N$-dimensional complex projective space equipped with a Fubini--Study metric with volume $(4\pi)^{N}/N!$. Given a connection $D$ on $\mathcal{O}(m)$ with curvature $F$, they showed that the spectrum of the Bochner Laplacian $\Delta_D=DD^{\dagger}+D^{\dagger}D$ acting on $(0,0)$-forms (scalars) is spanned by the following eigenvalues $\lambda_{n}$ with multiplicities $\ell_{n}$ for $n=0,1,\dots$:
\begin{align}
\lambda_{n} & =\left(n+\frac{|m|}{2}\right)\left(n+\frac{|m|}{2}+N\right)-\frac{m^{2}}{4},\\
\ell_{n} & =\begin{pmatrix}n+N-1\\
N-1
\end{pmatrix}\begin{pmatrix}n+|m|+N-1\\
N-1
\end{pmatrix}\frac{2n+|m|+N}{N}.
\end{align}

Looking back to Section \ref{subsec:The-Dolbeault-Laplacian}, from \eqref{eq:del_delb} we see that the Bochner Laplacian $\Delta_{D}\equiv\Delta_{\partial_{V}}+\Delta_{\bar{\partial}_{V}}$ acting on scalars is related to the $\bar{\partial}_{V}$-Laplacian as $\Delta_{D}=2\Delta_{\bar{\partial}_{V}}+\Lambda F$, where $\Lambda$ is contraction with the Kähler form on $X$. The fibre metric on $\mathcal{O}(m)$ is taken to be the unique Hermite--Einstein metric \eqref{eq:bundle_P3}, so that $F=\tfrac{1}{2}m\omega$, which then implies $\Lambda F=Nm/2$.\footnote{These somewhat unexpected coefficients come from the difference between the usual normalisation in algebraic geometry of $\int_{\bP^{N}}\omega^{N}=1$ vs $\int_{\bP^{N}}\omega^{N}=(4\pi)^{N}$ which is implied by the volume conventions of \cite{10.2748/tmj/1178228026,2302.11691}.} Thus, we expect the spectrum of the Dolbeault Laplacian to be given in terms of the Bochner Laplacian as
\begin{equation}
\Delta_{\bar{\partial}_{V}}=\frac{1}{2}\left(\Delta_{D}-\frac{Nm}{2}\right).
\end{equation}
Finally, our convention that $\Vol(\bP^{N})=1$ implies a rescaling of the eigenvalues by $4\pi/(N!)^{1/N}$. Putting this all together, we expect the $\Omega^{0,0}(\mathcal{O}(m))$ spectrum of the Dolbeault Laplacian to be given by
\begin{align}
\lambda_{n} & =\frac{2\pi}{(N!)^{1/N}}\left[n(n+N+|m|)+\frac{N(|m|-m)}{2}\right],\\
\ell_{n} & =\begin{pmatrix}n+N-1\\
N-1
\end{pmatrix}\begin{pmatrix}n+|m|+N-1\\
N-1
\end{pmatrix}\frac{2n+|m|+N}{N},
\end{align}
for $n\geq0$. As a check, we recall that the zero modes should appear with multiplicity predicted by \eqref{eq:Bott}. Indeed, for $n=0$ the above reduces to
\begin{equation}
\lambda_{0}=\frac{2\pi}{(N!)^{1/N}}\left[\frac{N(|m|-m)}{2}\right],\qquad\ell_{0}=\binom{N+|m|}{|m|},
\end{equation}
so that one has zero modes, $\lambda=0$, only for $m\geq0$ with multiplicities agreeing with the Bott formula~\eqref{eq:Bott}.

Using this exact expression, we give the first few eigenvalues in the spectrum for $N=3$ and $m\in\{-3,\dots,3\}$ in Table \ref{tab:P3_eigens}. The zero modes of $\Delta_{\bar{\partial}_{V}}$ are easy to understand -- they are the global holomorphic sections of $\mathcal{O}(m)$ given by symmetric monomials of the homogeneous coordinates. As we mentioned above, on $\bP^{N}$, there are $\binom{N+m}{m}$ of these, in agreement with the number of zero modes in Table \ref{tab:P3_eigens}. Furthermore, using the language of \cite{Ashmore:2020ujw}, the multiplicities of all the modes actually correspond to the dimensions of the $\SU{N+1}$ representation defined by the highest weight
\begin{equation}
(n,\underbrace{0,\dots,0}_{\text{\ensuremath{N-2} times}},n+|m|),
\end{equation}
and the eigenvalues themselves (up to the normalisation factor) are given by the difference of Casimir invariants for the weights $(n,0,\dots,0,n+|m|)$ and $(0,0,\dots,0,|m|)$. Note that the $m=0$ eigenvalues are exactly one-half of the eigenvalues of the de Rham Laplacian $\Delta$ calculated by Ikeda and Taniguchi~\cite{IKEDATANIGUCHI,Braun:2008jp,Ashmore:2020ujw}, which one expects since for $V=\mathcal{O}$ the Dolbeault Laplacian simplifies to $\Delta_{\bar{\partial}_{V}}=\tfrac{1}{2}\Delta$.
\noindent \begin{center}
\begin{table}
\noindent \begin{centering}
\scalebox{0.7}{%
\begin{tabular}{ccccccccccccccc}
\toprule 
$m$ & \multicolumn{2}{c}{$-3$} & \multicolumn{2}{c}{$-2$} & \multicolumn{2}{c}{$-1$} & \multicolumn{2}{c}{0} & \multicolumn{2}{c}{1} & \multicolumn{2}{c}{2} & \multicolumn{2}{c}{3}\tabularnewline
\midrule 
$n$ & $\lambda_{n}$ & $\ell_{n}$ & $\lambda_{n}$ & $\ell_{n}$ & $\lambda_{n}$ & $\ell_{n}$ & $\lambda_{n}$ & $\ell_{n}$ & $\lambda_{n}$ & $\ell_{n}$ & $\lambda_{n}$ & $\ell_{n}$ & $\lambda_{n}$ & $\ell_{n}$\tabularnewline
\midrule
\midrule 
0 & 31.1 & 20 & 20.7 & 10 & 10.4 & 4 & 0 & 1 & 0 & 4 & 0 & 10 & 0 & 20\tabularnewline
1 & 55.3 & 120 & 41.5 & 70 & 27.7 & 36 & 13.8 & 15 & 17.3 & 36 & 20.7 & 70 & 24.2 & 120\tabularnewline
2 & 86.4 & 420 & 69.2 & 270 & 51.9 & 160 & 34.6 & 84 & 41.5 & 160 & 48.4 & 270 & 55.3 & 420\tabularnewline
3 & 124 & 1120 & 104 & 770 & 83.0 & 500 & 62.2 & 300 & 72.6 & 500 & 83.0 & 770 & 93.4 & 1120\tabularnewline
4 & 169 & 2520 & 145 & 1820 & 121 & 1260 & 96.8 & 825 & 111 & 1260 & 124 & 1820 & 138 & 2520\tabularnewline
\bottomrule
\end{tabular}}
\par\end{centering}
\caption{Exact eigenvalues of $\Delta_{\bar{\partial}_{V}}$ and their multiplicities for $\mathcal{O}(m)$-valued scalars on $\protect\bP^{3}$. \label{tab:P3_eigens}}
\end{table}
\par\end{center}

In the rest of this section, we will lay out how to construct an approximate basis of bundle-valued forms which we use to compute matrix elements of the Dolbeault Laplacian. We will then compute the spectra of bundle-valued $(0,0)$- and $(0,1)$-forms, and compare these numerical results with the exact expressions given in Table \ref{tab:P3_eigens}. Note that we have not found exact expressions for the spectrum of $\mathcal{O}(m)$-valued $(0,1)$-forms on $\bP^{N}$ in the literature. Instead, as a check of this spectrum, we will appeal to Serre duality \eqref{eq:serre} which implies that the $\mathcal{O}(m)$-valued $(0,1)$-form spectrum should agree with the $K_{X}\otimes\mathcal{O}(-m)$-valued $(0,N-1)$-form spectrum. In particular, for $N=3$ we have $K_{X}\simeq\mathcal{O}(-4)$, so that the $\mathcal{O}(-m)$-valued $(0,1)$-form spectrum should agree with the honest $(0,2)$-form spectrum computed in  previous work~\cite{Ashmore:2020ujw}.

\subsection{An approximate basis\label{subsec:An-approximate-basis}}

We now want to find a basis of bundle-valued forms which can be used to approximate the space of eigenmodes of $\Delta_{\bar{\partial}_{V}}$ and calculate the spectrum via matrix elements such as \eqref{eq:gen_eigen}. We first consider bundle-valued scalars for $m\geq0$. Building on the work of \cite{Braun:2008jp,Ashmore:2020ujw}, we note that the set
\begin{equation}
\mathcal{F}_{k_{\phi}}^{0,0}(m)\equiv\frac{(\text{degree \ensuremath{k_{\phi}+m} monomials in \ensuremath{Z^{I}}})\otimes\overline{(\text{degree \ensuremath{k_{\phi}} monomials in \ensuremath{Z^{I}}})}}{(Z^{I}\bar{Z}_{I})^{k_{\phi}}},\label{eq:approx_basis}
\end{equation}
gives a finite set of $\mathcal{O}(m)$-valued scalar functions $\alpha_{A}$ on $\bP^{N}$, with the size of the set controlled by the non-negative integer parameter $k_{\phi}$. Under the scaling $Z^{I}\to \nu Z^{I}$, the scalars transform as $\alpha_{A}\to\nu^{m}\alpha_{A}$, and so they are naturally thought of as \emph{smooth} sections of $\mathcal{O}(m)$.
Upon increasing the degree $k_{\phi}$, one has a series of inclusions
\begin{equation}
\mathcal{F}_{0}^{0,0}(m)\subset F_{1}^{0,0}(m)\subset\dots\subset\Omega^{0,0}(\mathcal{O}(m)),
\end{equation}
where $\mathcal{F}_{0}^{0,0}(m)\simeq H^{0}(X,\mathcal{O}(m))$, so that larger values of $k_{\phi}$ better approximate the (infinite-dimensional) space of $\mathcal{O}(m)$-valued scalar functions on $\bP^{N}$, and so also the space of eigenfunctions of $\Delta_{\bar{\partial}_{V}}$. One recovers $\Omega^{0,0}(\mathcal{O}(m))$ only in the $k_\phi\to\infty$ limit. In fact, the eigenfunctions of $\Delta_{\bar{\partial}_{V}}$ on $\bP^{N}$ are given by finite linear combinations of these functions~\cite{10.2748/tmj/1178228026,2302.11691} at each degree, with $\mathcal{F}_{k_{\phi}}^{0,0}(m)$ spanning up to and including the $k_{\phi}$-th eigenspace. It is in this sense that an expansion in $\alpha_{A}\in\mathcal{F}_{k_{\phi}}^{0,0}(m)$ should be thought of as a spectral expansion on projective space.

As discussed elsewhere by one of the authors~\cite{Ashmore:2020ujw}, there is a generalisation of \eqref{eq:approx_basis} to give a finite set of $(p,q)$-forms at degree $k_{\phi}$ on $\bP^{N}$. As we review in Appendix \ref{sec:Differential-forms-on}, these are constructed by considering forms which are well defined on $\bP^{N}$ under both the $\bR^{+}$ and $\Uni 1$ action on the homogeneous coordinates. A simple extension of this produces a set $\mathcal{F}_{k_{\phi}}^{p,q}(m)$ of $\mathcal{O}(m)$-valued $(p,q)$-forms on $\bP^{N}$ for $m\geq0$:
\begin{equation}
\mathcal{F}_{k_{\phi}}^{p,q}(m)\equiv\frac{(\text{degree \ensuremath{k_{\phi}+m} \ensuremath{(p,0)}-forms in \ensuremath{Z^{I}}})\otimes\overline{(\text{degree \ensuremath{k_{\phi}} \ensuremath{(0,q)}-forms in \ensuremath{Z^{I}}})}}{(Z^{I}\bar{Z}_{I})^{k_{\phi}}},\label{eq:approx_basis_pq}
\end{equation}
where, for example, the degree-two $(1,0)$-forms are $\{Z^{0}\dd Z^{1}-Z^{1}\dd Z^{0},Z^{0}\dd Z^{2}-Z^{2}\dd Z^{0},\dots\}$ and so on. Unlike the scalars, there is some redundancy in this set, so one has to discard any $\alpha_{A}\in\mathcal{F}_{k_{\phi}}^{p,q}(m)$ which can be written as linear combinations of the remaining forms. Again, there is an inclusion of sets, $\mathcal{F}_{0}^{p,q}(m)\subset F_{1}^{p,q}(m)\subset\dots\subset\Omega^{p,q}(\mathcal{O}(m))$, so that larger values of $k_{\phi}$ will better approximate the space of eigenmodes of $\Delta_{\bar{\partial}_{V}}$.

What about for $m<0$? What kind of basis should we use in this case? Instead of simply writing it down, we note that the fibre metric \eqref{eq:bundle_P3} on $\mathcal{O}(1)$ pairs 
\begin{equation}
G\colon\mathcal{O}(1)\times\overline{\mathcal{O}(1)}\to\bC.
\end{equation}
Here, our notation is that sections of $\mathcal{O}(1)$, $\mathcal{O}(-1)$ and $\overline{\mathcal{O}(1)}$ transform with factors of $\nu$, $\nu^{-1}$ and $\bar{\nu}$ respectively under the scaling $Z^{I}\to\nu Z^{I}$ . There is also a natural pairing (without needing a metric) between $\mathcal{O}(1)$ and its dual bundle $\mathcal{O}(1)^{*}\simeq\mathcal{O}(-1)$ such that $\mathcal{O}(1)\times\mathcal{O}(-1)\to\bC.$ Combining these two, we see there is a map between smooth sections of $\mathcal{O}(-1)$ and $\overline{\mathcal{O}(1)}$ of the form
\begin{equation}
\begin{aligned}\overline{\mathcal{O}(1)} & \to\mathcal{O}(-1)\\
f(\bar{Z}) & \mapsto\frac{f(\bar{Z})}{Z^{I}\bar{Z}_{I}},
\end{aligned}
\end{equation}
where $G=(Z^I \bar{Z}_I)^{-1}$ is the Hermite--Einstein metric on $\mathcal{O}(1)$. Given $f(\bar{Z})\mapsto\bar{\nu}f(\bar{Z})$ under the scaling of homogeneous coordinates on $\bP^{N}$, we have $f(\bar{Z})(Z^{I}\bar{Z}_{I})^{-1}\mapsto\nu^{-1}f(\bar{Z})(Z^{I}\bar{Z}_{I})^{-1}$, and so it transforms as a (smooth) section of $\mathcal{O}(-1)$. Extending this logic to all $m<0$, this means the basis can be taken to be
\begin{equation}
\mathcal{F}_{k_{\phi}}^{p,q}(-|m|)\equiv\frac{(\text{degree \ensuremath{k_{\phi}} \ensuremath{(p,0)}-forms in \ensuremath{Z^{I}}})\otimes\overline{(\text{degree \ensuremath{k_{\phi}+|m|} \ensuremath{(0,q)}-forms in \ensuremath{Z^{I}}})}}{(Z^{I}\bar{Z}_{I})^{k_{\phi}+|m|}},\label{eq:approx_basis_pq_q<0}
\end{equation}
where again one should discard any elements that are linearly dependent on the remaining forms. Denoting this basis by $\mathcal{F}_{k_{\phi}}^{p,q}(m)\equiv\mathcal{F}_{k_{\phi}}^{p,q}(-|m|)$ for $m<0$, there is again an inclusion of sets, $\mathcal{F}_{0}^{p,q}(m)\subset F_{1}^{p,q}(m)\subset\dots\subset\Omega^{p,q}(\mathcal{O}(m))$, so that the eigenmodes of $\Delta_{\bar{\partial}_{V}}$ on $\bP^N$ for $m<0$ are again given by finite linear combinations of these forms.

For $(p,q)=(0,0)$ and $k_{\phi}=0$, the set $\mathcal{F}_{0}^{0,0}(-|m|)$ reduces to
\begin{equation}
\mathcal{F}_{0}^{0,0}(-|m|)=\frac{\overline{(\text{degree \ensuremath{|m|} monomials in \ensuremath{Z^{I}}})}}{(Z^{I}\bar{Z}_{I})^{|m|}},
\end{equation}
which are never holomorphic, consistent with the absence of zero modes for $m<0$ from \eqref{eq:Bott}. Similarly, for $(p,q)=(0,N)$ and $k_{\phi}=0$, $\mathcal{F}_{0}^{0,N}(-|m|)$ is spanned by
\begin{equation}
\mathcal{F}_{0}^{0,N}(-|m|)=\frac{\overline{(\text{degree \ensuremath{|m|} \ensuremath{(0,N)}-forms in \ensuremath{Z^{I}}})}}{(Z^{I}\bar{Z}_{I})^{|m|}},
\end{equation}
where one includes only those that are linearly independent on $\bP^{N}$. These are automatically $\bar{\partial}_{V}$-closed since there are no $(0,N+1)$-forms on a complex $N$-fold, and they actually give a basis for $H^{N}(\bP^{N},\mathcal{O}(-|m|))$ since they are also $\bar{\partial}_{V}^{\dagger}$-closed:
\begin{equation}
\bar{\partial}_{V}^{\dagger}\alpha_{A}\propto\star_{V}\partial\left((Z^{I}\bar{Z}_{I})^{|m|}\alpha_{A}\right)=0.
\end{equation}
The number of these forms on $\bP^{N}$ is $\dim\mathcal{F}_{0}^{0,N}(-|m|)=\binom{|m|-1}{|m|-N-1}$, again in agreement with the Bott formula.

As a check that our conventions for the Fubini--Study K\"ahler potential and so on are consistent with the exact results of Section \ref{subsec:Analytic-results}, in Appendix \ref{app:exact_O1} we use the basis constructed in \eqref{eq:approx_basis} to compute the first non-zero eigenvalue of the Dolbeault Laplacian for $\mathcal{O}(1)$-valued scalars. We find a perfect match between this calculation and the exact and numerical results of this section.

A final point: the attentive reader may have noticed that there has no been any mention of a local holomorphic frame for $\mathcal{O}(m)$. As we comment on in Appendix \ref{subsec:A-local-holomorphic}, one can introduce such a local frame and show that the integrals, matrix elements, etc.~are independent of the choice of frame. It turns out that line bundles on projective space are simple enough to write down expressions using \emph{global} sections, as in \eqref{eq:approx_basis}, with no need to work locally. This subtlety, however, cannot be sidestepped if one moves to non-abelian bundles for which global sections often do not exist.

\subsection{Numerical results}

Before presenting our numerical results, we recall the essential ingredients for computing the matrix elements $\Delta_{AB}$ and $O_{AB}$ in \eqref{O-AB} and \eqref{del-AB} that determine the generalised eigenvalue problem for the spectrum. Since descriptions of point sampling, discretisation and Monte Carlo methods for numerical metrics have appeared in many other works, we will be brief. More details can be found in the literature~\cite{Douglas:2006rr,Douglas:2006hz,Braun:2007sn,Braun:2008jp,Headrick:2009jz,Anderson:2010ke,Anderson:2011ed,Ashmore:2019wzb,Cui:2019uhy,Anderson:2020hux,Douglas:2020hpv,Jejjala:2020wcc,Ashmore:2020ujw,Larfors:2021pbb,Ashmore:2021ohf,Ashmore:2021qdf,Larfors:2022nep,Gerdes:2022nzr,Berglund:2022gvm,Cui:2023eqr,Ahmed:2023cnw}. 

One begins by choosing a truncated basis $\mathcal{F}_{k_{\phi}}^{0,0}(m)=\{\alpha_{A}\}$ of bundle-valued forms for some degree $k_{\phi}$. Larger values of $k_{\phi}$ will give larger matrices which better approximate the action of the Laplacian on the space of bundle-valued forms. The matrix elements $\Delta_{AB}$ and $O_{AB}$ are then computed relative to this basis by Monte Carlo integration on $\bP^{3}$, where integrals over projective space are approximated by summing over $N_\phi$ random points $p_{i}\in\mathbb{P}^3$ according to
\begin{equation}
\int_{\bP^3}f\vol\simeq\frac{1}{N_{\phi}}\sum_{i=1}^{N_\phi}f(p_{i}).\label{eq:discretise}
\end{equation}
Here, $\vol$ is the volume form associated to the Fubini--Study metric on $\bP^{3}$ and the distribution of the random points is chosen to reproduce this measure.\footnote{In a little more detail, if one picks random points distributed uniformly with respect to the $\SU 4$ action on $\bP^{3}$, the resulting measure is that of the Fubini--Study metric~\cite{Douglas:2006rr,Douglas:2006hz}. } With $\Delta_{AB}$ and $O_{AB}$ in hand, one computes the eigenvalues and eigenvectors using, for example, the Mathematica function \texttt{Eigensystem{[}\{Delta,O\}{]}}. The eigenvalues are the $\lambda$ which appear in \eqref{eq:eigenvalue_problem}, with the eigenvectors determining the eigenmodes $\phi$ in \eqref{eq:eigenvalue_problem} in the chosen basis $\{\alpha_{A}\}$.

Before moving to the results, we make a quick comment on the dependence on the number of integration points $N_\phi$ used to approximate integrals. The exact results in Section \ref{subsec:Analytic-results} showed that the eigenspaces of the Dolbeault Laplacian have dimensions given by $\SU 4$ representations. However, the finite point sampling explicitly breaks the $\SU 4$ symmetry. As we will see, this leads to eigenvalues which cluster around the analytic results but are not exactly degenerate. As the number of integration points is taken to infinity, the $\SU 4$ symmetry is restored and the spread of eigenvalues in a cluster decreases to zero, and so one expects that larger values of $N_\phi$ will better reproduce the exact degeneracies of the analytic results.

\subsubsection{The bundle-valued scalar spectrum}

\begin{figure}
\includegraphics{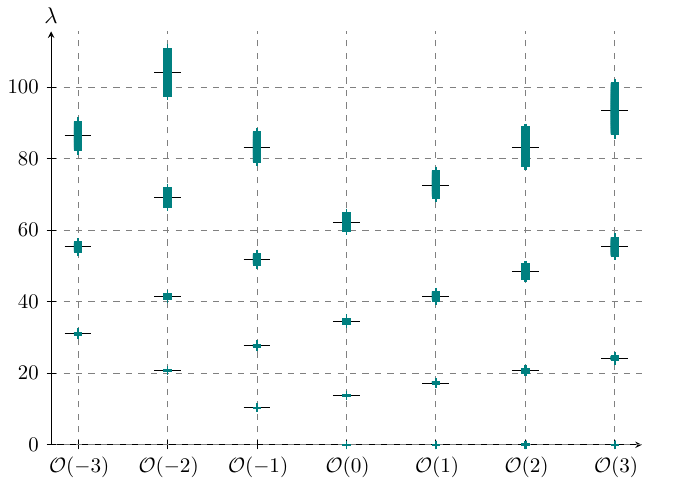}
\caption{Numerical eigenvalues $\lambda_{n}$ of $\Delta_{\bar{\partial}_{V}}$ on $\protect\bP^{3}$ acting on $\mathcal{O}(m)$-valued scalars for $m\in\{-3,\dots,3\}$. These were computed using the Fubini--Study metric on $\protect\bP^{3}$ and the associated Hermite--Einstein metric on $\mathcal{O}(m)$. Integrals were computed via Monte Carlo over $N_\phi={10}^{6}$ points. We used $k_{\phi}=3$ for the basis functions, giving us access to the first four eigenspaces. The horizontal black lines indicate the exact analytic values from Table \ref{tab:P3_eigens}. \label{fig:P3_scalar_graph}}
\end{figure}

We begin with a numerical calculation of the spectrum of bundle-valued eigenfunctions of $\Delta_{\bar{\partial}_{V}}$. The inputs are the exact Fubini--Study metric on $\bP^{3}$ determined by the Kähler potential in \eqref{eq:FS}, a bundle $\mathcal{O}(m)$ together with a hermitian structure determined by the Hermite--Einstein metric \eqref{eq:bundle_P3}, a choice of degree $k_{\phi}$ which determines the size of the approximate basis \eqref{eq:approx_basis} in which we expand the eigenfunctions, and the number of points $N_\phi$ that are used to discretise the integrals that appear in the matrix elements of the Laplacian. For the rest of this section, we fix $k_{\phi}=3$ and $N_\phi={10}^{6}$ and compute the spectrum for $m\in\{-3,\dots,3\}$. The results are shown in Table \ref{tab:P3_scalar_table} and Figure \ref{fig:P3_scalar_graph}.

We see that the numerical results in Table \ref{tab:P3_scalar_table} reproduce the exact results in Table \ref{tab:P3_eigens} with excellent precision and the correct multiplicities. In particular, the mean of the numerical eigenvalues in each cluster match the exact results to better than 1\% in all cases. One can also see this from Figure \ref{fig:P3_scalar_graph} which shows the numerical results and indicates the values of the exact eigenvalues; in all cases, the exact result is in the middle of the cluster of numerical eigenvalues. For $m=0$, these exactly match the numerical spectrum given by Ikeda and Taniguchi~\cite{IKEDATANIGUCHI,Braun:2008jp,Ashmore:2020ujw} after dividing by a factor of two to account for the difference between the de Rham Laplacian and the Dolbeault Laplacian. For $m>0$, one expects the zero modes of $\Delta_{\bar{\partial}_{V}}$ to be given by monomials of degree $m$ in the homogeneous $Z^{I}$ coordinates. The counting of these monomials, which is simply $\dim H^{0}(\bP^{3},\mathcal{O}(m))$, agrees with the number of zero modes we get in each case. For $m<0$, the numerical results indicate there are no zero modes, in agreement with $H^{0}(\bP^{3},\mathcal{O}(m))=\{0\}$ for $m<0$ from the Bott formula \eqref{eq:Bott}. 

As an additional check, since $\bar{\star}_V$ commutes with the Laplacian and the canonical bundle of $\bP^3$ is $K_{\bP^3}=\mathcal{O}(-4)$, the relations in \eqref{eq:serre} imply that the $\mathcal{O}(-4)$-valued scalar spectrum should agree with (one-half of) the honest $(0,3)$-form spectrum, which was calculated exactly in \cite{IKEDATANIGUCHI}. We have calculated this spectrum at $k_{\phi}=2$ and found the first three eigenvalues to be $(41.4\pm0.3,69\pm1,104\pm3)$, which agree well with the exact values of $(41.5,69.2,103.8)$ for the $(0,3)$-form spectrum; the multiplicities also match.

As we have explained, the degree $k_{\phi}$ controls the number of eigenvalues that one is computing, while the number of integration points $N_\phi$ controls how well we recover the $\SU 4$ symmetry of the underlying Fubini--Study metric. For larger values of $N_\phi$, the eigenvalues in Figure \ref{fig:P3_scalar_graph} become more tightly clustered, eventually becoming exactly degenerate in the $N_\phi\to\infty$ limit.

\begin{table}
\noindent \begin{centering}
\scalebox{0.7}{%
\begin{tabular}{ccccccccccccccc}
\toprule 
$m$ & \multicolumn{2}{c}{$-3$} & \multicolumn{2}{c}{$-2$} & \multicolumn{2}{c}{$-1$} & \multicolumn{2}{c}{0} & \multicolumn{2}{c}{1} & \multicolumn{2}{c}{2} & \multicolumn{2}{c}{3}\tabularnewline
\midrule 
$n$ & $\lambda_{n}$ & $\ell_{n}$ & $\lambda_{n}$ & $\ell_{n}$ & $\lambda_{n}$ & $\ell_{n}$ & $\lambda_{n}$ & $\ell_{n}$ & $\lambda_{n}$ & $\ell_{n}$ & $\lambda_{n}$ & $\ell_{n}$ & $\lambda_{n}$ & $\ell_{n}$\tabularnewline
\midrule
\midrule 
0 & $31.1\pm0.2$ & 20 & $20.7\pm0.1$ & 10 & $10.37\pm0.02$ & 4 & 0 & 1 & 0 & 4 & 0 & 10 & 0 & 20\tabularnewline
1 & $55.2\pm0.7$ & 120 & $41.4\pm0.4$ & 70 & $27.6\pm0.2$ & 36 & $13.8\pm0.1$ & 15 & $17.3\pm.1$ & 36 & $20.7\pm0.2$ & 70 & $24.2\pm0.3$ & 120\tabularnewline
2 & $86\pm2$ & 420 & $69\pm1$ & 270 & $51.8\pm0.8$ & 160 & $34.6\pm0.4$ & 84 & $41.5\pm0.6$ & 160 & $48.3\pm0.9$ & 270 & $55\pm1$ & 420\tabularnewline
3 & $125\pm5$ & 1120 & $104\pm3$ & 770 & $83\pm2$ & 500 & $62.3\pm1.3$ & 300 & $72.7\pm1.9$ & 500 & $83\pm3$ & 770 & $94\pm4$ & 1120\tabularnewline
\bottomrule
\end{tabular}}
\par\end{centering}
\caption{Numerical eigenvalues $\lambda_{n}$ of $\Delta_{\bar{\partial}_{V}}$ on $\protect\bP^{3}$ acting on $\mathcal{O}(m)$-valued scalars for $m\in\{-3,\dots,3\}$. We have also included their multiplicities $\ell_{n}$. The quoted eigenvalues are the mean of the eigenvalues in a cluster, with the error given by the standard deviation of the cluster. We used $k_{\phi}=3$ to allow us to compute the first four eigenspaces.\label{tab:P3_scalar_table}}
\end{table}

\subsubsection{The bundled-valued $(0,1)$-form spectrum}

\begin{figure}
\includegraphics{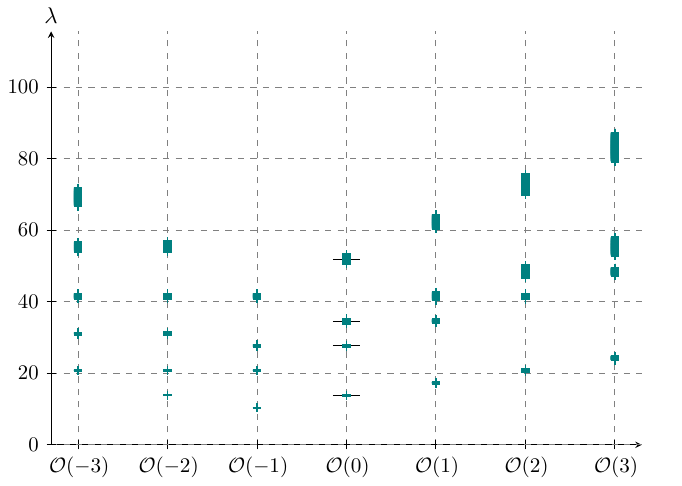}
\caption{Numerical eigenvalues $\lambda_{n}$ of $\Delta_{\bar{\partial}_{V}}$ on $\protect\bP^{3}$ acting on $\mathcal{O}(m)$-valued $(0,1)$-forms for $m\in\{-3,\dots,3\}$. These were computed using the Fubini--Study metric on $\protect\bP^{3}$ and the associated Hermite--Einstein metric on $\mathcal{O}(m)$. Integrals were computed via Monte Carlo over $N_\phi={10}^{6}$ points. We used $k_{\phi}=2$ for $m<0$ and $k_{\phi}=3$ for $m\protect\geq0$ to allow us to compute the first four eigenspaces. The horizontal black lines indicate the exact analytic values for the $\Omega^{0,1}(X)$ spectrum from \cite{IKEDATANIGUCHI}. \label{fig:P3_01_graph}}
\end{figure}

Next, we have the numerical calculation of the $\Omega^{0,1}(\mathcal{O}(m))$ spectrum. This follows the scalar calculation almost exactly apart from using an appropriate basis of bundle-valued $(0,1)$-forms from \eqref{eq:approx_basis_pq}. The results for $m\in\{-3,\dots,3\}$ are shown in Table \ref{tab:P3_01_table} and Figure \ref{fig:P3_01_graph}.

Unlike the scalar spectrum, we do not have complete exact results to compare with. For $m=0$, our results match the exact spectrum given in \cite{IKEDATANIGUCHI,Ashmore:2020ujw} after dividing by a factor of two to account for the difference between the de Rham Laplacian and the Dolbeault Laplacian. There are no zero modes for any values of $m$, in agreement with the Bott formula \eqref{eq:Bott}. As we observed for the scalar spectrum, the relations in \eqref{eq:serre} imply that the $\mathcal{O}(-4)$-valued $(0,1)$-form spectrum should agree with (one-half of) the honest $(0,2)$-form spectrum, calculated exactly in \cite{IKEDATANIGUCHI}. We have calculated this spectrum at $k_{\phi}=1$ and found the first three eigenvalues to be $(27.7\pm0.2,41.5\pm0.3,51.9\pm0.7)$, which agree well with the exact values of $(27.7,41.5,51.9)$ for the $(0,2)$-form spectrum; the multiplicities also agree.

As additional evidence that the spectra are correct, we recall that the $\bar{\partial}_{V}$ Hodge decomposition implies that a non-zero mode of $\Delta_{\bar{\partial}_{V}}$ must be either $\bar{\partial}_{V}$- or $\bar{\partial}_{V}^{\dagger}$-exact. Specifically, an $\mathcal{O}(m)$-valued $(0,1)$ eigenmode $\phi$ of the Laplacian with non-zero eigenvalue can be written as
\begin{equation}
\phi=\bar{\partial}_{V}\beta+\bar{\partial}_{V}^{\dagger}\gamma,\label{eq:hodge}
\end{equation}
where $\beta$ and $\gamma$ are $\mathcal{O}(m)$-valued scalar and $(0,2)$-forms respectively. Crucially, since $\Delta_{\bar{\partial}_{V}}$ commutes with both $\bar{\partial}_{V}$ and $\bar{\partial}_{V}^{\dagger}$, $\beta$ and $\gamma$ will also be eigenmodes of the Laplacian with the \emph{same }eigenvalue as $\phi$. From this we see that the $\Omega^{0,1}(\mathcal{O}(m))$ spectrum must be some combination of the $\Omega^{0,0}(\mathcal{O}(m))$ and $\Omega^{0,2}(\mathcal{O}(m))$ spectra. In fact, using a further Hodge decomposition for $\beta$ and $\gamma$, it is simple to see that the $\Omega^{0,1}(\mathcal{O}(m))$ spectrum should consist of the entire $\Omega^{0,0}(\mathcal{O}(m))$ non-zero mode spectrum plus the $\Omega^{0,2}(\mathcal{O}(m))$ eigenvalues whose eigenmodes are $\bar{\partial}_{V}$-exact. We then have one final simplification: since $\bar{\star}_{V}$ commutes with the Laplacian, \eqref{eq:serre} implies that the $\Omega^{0,2}(\mathcal{O}(m))$ and $\Omega^{0,1}(\mathcal{O}(-4-m))$ spectra should match. 

We can check these claims for the numerical spectrum we have calculated. For the $\bar{\partial}_{V}$-exact modes in \eqref{eq:hodge}, comparing Tables \ref{tab:P3_scalar_table} and \ref{tab:P3_01_table}, we see that, for example, for $m=-2$ the eigenvalues $(20.7,41.4)$ (to within numerical accuracy) appear in both the $(0,1)$ and $(0,0)$ spectra with the same multiplicities. A cursory glance at the rest of the results should assure the reader that this holds for the other values of $m$, with all the $(0,0)$ eigenvalues appearing in the $(0,1)$ spectra. For the $\bar{\partial}_{V}^{\dagger}$-exact modes in \eqref{eq:hodge}, for $m=-1$, for example, we expect that the remaining $\mathcal{O}(-1)$-valued $(0,1)$ eigenvalues should come from roughly half of $\mathcal{O}(-3)$-valued $(0,1)$ spectrum. Indeed, looking at Table \ref{tab:P3_01_table}, we see that both the contain the eigenvalues $(20.7,41.5)$ with the same multiplicities. Together, these constitute a non-trivial check that our numerical algorithm is correct for both the scalar and $(0,1)$ modes.

\begin{table}
\noindent \begin{centering}
\scalebox{0.7}{%
\begin{tabular}{ccccccccccccccc}
\toprule 
$m$ & \multicolumn{2}{c}{$-3$} & \multicolumn{2}{c}{$-2$} & \multicolumn{2}{c}{$-1$} & \multicolumn{2}{c}{0} & \multicolumn{2}{c}{1} & \multicolumn{2}{c}{2} & \multicolumn{2}{c}{3}\tabularnewline
\midrule 
$n$ & $\lambda_{n}$ & $\ell_{n}$ & $\lambda_{n}$ & $\ell_{n}$ & $\lambda_{n}$ & $\ell_{n}$ & $\lambda_{n}$ & $\ell_{n}$ & $\lambda_{n}$ & $\ell_{n}$ & $\lambda_{n}$ & $\ell_{n}$ & $\lambda_{n}$ & $\ell_{n}$\tabularnewline
\midrule
\midrule 
0 & $20.74\pm0.09$ & 20 & $13.83\pm0.03$ & 6 & $10.37\pm0.03$ & 4 & $13.8\pm0.1$ & 15 & $17.3\pm0.1$ & 36 & $20.7\pm0.2$ & 70 & $24.2\pm0.3$ & 120\tabularnewline
1 & $31.1\pm0.2$ & 20 & $20.74\pm0.09$ & 10 & $20.74\pm0.09$ & 20 & $27.7\pm0.2$ & 45 & $35.6\pm0.3$ & 84 & $41.4\pm0.4$ & 140 & $48.3\pm0.6$ & 216\tabularnewline
2 & $41.5\pm0.4$ & 140 & $31.1\pm0.2$ & 64 & $27.7\pm0.2$ & 36 & $34.6\pm0.4$ & 84 & $41.5\pm0.6$ & 160 & $48.4\pm0.9$ & 270 & $55\pm1$ & 420\tabularnewline
3 & $55.3\pm0.7$ & 120 & $41.5\pm0.4$ & 70 & $41.5\pm0.4$ & 140 & $51.9\pm0.7$ & 256 & $62\pm1$ & 420 & $73\pm2$ & 640 & $83\pm2$ & 924\tabularnewline
\bottomrule
\end{tabular}}
\par\end{centering}
\caption{Numerical eigenvalues $\lambda_{n}$ of $\Delta_{\bar{\partial}_{V}}$ on $\protect\bP^{3}$ acting on $\mathcal{O}(m)$-valued $(0,1)$-forms for $m\in\{-3,\dots,3\}$. We have also included their multiplicities $\ell_{n}$. The quoted eigenvalues are the mean of the eigenvalues in a cluster, with the error given by the standard deviation of the cluster. We used $k_{\phi}=2$ for $m<0$ and $k_{\phi}=3$ for $m\protect\geq0$.\label{tab:P3_01_table}}
\end{table}

\section{The torus as a Calabi--Yau one-fold\label{sec:The-torus-as-CY1}}

We now apply our numerical method to calculate the spectrum of bundle-valued scalars and $(0,1)$-forms on Calabi--Yau manifolds. As a warm-up, and as another example where we can check things analytically, we consider a Calabi--Yau one-fold (a torus) defined by a single cubic equation in $\bP^{2}$. As we discuss, the spectrum can be computed analytically and so provides a non-trivial check of our numerical results in the case of a hypersurface in projective space. Moving to Calabi--Yau three-folds is then just a matter of changing the dimension of the ambient projective space and the defining equation of the hypersurface (the algorithm does not change in any other way). With confidence that our algorithm is correct, in the next section we move to the more involved and physically relevant case of a Calabi--Yau three-fold.

The particular one-fold that we will study is the Fermat cubic hypersurface $X$ in $\bP^{2}$ defined by the vanishing locus of the equation\footnote{See \cite{Ahmed:2023cnw} for a nice discussion of the map between the description as a cubic hypersurface and a flat torus with complex structure $\tau$.}
\begin{equation}
Q\equiv Z_{0}^{3}+Z_{1}^{3}+Z_{2}^{3}=0.\label{eq:defining}
\end{equation}
This particular choice of defining equation corresponds to the ``equilateral torus''~\cite{B_rard_2016}, which will allow us to compare the spectrum with known results.

\subsection{Analytic results}

First, we note that since the canonical bundle of a Calabi--Yau is trivial, $K_{X}=\mathcal{O}$, the relations in \eqref{eq:serre} imply $\{\lambda\}_{V}^{0,0}=\{\lambda\}_{V^{*}}^{0,1}$. Thanks to this, once we compute the $\mathcal{O}(m)$-valued scalar spectrum for all $m$, we automatically have the bundle-valued $(0,1)$-form spectrum. With this in mind, let us review what is known about the scalar spectrum.

For $m=0$, since $\Delta_{\bar{\partial}_{V}}=\tfrac{1}{2}\Delta$, the scalar spectrum is exactly one-half of that for the de Rham Laplacian on the torus. This is given by~\cite{Milnor64}\footnote{See also \cite[Section 3.3]{1405.4944} for recent work.}
\begin{equation}
\lambda_{u,v}=4\pi^{2}b\left[\left(1+\frac{a^{2}}{b^{2}}\right)u^{2}-\frac{2a}{b^{2}}uv+\frac{v^{2}}{b^{2}}\right],\qquad u,v\in\bZ,
\end{equation}
where the complex structure $\tau\equiv a+\ii b$ is fixed to $\ee^{2\pi\ii/3}$ for the Fermat cubic/equilateral torus. The eigenvalue multiplicities match the dimensions of irreducible representations of the symmetry group of $X$ (automorphisms of $Q$ together with complex conjugation), which is $(S_{3}\times\bZ_{2})\rtimes(\bZ_{3}\times\bZ_{3})$~\cite{Ahmed:2023cnw}. 

For $m>0$, one expects the harmonic/zero modes of $\Delta_{\bar{\partial}_{V}}$, i.e.~those with $\lambda=0$, to be simply monomials of degree $m$ in the homogeneous $Z^{I}$ coordinates modulo $Q=0$. The counting of these monomials should agree with the number of zero modes -- this is indeed the case. For example, for $m=2$, the harmonic modes are linear combinations of $Z^{I}Z^{J}$, which span a six-dimensional space. For $m=3$, $\dim\{Z^{I}Z^{J}Z^{K}\}=10$, but one of these is linearly dependent thanks to $Q\equiv0$, so we are left with nine harmonic modes.

In fact, one can go further than this zero-mode analysis. For $m\neq0$, the exact scalar spectrum can be inferred from the results of Tejero Prieto~\cite{TEJEROPRIETO2006288}. There, they compute the eigenvalues and multiplicities for a Schrödinger-like operator
\begin{equation}
\hat{H}=\frac{\hbar^{2}}{2m}\Delta_D,
\end{equation}
where $D$ is a connection compatible with the hermitian metric on $V=\mathcal{O}(m)$, and $\Delta_D=D^{\dagger}D$ is the Bochner Laplacian for $V$. This can be related to the holomorphic structure on $V$ as follows. 

Given the Dolbeault operators $\partial_{V}$ and $\bar{\partial}_{V}$, where $D=\partial_{V}+\bar{\partial}_{V}$, \cite{TEJEROPRIETO2006288} gives the identity
\begin{equation}
\partial_{V}^{\dagger}\partial_{V}-\bar{\partial}_{V}^{\dagger}\bar{\partial}_{V}=\star F=\frac{e\hat{B}}{\hbar},
\end{equation}
where $F=eB/\hbar$ is the curvature of $D$ and $B=\hat{B}\vol$. This implies
\begin{align}
D^{\dagger}D & =2\bar{\partial}_{V}^{\dagger}\bar{\partial}_{V}+\frac{e\hat{B}}{\hbar},\label{eq:DDgaer}
\end{align}
where $\hat{B}$ is related to the degree of the line bundle $V$ by
\begin{equation}
\deg V=\frac{e\hat{B}}{2\pi\hbar}\Vol(X).
\end{equation}
Remembering that our eigenvalue problem is for the Dolbeault Laplacian $\Delta_{\bar{\partial}_{V}}=\bar{\partial}_{V}^{\dagger}\bar{\partial}_{V}$, we can use \eqref{eq:DDgaer} to relate the spectrum of $\hat{H}$ calculated in \cite{TEJEROPRIETO2006288} with the spectrum of $\Delta_{\bar{\partial}_{V}}$. From Section 4.2 of that work, the spectrum (with multiplicity $\ell$) of $\hat{H}$ is given by
\begin{align}
\op{spec}\hat{H} & =\left\{ E_{n}=\frac{2\pi\hbar^{2}}{M\Vol(X)}|\deg V|\left(n+\tfrac{1}{2}\right),\,n\geq0\right\} ,\\
\ell(E_{n}) & =|\deg V|.
\end{align}
Equation \eqref{eq:DDgaer} then implies
\begin{align}
\op{spec}\bar{\partial}_{V}^{\dagger}\bar{\partial}_{V} & =\left\{ \lambda_{n}=\frac{2\pi|\deg V|}{\Vol(X)}\left(n+\tfrac{1}{2}(1-\op{sign}\deg V)\right),\,n\geq0\right\} .
\end{align}
We then recall that for a line bundle $V=\mathcal{O}(m)$ on a torus, Riemann--Roch implies that the degree is given by $\deg V=h^{0}(V)-h^{0}(V^{*})$.\footnote{For a bundle $V$ over a complex genus-$g$ Riemann surface, the Riemann--Roch theorem implies
\[
h^{0}(V)-h^{1}(V)=\deg V-(1-g)\op{rank}V.
\]
For a line bundle over a torus, $g=1=\op{rank}V$ and the canonical bundle is trivial, so that $h^{1}(V)=h^{0}(V^{*})$.} Thus, for $m>0$, we have $\deg V=h^{0}(\mathcal{O}(m))$, while for $m<0$ we have $\deg V=-h^{0}(\mathcal{O}(|m|))$, with $h^{0}(\mathcal{O}(|m|))=3|m|$. Finally, remembering that we always normalise the volume of the Calabi--Yau to one, the eigenvalues and multiplicities of $\Delta_{\bar{\partial}_{V}}$ for $m\neq0$ should be
\begin{align}
\lambda_{n} & =\begin{cases}
6\pi mn & m>0,\\
6\pi|m|(n+1) & m<0,
\end{cases}\qquad n\geq0,\\
\ell_{n} & =3|m|.
\end{align}
The spectra for $m\in\{-3,\dots,3\}$ are given in Table \ref{tab:T2_eigens}. In particular, we notice that there are no zero modes for $m<0$, in agreement with $h^{0}(\mathcal{O}(m))=0$ for a negative-degree line bundle. The $\mathcal{O}(m)$-valued $(0,1)$-form spectra are then given by the $\mathcal{O}(-m)$-valued scalar spectra, corresponding to mirroring Table \ref{tab:T2_eigens} about the $m=0$ column. These are the exact results that we will compare our numerical calculations with.
\noindent \begin{center}
\begin{table}
\noindent \begin{centering}
\scalebox{0.7}{%
\begin{tabular}{ccccccccccccccc}
\toprule 
$m$ & \multicolumn{2}{c}{$-3$} & \multicolumn{2}{c}{$-2$} & \multicolumn{2}{c}{$-1$} & \multicolumn{2}{c}{0} & \multicolumn{2}{c}{1} & \multicolumn{2}{c}{2} & \multicolumn{2}{c}{3}\tabularnewline
\midrule 
$n$ & $\lambda_{n}$ & $\ell_{n}$ & $\lambda_{n}$ & $\ell_{n}$ & $\lambda_{n}$ & $\ell_{n}$ & $\lambda_{n}$ & $\ell_{n}$ & $\lambda_{n}$ & $\ell_{n}$ & $\lambda_{n}$ & $\ell_{n}$ & $\lambda_{n}$ & $\ell_{n}$\tabularnewline
\midrule
\midrule 
0 & $56.55$ & 9 & $37.70$ & 6 & $18.85$ & 3 & $0.0$ & 1 & $0.0$ & 3 & $0.0$ & 6 & $0.0$ & 9\tabularnewline
1 & $113.1$ & 9 & $75.40$ & 6 & $37.70$ & 3 & $22.79$ & 6 & $18.85$ & 3 & $37.70$ & 6 & $56.55$ & 9\tabularnewline
2 & $169.6$ & 9 & $113.1$ & 6 & $56.55$ & 3 & $68.38$ & 6 & $37.70$ & 3 & $75.40$ & 6 & $113.1$ & 9\tabularnewline
3 & $226.2$ & 9 & $150.8$ & 6 & $75.40$ & 3 & $91.17$ & 6 & $56.55$ & 3 & $113.1$ & 6 & $169.6$ & 9\tabularnewline
4 & $282.7$ & 9 & $188.5$ & 6 & $94.25$ & 3 & $159.6$ & 12 & $75.40$ & 3 & $150.8$ & 6 & $226.2$ & 9\tabularnewline
5 & $339.3$ & 9 & $226.2$ & 6 & $113.1$ & 3 & $205.1$ & 6 & $94.25$ & 3 & $188.5$ & 6 & $282.7$ & 9\tabularnewline
6 & $396.0$ & 9 & $263.9$ & 6 & $169.6$ & 3 & $273.5$ & 6 & $113.1$ & 3 & $226.2$ & 6 & $339.3$ & 9\tabularnewline
\bottomrule
\end{tabular}}
\par\end{centering}
\caption{Exact eigenvalues of $\Delta_{\bar{\partial}_{V}}$ and their multiplicities for $\mathcal{O}(m)$-valued scalars on the Fermat cubic. The spectrum of $\mathcal{O}(m)$-valued $(0,1)$-forms is given by reflecting the table about $m=0$.\label{tab:T2_eigens}}
\end{table}
\par\end{center}

\subsection{Numerical results}

Before presenting our numerical results, we quickly outline how the calculation on a Calabi--Yau hypersurface differs from that on projective space. More details can be found in, for example, \cite{Douglas:2006rr,Douglas:2006hz,Braun:2007sn,Braun:2008jp,Headrick:2009jz,Anderson:2010ke,Anderson:2011ed,Ashmore:2019wzb,Cui:2019uhy,Anderson:2020hux,Douglas:2020hpv,Jejjala:2020wcc,Ashmore:2020ujw,Larfors:2021pbb,Ashmore:2021ohf,Ashmore:2021qdf,Larfors:2022nep,Gerdes:2022nzr,Berglund:2022gvm,Cui:2023eqr,Ahmed:2023cnw}. Practically, the salient differences are:
\begin{itemize}
\item The metric on the Calabi--Yau is not known analytically, but must be computed numerically. We compute the Calabi--Yau using the ``energy functional'' approach introduced by Headrick and Nassar~\cite{Headrick:2009jz}. In the case of the torus, the Calabi--Yau metric is simply the flat metric associated to the presentation of the torus as a quotient of $\bC$. However, this metric looks non-trivial in the coordinates inherited from the ambient projective space. Thanks to this, and also to mimic the higher-dimensional case where there are no analytic results, we will compute the metric numerically.
\item The set $F_{k_{\phi}}^{p,q}(m)$ defined in \eqref{eq:approx_basis_pq} is pulled back to the hypersurface to give an approximate basis on the Calabi--Yau. The set may be overcomplete in the sense that some elements are linearly dependent when restricted to the hypersurface. In practice, this means removing elements of $F_{k_{\phi}}^{p,q}(m)$ that are related by $Q=0$. Choosing larger values of $k_{\phi}$ corresponds to using a larger basis of forms with which to approximate the eigenmodes of the Laplacian.
\item The random points used to discretise integrals as in \eqref{eq:discretise} should be distributed according to the Calabi--Yau measure rather than the Fubini--Study measure. This problem was solved for Calabi--Yau hypersurfaces by Douglas et al.~\cite{Douglas:2006rr} and Braun et al.~\cite{Braun:2007sn}.
\end{itemize}
The metric on $X$ is given by a choice of complex structure, via the defining equation \eqref{eq:defining}, and a choice of Kähler potential. As usual, this is approximated by an ``algebraic metric''~\cite{Tian,math/0512625} with K\"ahler potential
\begin{equation}
K=\frac{1}{\pi k_{h}}\log s_{\alpha}h^{\alpha\bar{\beta}}\overline{s_{\beta}},\label{eq:approx_Kahler}
\end{equation}
where $h^{\alpha\bar{\beta}}$ is a hermitian matrix of parameters and $\{s_{\alpha}\}$ are a basis for the degree-$k_{h}$ polynomials (sections of $\mathcal{O}(k_{h})$) on $\bP^{2}$ restricted to the hypersurface. Here, $k_h$ is a positive integer parameter which controls the complexity of the ansatz \eqref{eq:approx_Kahler} -- larger values of $k_h$ should be thought of as including higher Fourier modes to better approximate the honest Calabi--Yau metric on $X$. The corresponding Kähler metric is $g_{i\bar{j}}=\partial_{i}\bar{\partial}_{\bar{j}}K$, where a pullback to the hypersurface on the $i,\bar{j}$ indices is implicit. 

The bundles we consider are line bundles $V=\mathcal{O}(m)$ on the torus $X$ for integer values of $m$. Since the approximate Calabi--Yau metric is defined by \eqref{eq:approx_Kahler}, similar to \eqref{eq:bundle_P3}, a Hermite--Einstein metric on the fibres of $\mathcal{O}(m)$ is given by
\begin{equation}
G=\bigl(s_{\alpha}h^{\alpha\bar{\beta}}\overline{s_{\beta}}\bigr)^{-m/k_{h}}.\label{eq:bundle_CY_metric}
\end{equation}
Again, one can check that this choice satisfies the hermitian Yang--Mills equation on $X$ with the Kähler metric determined by \eqref{eq:approx_Kahler}. With these ingredients, we can now compute the numerical spectrum of the Dolbeault Laplacian on our first example of a Calabi--Yau hypersurface. In what follows, we computed the approximate Calabi--Yau metric at $k_{h}=10$ corresponding to a ``$\sigma$-measure'' of $\sigma\approx2\times{10}^{-15}$~\cite{Douglas:2006rr}. Integrals were computed via Monte Carlo using $N_\phi={10}^{6}$ points.

\subsubsection{The bundle-valued scalar spectrum}

\begin{figure}
\includegraphics{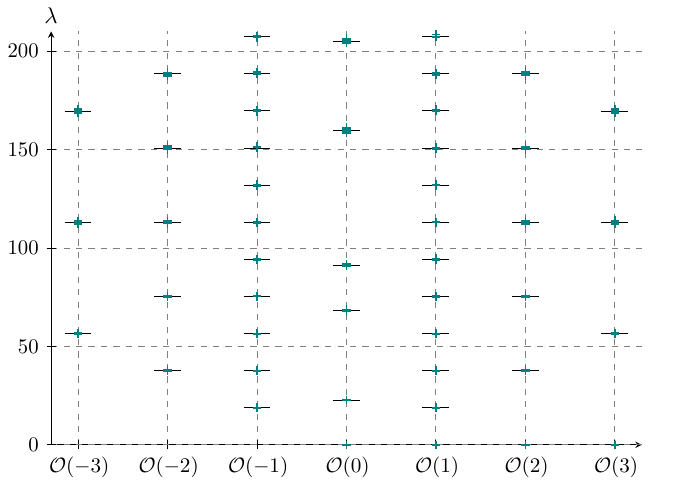}
\caption{Numerical eigenvalues $\lambda_{n}$ of $\Delta_{\bar{\partial}_{V}}$ on the Fermat cubic acting on $\mathcal{O}(m)$-valued scalars for $m\in\{-3,\dots,3\}$. These were computed using a numerical Calabi--Yau metric computed at $k_{h}=10$ and the associated Hermite--Einstein metric on $\mathcal{O}(m)$. Integrals were computed via Monte Carlo over $N_\phi={10}^{6}$ points. We used $k_{\phi}=3$ for the basis functions. The horizontal black lines indicate the exact analytic values from Table \ref{tab:T2_eigens}. \label{fig:T2_scalar_graph}}
\end{figure}

We begin with a numerical calculation of the spectrum of bundle-valued eigenfunctions of $\Delta_{\bar{\partial}_{V}}$. The inputs are the approximate Calabi--Yau metric on $X$ determined by the Kähler potential in \eqref{eq:approx_Kahler} with the parameters fixed by the ``energy functional'' approach~\cite{Headrick:2009jz}, a bundle $\mathcal{O}(m)$ together with a Hermite--Einstein metric \eqref{eq:bundle_CY_metric}, a choice of degree $k_{\phi}$ which determines the size of the approximate basis \eqref{eq:approx_basis} in which we expand the eigenfunctions, and the number of points $N_\phi={10}^{6}$ that are used to discretise the integrals that appear in matrix elements of the Laplacian. For the rest of this section, we fix $k_{\phi}=3$ and compute the spectrum for $m\in\{-3,\dots,3\}$. Our numerical results are shown in Table \ref{tab:T2_scalar_table} and Figure \ref{fig:T2_scalar_graph}.

The numerical results in Table \ref{tab:T2_scalar_table} reproduce the exact results in Table \ref{tab:T2_eigens} with excellent precision and the correct multiplicities. This is also visible in Figure \ref{fig:T2_scalar_graph} which shows the numerical results and indicates the values of the exact eigenvalues; in all cases, the exact result lies in the middle of the cluster of numerical eigenvalues. For larger values of $N_\phi$, the eigenvalues in Figure \ref{fig:T2_scalar_graph} become more tightly clustered. In the $N_\phi\to\infty$ limit, one recovers the $(S_{3}\times\bZ_{2})\rtimes(\bZ_{3}\times\bZ_{3})$ symmetry of $X$, and the eigenvalues become exactly degenerate.

\begin{table}
\noindent \begin{centering}
\scalebox{0.7}{%
\begin{tabular}{ccccccccccccccc}
\toprule 
$m$ & \multicolumn{2}{c}{$-3$} & \multicolumn{2}{c}{$-2$} & \multicolumn{2}{c}{$-1$} & \multicolumn{2}{c}{0} & \multicolumn{2}{c}{1} & \multicolumn{2}{c}{2} & \multicolumn{2}{c}{3}\tabularnewline
\midrule 
$n$ & $\lambda_{n}$ & $\ell_{n}$ & $\lambda_{n}$ & $\ell_{n}$ & $\lambda_{n}$ & $\ell_{n}$ & $\lambda_{n}$ & $\ell_{n}$ & $\lambda_{n}$ & $\ell_{n}$ & $\lambda_{n}$ & $\ell_{n}$ & $\lambda_{n}$ & $\ell_{n}$\tabularnewline
\midrule
\midrule 
0 & $56.5\pm0.2$ & 9 & $37.7\pm0.1$ & 6 & $18.85\pm0.05$ & 3 & $0.0$ & 1 & $0.0$ & 3 & $0.0$ & 6 & $0.0$ & 9\tabularnewline
1 & $113.1\pm0.5$ & 9 & $75.4\pm0.3$ & 6 & $37.7\pm0.1$ & 3 & $22.80\pm0.08$ & 6 & $18.85\pm0.05$ & 3 & $37.7\pm0.1$ & 6 & $56.5\pm0.2$ & 9\tabularnewline
2 & $169.6\pm0.7$ & 9 & $113.1\pm0.5$ & 6 & $56.4\pm0.2$ & 3 & $68.4\pm0.2$ & 6 & $37.7\pm0.1$ & 3 & $75.4\pm0.3$ & 6 & $113.1\pm0.5$ & 9\tabularnewline
3 & $226\pm1$ & 9 & $150.8\pm0.7$ & 6 & $75.6\pm0.2$ & 3 & $91.2\pm0.4$ & 6 & $56.6\pm0.1$ & 3 & $113.1\pm0.5$ & 6 & $169.6\pm0.7$ & 9\tabularnewline
4 & $283\pm2$ & 9 & $188.5\pm0.6$ & 6 & $94.3\pm0.3$ & 3 & $159.7\pm0.9$ & 12 & $75.3\pm0.1$ & 3 & $150.8\pm0.6$ & 6 & $226\pm1$ & 9\tabularnewline
5 & $340\pm2$ & 9 & $226.4\pm0.8$ & 6 & $113.0\pm0.3$ & 3 & $205.3\pm0.9$ & 6 & $94.3\pm0.3$ & 3 & $188.6\pm0.7$ & 6 & $283\pm1$ & 9\tabularnewline
6 & $396\pm2$ & 9 & $264.0\pm0.9$ & 6 & $131.9\pm0.4$ & 3 & $274\pm1$ & 6 & $113.1\pm0.2$ & 3 & $226.1\pm0.9$ & 6 & $339\pm1$ & 9\tabularnewline
\bottomrule
\end{tabular}}
\par\end{centering}
\caption{Numerical eigenvalues $\lambda_{n}$ of $\Delta_{\bar{\partial}_{V}}$ on the Fermat cubic acting on $\mathcal{O}(m)$-valued scalars for $m\in\{-3,\dots,3\}$ with $k_{\phi}=3$. We have also included their multiplicities $\ell_{n}$. These were computed using a numerical Calabi--Yau metric computed at $k_{h}=10$ and the associated Hermite--Einstein metric on $\mathcal{O}(m)$. Integrals were computed via Monte Carlo over $N_\phi={10}^{6}$ points. The quoted eigenvalues are the mean of the eigenvalues in a cluster, with the error given by the standard deviation of the cluster.\label{tab:T2_scalar_table}}
\end{table}

\subsubsection{The bundled-valued $(0,1)$-form spectrum}

\begin{figure}
\includegraphics{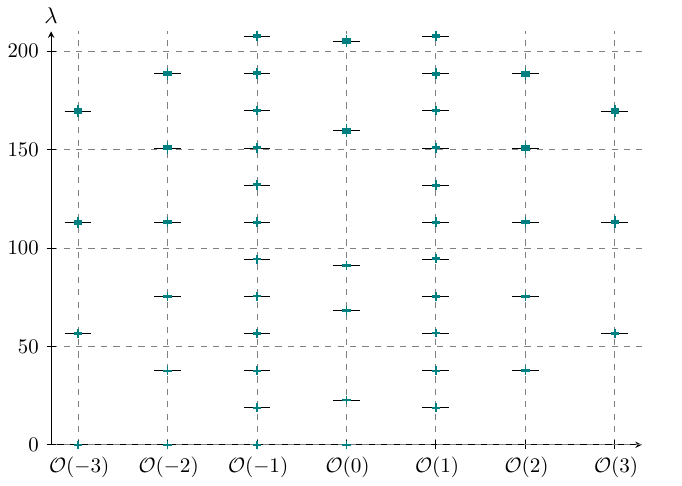}
\caption{Numerical eigenvalues $\lambda_{n}$ of $\Delta_{\bar{\partial}_{V}}$ on the Fermat cubic acting on $\mathcal{O}(m)$-valued $(0,1)$-forms for $m\in\{-3,\dots,3\}$. These were computed using a numerical Calabi--Yau metric computed at $k_{h}=10$ and the associated Hermite--Einstein metric on $\mathcal{O}(m)$. Integrals were computed via Monte Carlo over $N_\phi={10}^{6}$ points. We used $k_{\phi}=3$ for the basis functions. The horizontal black lines indicate the exact analytic values inferred from Table \ref{tab:T2_eigens} and the identity \eqref{eq:serre}. \label{fig:T2_01_graph}}
\end{figure}

Next, we have the numerical calculation of the $\Omega^{0,1}(\mathcal{O}(m))$ spectrum. Again, this follows the scalar calculation in the previous subsection almost exactly, apart from using an appropriate basis of bundle-valued $(0,1)$-forms from \eqref{eq:approx_basis_pq}. The results for $m\in\{-3,\dots,3\}$ are shown in Table \ref{tab:T2_01_table} and Figure \ref{fig:T2_01_graph}.

As we mentioned at the start of this section, since $X$ is Calabi--Yau, its canonical bundle is trivial, $K_{X}=\mathcal{O}$. The identity \eqref{eq:serre} then implies that the $\mathcal{O}(m)$-valued $(0,1)$-form spectrum should match the $\mathcal{O}(-m)$-valued scalar spectrum. Comparing Tables \ref{tab:T2_scalar_table} and \ref{tab:T2_01_table}, we see this is indeed the case up to numerical accuracy. This is also apparent in Figure \ref{fig:T2_01_graph}, where we have plotted the numerical $(0,1)$ eigenvalues and indicated the values that one infers from the exact $\mathcal{O}(m)$ scalar spectrum with black lines. In all cases, we see the two agree.

\begin{table}
\noindent \begin{centering}
\scalebox{0.7}{%
\begin{tabular}{ccccccccccccccc}
\toprule 
$m$ & \multicolumn{2}{c}{$-3$} & \multicolumn{2}{c}{$-2$} & \multicolumn{2}{c}{$-1$} & \multicolumn{2}{c}{0} & \multicolumn{2}{c}{1} & \multicolumn{2}{c}{2} & \multicolumn{2}{c}{3}\tabularnewline
\midrule 
$n$ & $\lambda_{n}$ & $\ell_{n}$ & $\lambda_{n}$ & $\ell_{n}$ & $\lambda_{n}$ & $\ell_{n}$ & $\lambda_{n}$ & $\ell_{n}$ & $\lambda_{n}$ & $\ell_{n}$ & $\lambda_{n}$ & $\ell_{n}$ & $\lambda_{n}$ & $\ell_{n}$\tabularnewline
\midrule
\midrule 
0 & $0.0$ & 9 & $0.0$ & 6 & $0.0$ & 3 & $0.04$ & 1 & $18.85\pm0.05$ & 3 & $37.7\pm0.1$ & 6 & $56.5\pm0.2$ & 9\tabularnewline
1 & $56.5\pm0.2$ & 9 & $37.7\pm0.1$ & 6 & $18.85\pm0.06$ & 3 & $22.79\pm0.08$ & 6 & $37.7\pm0.1$ & 3 & $75.4\pm0.3$ & 6 & $113.1\pm0.5$ & 9\tabularnewline
2 & $113.1\pm0.5$ & 9 & $75.4\pm0.2$ & 6 & $37.7\pm0.1$ & 3 & $68.4\pm0.2$ & 6 & $56.4\pm0.2$ & 3 & $113.1\pm0.5$ & 6 & $169.6\pm0.7$ & 9\tabularnewline
3 & $169.6\pm0.7$ & 9 & $113.1\pm0.4$ & 6 & $56.5\pm0.1$ & 3 & $91.2\pm0.2$ & 6 & $75.6\pm0.2$ & 3 & $150.8\pm0.7$ & 6 & $226\pm1$ & 9\tabularnewline
4 & $226\pm1$ & 9 & $150.8\pm0.7$ & 6 & $75.4\pm0.1$ & 3 & $159.5\pm0.7$ & 12 & $94.3\pm0.3$ & 3 & $188.5\pm0.6$ & 6 & $283\pm2$ & 9\tabularnewline
5 & $283\pm1$ & 9 & $188.6\pm0.6$ & 6 & $94.3\pm0.2$ & 3 & $205.2\pm0.7$ & 6 & $113.0\pm0.3$ & 3 & $226.4\pm0.8$ & 6 & $340\pm2$ & 9\tabularnewline
6 & $340\pm1$ & 9 & $226.3\pm0.8$ & 6 & $113.1\pm0.3$ & 3 & $274\pm1$ & 6 & $131.9\pm0.4$ & 3 & $264.0\pm0.9$ & 6 & $396\pm2$ & 9\tabularnewline
\bottomrule
\end{tabular}}
\par\end{centering}
\caption{Numerical eigenvalues $\lambda_{n}$ of $\Delta_{\bar{\partial}_{V}}$ on the Fermat cubic acting on $\mathcal{O}(m)$-valued $(0,1)$-forms for $m\in\{-3,\dots,3\}$ with $k_{\phi}=3$ ($k_{\phi}=4$ for $m=0$). We have also included their multiplicities $\ell_{n}$. These were computed using a numerical Calabi--Yau metric computed at $k_{h}=10$ and the associated Hermite--Einstein metric on $\mathcal{O}(m)$. Integrals were computed via Monte Carlo over $N_\phi={10}^{6}$ points. The quoted eigenvalues are the mean of the eigenvalues in a cluster, with the error given by the standard deviation of the cluster. Thanks to \eqref{eq:serre}, these eigenvalues should be related to those of Table \ref{tab:T2_scalar_table} by $\{\lambda\}_{\mathcal{O}(m)}^{(0,0)}=\{\lambda\}_{\mathcal{O}(-m)}^{(0,1)}$, which simply reflects the table about $m=0$. \label{tab:T2_01_table}}
\end{table}

\section{Quintic Calabi--Yau three-folds\label{sec:Quintic-Calabi=002013Yau-three-folds}}

In the previous section, we extended the numerical calculation of the bundle-valued scalar and $(0,1)$-form spectra to a torus defined as a hypersurface in projective space. From this toy example, it is simple to generalise to higher-dimensional Calabi--Yau manifolds defined as hypersurfaces. The particular example that we focus on is that of the Fermat quintic three-fold $X$ defined as the vanishing locus in $\bP^{4}$ of the equation
\begin{equation}
Q\equiv Z_{0}^{5}+Z_{1}^{5}+Z_{2}^{5}+Z_{3}^{5}+Z_{4}^{5}=0.
\end{equation}
Unlike the previous examples, there are no analytic results to match to other than the dimensions of certain bundle-valued cohomologies which count zero modes. The results we present below are thus the first calculation of the spectrum of a bundle-valued Laplacian on a non-trivial Calabi--Yau manifold.

Before moving to the numerical results, we describe various constraints on bundle cohomologies on general Calabi--Yau manifolds.\footnote{See, for example, \cite{1906.00392} and references therein for a nice review of these vanishing theorems.} These will provide a consistency check for the count of zero modes. First, Serre duality relates the sheaf cohomologies as
\begin{equation}
H^{p}(X,V)=H^{n-p}(X,V^{*}).
\end{equation}
Second, the Kodaira vanishing theorem states that on a Calabi--Yau $X$
\begin{equation}
H^{p}(X,V)=\{0\}\quad\text{for \ensuremath{p>0} if \ensuremath{V} is positive},
\end{equation}
where for manifolds with Picard rank one (such as the hypersurfaces in a single projective space that we consider), positive just means line bundles $V=\mathcal{O}(m)$ with $m>0$. These constraints imply that on a three-fold, such as the Fermat quintic, the only non-vanishing cohomologies are $h^{0}(\mathcal{O}(m))$ and $h^{3}(\mathcal{O}(-m))$ for $m>0$. The scalar zero modes of $\Delta_{\bar{\partial}_{V}}$, counted by $h^{0}(\mathcal{O}(m))$, are the degree-$m$ holomorphic monomials of the $Z^{I}$ coordinates on $\bP^{4}$ pulled back to the hypersurface. Since $h^{1}(\mathcal{O}(m))=0$ for all $m$, there are no bundle-valued $(0,1)$-form zero modes. We will see this counting reflected in the numerical results in the next subsection.

\subsection{Numerical results}

Since there are no known explicit expressions for either the Calabi--Yau metric on the quintic nor Hermite--Einstein metrics on bundles over it, we must compute these numerically. The ansatz for the Kähler potential is again of the form \eqref{eq:approx_Kahler}, but now with degree-$k_{h}$ polynomials on $\bP^{4}$ restricted to the hypersurface. Similarly, the Hermite--Einstein metric on the fibres of $\mathcal{O}(m)$ over the quintic is given by \eqref{eq:bundle_CY_metric}. For what follows, we will use an approximate Calabi--Yau metric on the quintic computed at $k_{h}=6$ using the ``energy functional'' approach of Headrick and Nassar~\cite{Headrick:2009jz}, with a $\sigma$-measure of $\sigma\approx2\times{10}^{-4}$~\cite{Douglas:2006rr}. The numerical integrations were carried out using $N_\phi=5\times{10}^{6}$ points. Unless otherwise stated, the spectra were computed using an approximate basis $\mathcal{F}_{k_{\phi}}^{p,q}(m)$ at $k_{\phi}=3$.

\subsubsection{The bundle-valued scalar spectrum}

\begin{figure}
\includegraphics{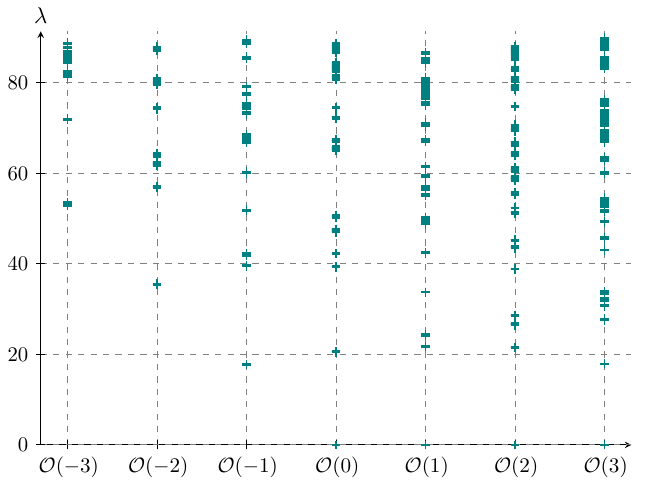}
\caption{Numerical eigenvalues $\lambda_{n}$ of $\Delta_{\bar{\partial}_{V}}$ on the Fermat quintic acting on $\mathcal{O}(m)$-valued scalars for $m\in\{-3,\dots,3\}$. These were computed using a numerical Calabi--Yau metric computed at $k_{h}=6$ and the associated Hermite--Einstein metric on $\mathcal{O}(m)$. Integrals were computed via Monte Carlo over $N_\phi=5\times{10}^{6}$ points. We used $k_{\phi}=3$ for the basis functions, except for $m=\pm3$ which were computed at $k_{\phi}=2$. \label{fig:quintic_scalar_graph}}
\end{figure}

We have computed numerically the $\mathcal{O}(m)$-valued scalar spectrum of $\Delta_{\bar{\partial}_{V}}$ for $m\in\{-3,\dots,3\}$ on the Fermat quintic three-fold. The results are shown in Table \ref{tab:quintic_scalar_table} and Figure \ref{fig:quintic_scalar_graph}. For $m=0$, the eigenvalues are one-half of those computed in \cite{Ashmore:2020ujw}, as expected from the identity $\Delta_{\bar{\partial}_{V}}=\tfrac{1}{2}\Delta$ when $V=\mathcal{O}$. For $m>0$, the zero modes of $\Delta_{\bar{\partial}_{V}}$ should be monomials of degree $m$ in the homogeneous $Z^{I}$ coordinates modulo the defining equation, $Q=0$. The counting of these monomials, given by $h^{0}(\mathcal{O}(m))=\binom{4+m}{m}$ for $0<m<5$, agrees with the number of zero modes in Table \ref{tab:quintic_scalar_table}. For $m<0$, the numerical results indicate there are no zero modes, in agreement with the vanishing of the relevant cohomologies that we mentioned above.

\begin{table}
\noindent \begin{centering}
\scalebox{0.7}{%
\begin{tabular}{ccccccccccccccc}
\toprule 
$m$ & \multicolumn{2}{c}{$-3$} & \multicolumn{2}{c}{$-2$} & \multicolumn{2}{c}{$-1$} & \multicolumn{2}{c}{0} & \multicolumn{2}{c}{1} & \multicolumn{2}{c}{2} & \multicolumn{2}{c}{3}\tabularnewline
\midrule 
$n$ & $\lambda_{n}$ & $\ell_{n}$ & $\lambda_{n}$ & $\ell_{n}$ & $\lambda_{n}$ & $\ell_{n}$ & $\lambda_{n}$ & $\ell_{n}$ & $\lambda_{n}$ & $\ell_{n}$ & $\lambda_{n}$ & $\ell_{n}$ & $\lambda_{n}$ & $\ell_{n}$\tabularnewline
\midrule
\midrule 
0 & $53.2\pm0.2$ & 35 & $35.4\pm0.1$ & 15 & $17.73\pm0.05$ & 5 & $0.0$ & 1 & $0.0$ & 5 & $0.0$ & 15 & $0.0$ & 35\tabularnewline
1 & $71.9\pm0.1$ & 10 & $57.0\pm0.1$ & 20 & $39.5\pm0.1$ & 20 & $20.56\pm0.07$ & 20 & $21.79\pm0.07$ & 20 & $21.50\pm0.06$ & 20 & $17.94\pm0.04$ & 10\tabularnewline
2 & $82.0\pm0.4$ & 60 & $62.6\pm0.9$ & 80 & $42.0\pm0.2$ & 30 & $39.4\pm0.1$ & 20 & $24.3\pm0.1$ & 30 & $26.6\pm0.1$ & 60 & $27.7\pm0.1$ & 60\tabularnewline
3 & $85.8\pm0.8$ & 90 & $74.4\pm0.1$ & 15 & $51.9\pm0.1$ & 10 & $42.28\pm0.06$ & 4 & $33.75\pm0.06$ & 10 & $28.6\pm0.1$ & 20 & $30.7\pm0.1$ & 30\tabularnewline
4 & $87.8\pm0.1$ & 5 & $80.2\pm0.5$ & 50 & $60.1\pm0.1$ & 15 & $47.3\pm0.2$ & 60 & $42.43\pm0.08$ & 15 & $38.8\pm0.07$ & 15 & $32.0\pm0.2$ & 60\tabularnewline
\bottomrule
\end{tabular}}
\par\end{centering}
\caption{Numerical eigenvalues $\lambda_{n}$ of $\Delta_{\bar{\partial}_{V}}$ on the Fermat quintic acting on $\mathcal{O}(m)$-valued scalars for $m\in\{-3,\dots,3\}$. These were computed using a numerical Calabi--Yau metric computed at $k_{h}=6$ and the associated Hermite--Einstein metric on $\mathcal{O}(m)$. Integrals were computed via Monte Carlo over $N_\phi=5\times{10}^{6}$ points. The approximate basis used $k_{\phi}=3$, except for $m=\pm3$ which were computed at $k_{\phi}=2$. We have also included their multiplicities $\ell_{n}$. The quoted eigenvalues are the mean of the eigenvalues in a cluster, with the error given by the standard deviation of the cluster.\label{tab:quintic_scalar_table}}
\end{table}

\subsubsection{The bundle-valued $(0,1)$-form spectrum}\label{subsec:The-bundled-valued--form}

\begin{figure}
\includegraphics{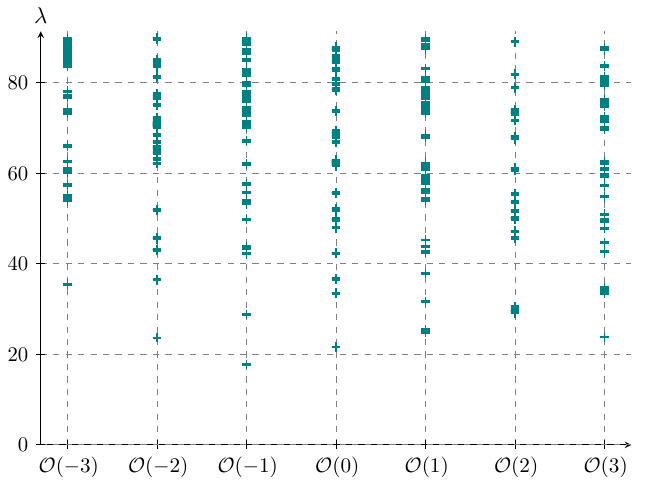}
\caption{Numerical eigenvalues $\lambda_{n}$ of $\Delta_{\bar{\partial}_{V}}$ on the Fermat quintic acting on $\mathcal{O}(m)$-valued $(0,1)$-forms for $m\in\{-3,\dots,3\}$. These were computed using a numerical Calabi--Yau metric computed at $k_{h}=6$ and the associated Hermite--Einstein metric on $\mathcal{O}(m)$. Integrals were computed via Monte Carlo over $N_\phi=5\times{10}^{6}$ points. We used $k_{\phi}=3$ for the basis functions for $m=0,\pm1$ and $k_{\phi}=2$ for $m=\pm2,\pm3$. \label{fig:quintic_01_graph}}
\end{figure}

Finally, we have the numerical calculation of the $\Omega^{0,1}(\mathcal{O}(m))$ spectrum on the Fermat quintic. Our results for $m\in\{-3,\dots,3\}$ are shown in Table \ref{tab:quintic_01_table} and Figure \ref{fig:quintic_01_graph}. We first note that there are no zero modes for any values of $m$, in agreement with the constraints from Serre duality and the Kodaira vanishing theorem. As additional evidence that the spectra are consistent, we can again appeal to the $\bar{\partial}_{V}$ Hodge decomposition. This discussion mirrors that for projective space given around Equation \eqref{eq:hodge}. Since there are no zero modes, all $(0,1)$-form eigenmodes of $\Delta_{\bar{\partial}_{V}}$ must be either $\bar{\partial}_{V}$- or $\bar{\partial}_{V}^{\dagger}$-exact. The $\bar{\partial}_{V}$-exact eigenmodes must be of the form $\bar{\partial}_{V}\beta$, where $\beta$ is an $\mathcal{O}(m)$-valued scalar eigenmode, while the $\bar{\partial}_{V}^{\dagger}$-exact modes are of the form $\bar{\partial}_{V}^{\dagger}\gamma$, where $\gamma$ is $\bar{\partial}_{V}$-exact $\mathcal{O}(m)$-valued $(0,2)$ eigenmode. Since $\Delta_{\bar{\partial}_{V}}$ commutes with $\bar{\star}_{V}$ and the canonical bundle of a Calabi--Yau is trivial, the spectrum of $\mathcal{O}(m)$-valued $(0,2)$ eigenmodes agrees with the $\mathcal{O}(-m)$-valued $(0,1)$ spectrum. Putting this together, the spectrum of the Laplacian acting on $\Omega^{0,1}(\mathcal{O}(m))$ should be the union of the entire $\Omega^{0,0}(\mathcal{O}(m))$ spectrum and roughly half of the $\Omega^{0,1}(\mathcal{O}(-m))$ spectrum.

Comparing Tables \ref{tab:quintic_scalar_table} and \ref{tab:quintic_01_table}, we see this appears to be the case, though with worse accuracy than we achieved for $\bP^{3}$. For example, for $m=1$, the $\Omega^{0,1}(\mathcal{O}(1))$ modes with eigenvalue $25.2$ and multiplicity $50$ originate from the $\Omega^{0,0}(\mathcal{O}(1))$ modes with eigenvalues $(21.8,24.3)$ whose multiplicities sum to $50$. (It appears that either the truncated basis of forms or the number of integration points was not sufficient for the $(0,1)$ modes to be properly resolved.) Moving up the spectrum, the $\Omega^{0,1}(\mathcal{O}(1))$ modes with eigenvalue $31.7$ and multiplicity $30$ likely come from the $\Omega^{0,1}(\mathcal{O}(-1))$ modes with eigenvalue $28.8$ and the same multiplicity. Similarly, the $\Omega^{0,1}(\mathcal{O}(1))$ modes with eigenvalue $37.8$ and multiplicity $10$ likely come from the $\Omega^{0,0}(\mathcal{O}(1))$ modes with eigenvalue $33.8$ and the same multiplicity. 

A glance at the other results should convince the reader that this decomposition holds more generally, though the match is not perfect. This is likely due to inaccuracies introduced by the truncation at $k_{\phi}=3$ to a finite-dimensional basis of forms. Recall that on projective space, the basis $\mathcal{F}_{k_{\phi}}^{p,q}(m)$ exactly spans the first $k_{\phi}$ eigenspaces of $\Delta_{\bar{\partial}_{V}}$. However, since the Calabi--Yau metric is not simply the pullback of Fubini--Study, the eigenspaces of the Laplacian are not exactly spanned by $\mathcal{F}_{k_{\phi}}^{p,q}(m)$ for finite $k_{\phi}$, nor there is not a direct map between $(0,0)$ and $(0,1)$ modes at each degree $k_{\phi}$. Instead, the approximate eigenmodes computed at some finite degree will receive corrections as $k_{\phi}$ is increased and the basis of forms is enlarged. We believe that upon moving to larger values of $k_{\phi}$ and increasing the number of integration points, the match between the $\Omega^{0,1}(\mathcal{O}(m))$ and the $\Omega^{0,0}(\mathcal{O}(m))$ and $\Omega^{0,1}(\mathcal{O}(-m))$ spectra will improve. 

Regardless of this, one should remember that the lower-dimensional physics of a string compactification is determined by properties of harmonic/zero modes on the compactification manifold. These zero modes are, by definition, long wavelength and slowly varying, and likely to be very well approximated already at the modest values of $k_{\phi}$ that we have used. The same is certainly not true for massive modes higher up the spectrum; thankfully, these modes seem to be less relevant for low-energy physics questions.

\begin{table}
\noindent \begin{centering}
\scalebox{0.7}{%
\begin{tabular}{ccccccccccccccc}
\toprule 
$m$ & \multicolumn{2}{c}{$-3$} & \multicolumn{2}{c}{$-2$} & \multicolumn{2}{c}{$-1$} & \multicolumn{2}{c}{0} & \multicolumn{2}{c}{1} & \multicolumn{2}{c}{2} & \multicolumn{2}{c}{3}\tabularnewline
\midrule 
$n$ & $\lambda_{n}$ & $\ell_{n}$ & $\lambda_{n}$ & $\ell_{n}$ & $\lambda_{n}$ & $\ell_{n}$ & $\lambda_{n}$ & $\ell_{n}$ & $\lambda_{n}$ & $\ell_{n}$ & $\lambda_{n}$ & $\ell_{n}$ & $\lambda_{n}$ & $\ell_{n}$\tabularnewline
\midrule
\midrule 
0 & $35.5\pm0.1$ & 40 & $23.7\pm0.1$ & 10 & $17.76\pm0.04$ & 5 & $21.60\pm0.06$ & 20 & $25.2\pm0.1$ & 50 & $29.7\pm0.4$ & 110 & $23.83\pm0.05$ & 20\tabularnewline
1 & $53.3\pm0.4$ & 957 & $36.5\pm0.1$ & 15 & $28.77\pm0.08$ & 30 & $33.50\pm0.08$ & 30 & $31.7\pm0.1$ & 30 & $45.7\pm0.1$ & 15 & $33.9\pm0.4$ & 155\tabularnewline
2 & $60.6\pm0.3$ & 30 & $43.1\pm0.1$ & 60 & $43.3\pm0.5$ & 110 & $36.7\pm0.1$ & 30 & $37.8\pm0.1$ & 10 & $47.1\pm0.1$ & 20 & $42.7\pm0.1$ & 40\tabularnewline
3 & $62.6\pm0.1$ & 120 & $45.6\pm0.1$ & 40 & $49.7\pm0.1$ & 10 & $42.3\pm0.1$ & 34 & $42.8\pm0.5$ & 75 & $50.0\pm0.2$ & 60 & $44.73\pm0.09$ & 10\tabularnewline
4 & $66.1\pm0.1$ & 15 & $51.9\pm0.1$ & 30 & $53.6\pm0.2$ & 115 & $48.0\pm0.1$ & 20 & $45.3\pm0.1$ & 10 & $51.8\pm0.1$ & 30 & $47.8\pm0.1$ & 15\tabularnewline
\bottomrule
\end{tabular}}
\par\end{centering}
\caption{Numerical eigenvalues $\lambda_{n}$ of $\Delta_{\bar{\partial}_{V}}$ on the Fermat quintic acting on $\mathcal{O}(m)$-valued $(0,1)$-forms for $m\in\{-3,\dots,3\}$. These were computed using a numerical Calabi--Yau metric computed at $k_{h}=6$ and the associated Hermite--Einstein metric on $\mathcal{O}(m)$. Integrals were computed via Monte Carlo over $N_\phi=5\times{10}^{6}$ points. We used $k_{\phi}=3$ for the basis functions for $m=0,\pm1$ and $k_{\phi}=2$ for $m=\pm2,\pm3$. We have also included their multiplicities $\ell_{n}$. The quoted eigenvalues are the mean of the eigenvalues in a cluster, with the error given by the standard deviation of the cluster. \label{tab:quintic_01_table}}
\end{table}

\subsection{Application: computing a superpotential}

As we outlined in Section \ref{subsec:Supersymmetry,-Yukawa-couplings}, the low-energy $\mathcal{N}=1$ physics of a Calabi--Yau compactification is controlled by a superpotential and a Kähler potential. In particular, the matter sector is determined, to lowest order, by integrals of harmonic modes on the Calabi--Yau. In principle, the numerical method that we have presented gives us direct access to the data needed to compute all of this information. In practice, however, the line bundle and three-fold we have considered are too simple to admit non-vanishing superpotential couplings. Let us see why this is the case.

The Fermat quintic three-fold was constructed as a hypersurface in a single ambient projective space. This implies that the rank of the Picard lattice is one and so line bundles on this quintic are of the form $\mathcal{O}(m)$ for some integer $m$.\footnote{In fact, this is obvious from $h^{1,1}=1$ for the Fermat quintic. However, the following argument holds for any hypersurface in a single projective space.} Now imagine trying to write down a non-vanishing superpotential coupling as 
\begin{equation}
\lambda_{IJK}(m_{1},m_{2},m_{3})=\int_{X}\Omega\wedge\psi_{m_{1}}^{I}\wedge\psi_{m_{2}}^{J}\wedge\psi_{m_{3}}^{K},\label{eq:integrand-1}
\end{equation}
where $\psi_{m_{1}}^{I}\in H^{1}(X,\mathcal{O}(m_{1}))$ is an $\mathcal{O}(m_{1})$-valued harmonic $(0,1)$-form, and we have dropped a trace compared with \eqref{eq:superpotential} since the relevant group is abelian. Since $\Omega$ is an honest three-form, this integral vanishes whenever the degrees of the relevant line bundles do not sum to zero:
\begin{equation}
\lambda_{IJK}(m_{1},m_{2},m_{3})=0\quad\text{if}\quad m_{1}+m_{2}+m_{3}\neq0.
\end{equation}
In other words, the charges of the harmonic modes must sum to zero so that the integrand of \eqref{eq:integrand-1} is an honest top-form. Thus, for a non-vanishing superpotential contribution, all the charges must be zero or at least one of them is negative. However, it is simple to argue that the requisite harmonic $(0,1)$ modes are not present in either case. When the charges are zero, we need harmonic $(0,1)$-forms. These are counted by the Hodge number $h^{0,1}$ which vanishes on the quintic (and any Calabi--Yau three-fold with irreducible holonomy), so the $(0,1)$ modes are not present. When there are both positive and negative charges, we can appeal to Serre duality and the Kodaira vanishing theorem. The first of these gives $h^{0,1}(\mathcal{O}(-m))=h^{0,2}(\mathcal{O}(m))$, while the second implies $h^{0,1}(\mathcal{O}(m))=h^{0,2}(\mathcal{O}(m))=0$ if $m>0$. Together, these imply that there are no harmonic $\mathcal{O}(m)$-valued $(0,1)$-forms for either sign of $m$, in complete agreement with our numerical results in Table \ref{tab:quintic_01_table}. We conclude that there are no non-vanishing superpotential couplings for matter coming from line bundles on the Fermat quintic.

There are a number of ways to generalise our set-up to allow for interesting superpotential couplings. First is simply moving from line bundles to non-abelian bundles, such as the examples given in \cite{Douglas:2006hz,Anderson:2010ke,Anderson:2011ed}, where instead of the charges summing to zero, one requires a singlet in the antisymmetric product of the three representations appearing in the cubic coupling. Second, we can stay with line bundles but move to Calabi--Yau manifolds with higher-rank Picard lattices. Line bundles on these spaces are labelled by a vector of charges $\boldsymbol{m}$ and the corresponding vanishing theorems are less restrictive. In practice, this means moving to, for example, complete intersection Calabi--Yau (CICY) manifolds given as hypersurfaces in products of projective spaces, such as those used for finding heterotic line bundle models~\cite{hep-th/0512177,hep-th/0502155,Braun:2005ux,Bouchard:2005ag,Anderson:2009mh,1112.1097,Anderson:2011ns,Anderson:2012yf,Anderson:2013xka,GrootNibbelink:2015dvi,GrootNibbelink:2015lme,1007.0203,hep-th/9903052,Otsuka:2018oyf,Otsuka:2018rki}.

\subsection*{Acknowledgements}

It is a pleasure to thank Clay C\'ordova and Edward Mazenc for useful discussions. AA is supported in part by NSF Grant No.\,PHY2014195 and in part by the Kadanoff Center for Theoretical Physics. AA also acknowledges the support of the European Union’s Horizon 2020 research and innovation program under the Marie Sk\l{}odowska-Curie grant agreement No.\,838776. YHH would like to thank STFC for grant ST/J00037X/2. EH would like to thank SMCSE at City, University of London for the PhD studentship, as well as the Jersey Government for a postgraduate grant. BAO is supported in part by both the research grant DOE No.\,DESC0007901 and SAS Account 020-0188-2-010202-6603-0338. This work was completed in part with resources provided by the University of Chicago Research Computing Center.

\appendix

\section{Useful calculations}

Here we collect a few useful calculations which we refer to in the main text.

\subsection{The slope \texorpdfstring{$\mu$}{mu}\label{subsec:The-slope}}

Following \cite[Appendix C]{Blesneag:2021wdf}, let us compute an expression for the slope $\mu$ on $\bP^{N}$. It is cleanest to work in conventions where $\int_{\bP^{N}}\omega^{N}=1.$ The Fubini--Study Kähler potential is simply
\begin{equation}
K=\frac{1}{2\pi}\log\kappa,
\end{equation}
where $\kappa$ restricts to $1+z^{i}\bar{z}_{i}$ on the patch $U_{0}=\{Z^{0}=1\}$ with $Z^{I}=(1,z^{i})$. The corresponding Kähler form and metric are then
\begin{equation}
\omega=\ii\partial\bar{\partial}K,\qquad g_{i\bar{j}}=\partial_{i}\bar{\partial}_{\bar{j}}K.
\end{equation}
The bundle metric on $\mathcal{O}(m)$ is given by $G=\kappa^{-m}$, with gauge field and curvature
\begin{equation}
A=\partial\log G=-2\pi m\partial K,\qquad F=\bar{\partial}A=-2\pi\ii m\omega.
\end{equation}
The Chern class of $\mathcal{O}(m)$ is then
\begin{equation}
c_{1}(\mathcal{O}(m))=\frac{\ii}{2\pi}F=m\omega.
\end{equation}
Using the expression \eqref{eq:slope} for the slope and the volume normalisation above, one finds
\begin{equation}
\mu(\mathcal{O}(m))=\int_{\bP^{N}}c_{1}(\mathcal{O}(m))\wedge\omega^{N-1}=m,
\end{equation}
as expected.

\subsection{Matrix elements of the Laplacian\label{sec:Matrix-elements-of}}

As in \cite{Ashmore:2020ujw}, we denote real coordinate indices by $\{a,b,\ldots\}$ and complex coordinates by $\{i,j,\ldots\}$ and $\{\bar{i},\bar{j},\ldots\}$. As discussed in the main text, the two matrices that we need to compute to find the spectrum of the Laplacian are $\Delta_{AB}\equiv\langle\alpha_{A},\Delta_{\bar{\partial}_{V}}\alpha_{B}\rangle$ and $O_{AB}\equiv\langle\alpha_{A},\alpha_{B}\rangle$, where $\{\alpha_{A}\}$ is a finite set of bundle-valued $(p,q)$-forms. The second of these can be computed straightforwardly using the inner product elements of $\Omega^{p,q}(\mathcal{O}(m))$:
\begin{equation}
\langle v,w\rangle=\int\star_{V}\bar{v}\wedge w=\frac{1}{p!}\int\vol\,\frac{1}{(Z^{I}\bar{Z}_{I})^{m}}\,g^{a_{1}b_{1}}\ldots g^{a_{p}b_{p}}v_{a_{1}\ldots a_{p}}^{*}w_{b_{1}\ldots b_{p}}.
\end{equation}
For example, for scalars, this is simply
\begin{equation}
\langle\alpha_{A},\alpha_{B}\rangle=\int\vol\,\frac{1}{(Z^{I}\bar{Z}_{I})^{m}}\,(\alpha_{A})^{*}\alpha_{B},
\end{equation}
whereas for $(0,1)$-forms, we have
\begin{equation}
\langle\alpha_{A},\alpha_{B}\rangle=\int\vol\,\frac{1}{(Z^{I}\bar{Z}_{I})^{m}}\,g^{ab}(\alpha_{A})_{a}^{*}(\alpha_{B})_{b}=\int\vol\,\frac{1}{(Z^{I}\bar{Z}_{I})^{m}}\,g^{i\bar{j}}(\alpha_{A})_{i}^{*}(\alpha_{B})_{\bar{j}}.
\end{equation}
The matrix element of the Dolbeault Laplacian can be computed similarly. For scalars, it is given by
\begin{equation}
\begin{aligned}\langle\alpha_{A},\Delta_{\bar{\partial}_{V}}\alpha_{B}\rangle=\langle\bar{\partial}_{V}\alpha_{A},\bar{\partial}_{V}\alpha_{B}\rangle & =\int\vol\,\frac{1}{(Z^{I}\bar{Z}_{I})^{m}}\,g^{ab}(\bar{\partial}_{V}\alpha_{A})_{a}^{*}(\bar{\partial}_{V}\alpha_{B})_{b}\\
 & =\int\vol\,\frac{1}{(Z^{I}\bar{Z}_{I})^{m}}\,g^{i\bar{j}}(\bar{\partial}_{V}\alpha_{A})_{i}^{*}(\bar{\partial}_{V}\alpha_{B})_{\bar{j}},
\end{aligned}
\label{eq:scalar}
\end{equation}
with
\begin{equation}
(\bar{\partial}_{V}\alpha_{B})_{\bar{j}}=\bar{\partial}_{\bar{j}}\alpha_{B}.
\end{equation}
For $(0,1)$-forms, we have
\begin{equation}
\begin{aligned}\langle\alpha_{A},\Delta_{\bar{\partial}_{V}}\alpha_{B}\rangle & =\langle\bar{\partial}_{V}\alpha_{A},\bar{\partial}_{V}\alpha_{B}\rangle+\langle\bar{\partial}_{V}^{\dagger}\alpha_{A},\bar{\partial}_{V}^{\dagger}\alpha_{B}\rangle\\
 & =\frac{1}{2!}\int\vol\,\frac{1}{(Z^{I}\bar{Z}_{I})^{m}}\,g^{a_{1}b_{1}}g^{a_{2}b_{2}}(\bar{\partial}_{V}\alpha_{A})_{a_{1}a_{2}}^{*}(\bar{\partial}_{V}\alpha_{B})_{b_{1}b_{2}}\\
 & \eqspace+\int\vol\,\frac{1}{(Z^{I}\bar{Z}_{I})^{m}}\,(\bar{\partial}_{V}^{\dagger}\alpha_{A})^{*}\bar{\partial}_{V}^{\dagger}\alpha_{B}\\
 & =\frac{1}{2}\int\vol\,\frac{1}{(Z^{I}\bar{Z}_{I})^{m}}\,g^{i\bar{k}}g^{j\bar{l}}(\bar{\partial}_{V}\alpha_{A})_{ij}^{*}(\bar{\partial}_{V}\alpha_{B})_{\bar{k}\bar{l}}\\
 & \eqspace+\int\vol\,\frac{1}{(Z^{I}\bar{Z}_{I})^{m}}\,(\bar{\partial}_{V}^{\dagger}\alpha_{A})^{*}\bar{\partial}_{V}^{\dagger}\alpha_{B},
\end{aligned}
\end{equation}
with
\begin{equation}
(\bar{\partial}_{V}\alpha)_{\bar{k}\bar{l}}=\bar{\partial}_{\bar{k}}\alpha_{\bar{l}}-\bar{\partial}_{\bar{l}}\alpha_{\bar{k}},\qquad\bar{\partial}_{V}^{\dagger}\alpha=-g^{i\bar{j}}D_{i}\alpha_{\bar{j}}=-g^{i\bar{j}}(\partial_{i}\alpha_{\bar{j}}+A_{i}\alpha_{\bar{j}}),
\end{equation}
where we have used \eqref{eq:adjoint} to write $\bar{\partial}_{V}^{\dagger}\alpha=-\imath_{D}\alpha$ with $D=\nabla+A$, $\nabla_{i}\alpha_{\bar{j}}=\partial_{i}\alpha_{\bar{j}}$ and the connection is given by $A_{i}=-\partial_{i}G^{-1}\,G=G^{-1}\partial_{i}G$.

\subsection{A local holomorphic frame\label{subsec:A-local-holomorphic}}

As we mentioned in Section \ref{subsec:An-approximate-basis}, for line bundles on projective space (or hypersurfaces therein), one does not need to choose a local holomorphic frame and instead one can work with global objects. Here, we collect a few relevant comments to this effect. 

As an example, we focus on $\bP^{2}$. The homogeneous $Z^{I}$ coordinates are global holomorphic sections of $\mathcal{O}(1)$ where we specify that in each patch we have $U_{A}=\{Z^{A}=1\}$. On $\bP^{2}$ we have three such patches:
\begin{equation}
\begin{gathered}
U_{0}=\{Z^{0}=1,Z^{1}=z^{1},Z^{2}=z^{2}\},\qquad U_{1}=\{Z^{0}=w^{1},Z^{1}=1,Z^{2}=w^{2}\},\\ U_{2}=\{Z^{0}=x^{1},Z^{1}=x^{2},Z^{2}=1\}.
\end{gathered}
\end{equation}
When $\bP^{2}$ is endowed with the standard Fubini--Study metric, the Hermite--Einstein metric on $\mathcal{O}(1)$ is given by $G=({Z^{I}\bar{Z}_{I}})^{-1}$. In each patch, this restricts to
\begin{equation}
G|_{U_{0}}=\frac{1}{1+z^{i}\bar{z}_{i}},\qquad G|_{U_{1}}=\frac{1}{1+w^{i}\bar{w}_{i}},\qquad G|_{U_{2}}=\frac{1}{1+x^{i}\bar{x}_{i}}.
\end{equation}
Crucially, these expressions are valid when working with the global form of both the sections and $G$. Instead, as in Section \ref{subsec:An-approximate-basis}, let us introduce a local holomorphic frame $E_{a}$ for $\mathcal{O}(1)$ as
\begin{equation}
E_{1}=Z^{0}.
\end{equation}
One could in principle pick any linear combination of the $Z^{I}$ -- gauge-invariant quantities will not be affected by the choice. In each patch, the frame restricts to
\begin{equation}
E_{1}|_{U_{0}}=1,\qquad E_{1}|_{U_{1}}=w^{1},\qquad E_{1}|_{U_{2}}=x^{1}.
\end{equation}
Expressing $G$ relative to our choice of frame gives
\begin{equation}
G=\frac{1}{Z^{I}\bar{Z}_{I}}=G_{1\bar{1}}E^{1}\otimes\bar{E}^{1}=G_{1\bar{1}}\frac{1}{Z^{0}\bar{Z}_{0}},
\end{equation}
where $E^a$ is a frame for $\mathcal{O}(1)^*\simeq\mathcal{O}(-1)$, such that $E^a(E_b)=\delta^a_b$. We see that the components of $G$ are given by
\begin{equation}
G_{1\bar{1}}=\frac{1}{Z^{I}\bar{Z}_{I}}Z^{0}\bar{Z}_{0}.\label{eq:G11}
\end{equation}
In each patch we have
\begin{equation}
G_{1\bar{1}}|_{U_{0}}=\frac{1}{1+z^{i}\bar{z}_{i}},\qquad G_{1\bar{1}}|_{U_{1}}=\frac{w^{1}\bar{w}_{1}}{1+w^{i}\bar{w}_{i}},\qquad G_{1\bar{1}}|_{U_{2}}=\frac{x^{1}\bar{x}_{1}}{1+x^{i}\bar{x}_{i}}.\label{eq:metric_frame}
\end{equation}
Recall that $G_{1\bar{1}}$ should be invariant under the rescaling $Z^{I}\mapsto\nu Z^{I}$ (it is a scalar for the $\bC^{*}$ action, though a tensor for $\GL{1,\bC}$ changes of frame). Thankfully, this is obvious from \eqref{eq:G11} or can be checked explicitly on the overlaps of patches $U_{A}\cap U_{B}$.

With these observations in mind, we can check that the matrix elements computed in the previous appendix do not depend on whether one picks a local frame or works with global objects. For example, acting on bundle-valued scalars, the relevant matrix element \eqref{eq:scalar} is
\begin{align}
\langle\alpha_{A},\Delta_{\bar{\partial}_{V}}\alpha_{B}\rangle=\int_{\bP^{2}}\vol\,\frac{1}{(Z^{I}\bar{Z}_{I})^{m}}\,g^{i\bar{j}}\partial_{i}\bar{\alpha}_{A}\bar{\partial}_{\bar{j}}\alpha_{B} & ,\label{eq:integrand}
\end{align}
where the $\alpha_{A}\in\mathcal{F}_{k_{\phi}}^{0,0}(m)$ transform as sections of $\mathcal{O}(m)$ as defined in \eqref{eq:approx_basis}. For the example of $k_{\phi}=m=1$, the basis is spanned by
\begin{equation}
\{\alpha_{A}\}=\frac{(Z_{0}^{2},Z_{0}Z_{1},Z_{1}^{2},Z_{1}Z_{2},Z_{2}^{2},Z_{0}Z_{2})\otimes(\bar{Z}_{0},\bar{Z}_{1},\bar{Z}_{2})}{Z^{I}\bar{Z}_{I}}.\label{eq:basis_appendix}
\end{equation}
In the $U_{0}$ and $U_{1}$ patches, we have
\begin{equation}
\begin{aligned}
\{\alpha_{A}\}|_{U_{0}}&=\frac{(1,z_{1},z_{1}^{2},z_{1}z_{2},z_{2}^{2},z_{2})\otimes(1,\bar{z}_{1},\bar{z}_{2})}{1+z^{i}\bar{z}_{i}},\\
\{\alpha_{A}\}|_{U_{1}}&=\frac{(w_{1}^{2},w_{1},1,w_{2},w_{2}^{2},w_{1}w_{2})\otimes(\bar{w}_{1},1,\bar{w}_{2})}{1+w^{i}\bar{w}_{i}}.
\end{aligned}
\end{equation}
The integrands of \eqref{eq:integrand} in each case become
\begin{equation}
\begin{aligned}U_{0}\colon & \frac{1}{1+z^{i}\bar{z}_{i}}\,g^{i\bar{j}}(1,z_{1},z_{1}^{2},z_{1}z_{2},z_{2}^{2},z_{2})\otimes\bar{\partial}_{\bar{j}}\left(\frac{(1,\bar{z}_{1},\bar{z}_{2})}{1+z^{i}\bar{z}_{i}}\right)(\dots)^{*},\\
U_{1}\colon & \frac{1}{1+w^{i}\bar{w}_{i}}\,g^{i\bar{j}}(w_{1}^{2},w_{1},1,w_{2},w_{2}^{2},w_{1}w_{2})\otimes\bar{\partial}_{\bar{j}}\left(\frac{(\bar{w}_{1},1,\bar{w}_{2})}{1+w^{i}\bar{w}_{i}}\right)(\dots)^{*},
\end{aligned}
\end{equation}
where $(\dots)^{*}$ denotes the conjugate of the tensor product of sections. Using \eqref{eq:metric_frame}, we then convert the bundle metric pre-factors into the components in the frame $E_{1}=Z^{0}$:
\begin{equation}
\begin{aligned}U_{0}\colon & G_{1\bar{1}}\,g^{i\bar{j}}(1,z_{1},z_{1}^{2},z_{1}z_{2},z_{2}^{2},z_{2})\otimes\bar{\partial}_{\bar{j}}\left(\frac{(1,\bar{z}_{1},\bar{z}_{2})}{1+z^{i}\bar{z}_{i}}\right)(\dots)^{*},\\
U_{1}\colon & \frac{1}{w^{1}\bar{w}_{1}}G_{1\bar{1}}\,g^{i\bar{j}}(w_{1}^{2},w_{1},1,w_{2},w_{2}^{2},w_{1}w_{2})\otimes\bar{\partial}_{\bar{j}}\left(\frac{(\bar{w}_{1},1,\bar{w}_{2})}{1+w^{i}\bar{w}_{i}}\right)(\dots)^{*}.
\end{aligned}
\end{equation}
Then move the denominator of $w^{1}\bar{w}_{1}$ into the sections $(w_{1}^{2},w_{1},1,\dots)$ and the conjugate terms:
\begin{equation}
\begin{aligned}U_{0}\colon & G_{1\bar{1}}\,g^{i\bar{j}}(1,z_{1},z_{1}^{2},z_{1}z_{2},z_{2}^{2},z_{2})\otimes\bar{\partial}_{\bar{j}}\left(\frac{(1,\bar{z}_{1},\bar{z}_{2})}{1+z^{i}\bar{z}_{i}}\right)(\dots)^{*},\\
U_{1}\colon & G_{1\bar{1}}\,g^{i\bar{j}}(w_{1},1,w_{1}^{-1},w_{1}^{-1}w_{2},w_{1}^{-1}w_{2}^{2},w_{2})\otimes\bar{\partial}_{\bar{j}}\left(\frac{(\bar{w}_{1},1,\bar{w}_{2})}{1+w^{i}\bar{w}_{i}}\right)(\dots)^{*}.
\end{aligned}
\end{equation}
We see that the sections that appear are simply those of \eqref{eq:basis_appendix} written relative to the local frame, i.e.~on $U_{0}\cap U_{1}$
\begin{equation}
\begin{aligned}(Z_{0}^{2},Z_{0}Z_{1},Z_{1}^{2},Z_{1}Z_{2},Z_{2}^{2},Z_{0}Z_{2}) & =(Z_{0},Z_{1},Z_{0}^{-1}Z_{1}^{2},Z_{0}^{-1}Z_{1}Z_{2},Z_{0}^{-1}Z_{2}^{2},Z_{2})E_{1}\\
 & \equiv(1,z_{1},z_{1}^{2},z_{1}z_{2},z_{2}^{2},z_{2})E_{1}\\
 & \equiv(w_{1},1,w_{1}^{-1},w_{1}^{-1}w_{2},w_{1}^{-1}w_{2}^{2},w_{2})E_{1}.
\end{aligned}
\end{equation}
The upshot of this is that if you compute $G_{a\bar{b}}$ on each patch relative to some choice of frame, the sections that appear in the integrals should also be written in that frame. For line bundles, you can just use the global form of the objects instead. This will not be true for higher-rank bundles.

\subsection{\texorpdfstring{$\mathcal{O}(1)$}{O(1)}-valued scalar spectrum of the Dolbeault Laplacian on \texorpdfstring{$\mathbb{P}^{3}$}{P3}}\label{app:exact_O1}

As a check that our conventions and normalisations are consistent, we compute explicitly the first non-trivial eigenvalue of the Dolbeault Laplacian acting on $\mathcal{O}(1)$-valued scalars on $\mathbb{P}^{3}$. Acting on scalars, the operator reduces to $\Delta_{\bar{\partial}_{V}} \equiv \bar{\partial}_{V}^{\dagger}\bar{\partial}_{V}$. Therefore, the eigenvalue problem we wish to solve is
\begin{equation}
    \bar{\partial}_{V}^{\dagger}\bar{\partial}_{V}\phi = \lambda\phi,
\end{equation}
for $\phi$ an $\mathcal{O}(1)$-valued $(0,0)$-form. 
We work in the patch $U_{0}=\{Z^{0}=1,Z^{i}=z^{i}\}$ with $i=1,2,3$.
Using the definition of the Kähler potential given in \eqref{eq:FS}, the inverse Fubini--Study metric is
\begin{equation}
    g^{\bar{j}i} = 
    \frac{2\pi}{\sqrt[3]{6}} (1+z^{i}\bar{z}_{i})
    \begin{pmatrix}
        1 + z^{1}\bar{z}_{1} & z^{2}\bar{z}_{1} & z^{3}\bar{z}_{1} \\
        z^{1}\bar{z}_{2} & 1 + z^{2}\bar{z}_{2} & z^{3}\bar{z}_{2} \\
        z^{1}\bar{z}_{3} & z^{2}\bar{z}_{3} & 1 + z^{3}\bar{z}_{3}
    \end{pmatrix},
\end{equation}
and the Hermite--Einstein metric on $\mathcal{O}(1)$, restricted to the patch $U_{0}$, is simply
\begin{equation}
    G = \frac{1}{1 + z^{i}\bar{z}_{i}}.
\end{equation}
Finally, from \eqref{eq:approx_basis}, a basis of sections of $\mathcal{O}(1)$ at $k_\phi=1$ is given by
\begin{equation}\label{eq:explicit_basis}
    \{ \alpha_{A} \} = \frac{(1,z_{1},z_{2},z_{3},z_{1}^{2},z_{1}z_{2},z_{1}z_{3},z_{2}^{2},z_{2}z_{3},z_{3}^{2}) \otimes (1,\bar{z}_{1},\bar{z}_{2},\bar{z}_{3})}{1+z^{i}\bar{z}_{i}}.
\end{equation}
As discussed in Section \ref{subsec:The-Dolbeault-Laplacian}, the action of $\bar{\partial}_{V}$ on $\phi$ is simply 
\begin{equation}
    \bar{\partial}_{V}\phi = \bar{\partial}\phi,
\end{equation}
where $\bar{\partial}$ is the usual Dolbeault differential. The action of $\bar{\partial}^{\dagger}_{V}$ on $\bar{\partial}_{V}\phi$ is then simply the contraction of the connection $D$ into the one-form component of $\bar{\partial}_{V}\phi$. That is, if $\alpha$ is a bundle-valued $(0,1)$-form, $\bar{\partial}^\dagger$ acts as
\begin{equation}
    \bar{\partial}_{V}^{\dagger} \alpha = -\imath_D \alpha=-g^{i\bar{j}} D_{i} \alpha_{\bar{j}} = -g^{i\bar{j}}(\partial_{i}\alpha_{\bar{j}} + A_{i}\alpha_{\bar{j}}),
\end{equation}
where $A$ is the connection one-form defined in \eqref{eq:A}. In our case, we have
\begin{equation}
    A_{i} = -G^{-1} \partial_{i} G 
    = -\frac{\bar{z}_{i}}{1 +z^{j}\bar{z}_{j}}.
\end{equation}
Replacing $\alpha$ by $\bar{\partial}_{V}\phi$ then gives an explicit expression for the action of the Dolbeault Laplacian on $\phi$:
\begin{equation}
    \bar{\partial}_{V}^{\dagger}\bar{\partial}_{V}\phi = -g^{i\bar{j}}(\partial_{i}\bar{\partial}_{\bar{j}}\phi + A_{i}\bar{\partial}_{\bar{j}}\phi).
\end{equation}
Using the expressions for $A_{i}$ and $g^{i\bar{j}}$ from above, we then compute the action of $\bar{\partial}_{V}^{\dagger}\bar{\partial}_{V}$ on each of the elements in $\{\alpha_{A}\}$. 
For example, for $\alpha_{1}=\frac{1}{1+z^{i}\bar{z}_{i}}$, using Mathematica it is simple to check that
\begin{equation}
   \bar{\partial}_{V}^{\dagger}\bar{\partial}_{V}\alpha_{1} = -\frac{2\pi}{\sqrt[3]{6}} \frac{-3+2z^{i}\bar{z}_{i}}{1+z^{j}\bar{z}_{j}}.
\end{equation}
The right-hand side can be written in terms of the basis \eqref{eq:explicit_basis} as
\begin{equation}
    -\frac{2\pi}{\sqrt[3]{6}} (-3\alpha_{1} + 2\alpha_{8} + 2\alpha_{12} + 2\alpha_{16}).
\end{equation}
Repeating this procedure for the other basis elements, we can write the action of $\Delta_{\bar{\partial}_V}$ in terms of a matrix acting on the $\{\alpha_{A}\}$ basis. The eigenvalues of this matrix are then the eigenvalues of $\Delta_{\bar{\partial}_V}$. Explicitly, we find that the exact eigenvalues are
\begin{equation}
    \biggl(0,0,0,0,\underbrace{\frac{10\pi}{\sqrt[3]{6}},\dots,\frac{10\pi}{\sqrt[3]{6}}}_{\text{36 times}}\biggr),
\end{equation}
where $10\pi/\sqrt[3]{6}\approx 17.3$. This agrees with both the exact and numerical results given in Tables \ref{tab:P3_eigens} and \ref{tab:P3_scalar_table} respectively.

\section{Differential forms on projective space\label{sec:Differential-forms-on}}

For numerical calculations, it is useful to have an explicit construction of differential forms on projective space. In doing this, we will make explicit what was left implicit in the construction of the basis of $(p,q)$-forms in \cite{Ashmore:2020ujw}. Much of this discussion follows the recent textbook by Tomasiello~\cite[Chapter 6]{tomasiello_2022}.

Complex projective space can be defined as
\begin{equation}
\bP^{N}\equiv\frac{\bC^{N+1}-\{\boldsymbol{0}\}}{\bC^{*}},
\end{equation}
or equivalently as the base space of a $\bC^{*}$-bundle with total space $\bC^{N+1}-\{\boldsymbol{0}\}$, where the $\bC^{*}$ action acts as
\begin{equation}
\nu\cdot(Z^{0},\dots,Z^{N})=(\nu Z^{0},\dots,\nu Z^{N}),\label{eq:C*}
\end{equation}
and $Z^{I}$ are the homogeneous coordinates on $\bP^{N}$. As usual, one can cover $\bP^{N}$ with charts $U_{A}=\{Z^{A}\neq0\}$, $A=0,\dots,N$, with coordinates
\begin{equation}
\left\{ z_{(A)}^{1}=\frac{Z^{0}}{Z^{A}},\dots,z_{(A)}^{i}=\frac{Z^{i+1}}{Z^{A}},\dots\right\} ,
\end{equation}
where the coordinate $z_{(A)}^{A}$ is skipped, since it is equal to one. These coordinates are invariant under the $\bC^{*}$ action and so are good coordinates on $\bP^{N}$.

Constructing functions or forms on $\bP^{N}$ is somewhat subtle thanks to the $\bC^{*}$ identification. One way to proceed is to construct them on a space that we understand, say a sphere, and then project down to projective space. For example, $\bP^{3}$ is equivalent to $\text{S}^{7}/\Uni 1$, so functions on $\bP^{3}$ are functions on the seven-sphere that are also invariant under the $\Uni 1$ action. Formally, this means we think of taking the $\bC^{*}=\bR^{+}\times\Uni 1$ quotient in two steps: first quotienting by the $\bR^{+}$ to give the $(2N+1)$-sphere, and then by the $\Uni 1$. The first of these is realised by

\begin{equation}\label{eq:R+_bundle}
\begin{tikzcd}
      \mathbb{R}^+ \arrow[r, hook] & \mathbb{C}^{N+1}-\{\boldsymbol{0}\} \arrow[d] \\
     {} & \text{S}^{2N+1} 
\end{tikzcd}
\end{equation}so that $\bC^{N+1}-\{\boldsymbol{0}\}$ is an $\bR^{+}$-bundle over the sphere. One can perform the $\bR^{+}$ quotient simply by choosing a sphere $\text{S}^{2N+1}\subset\bC^{N+1}$ of fixed radius $r$:
\begin{equation}
r^{2}\equiv Z^{I}\bar{Z}_{I}.\label{eq:radius}
\end{equation}
Of the initial $\bC^{*}$ action, this choice is left invariant by a residual $\Uni 1$, given by $\nu=\ee^{\ii\varphi}$ with $\varphi\in\bR$. The sphere is then the total space of a circle bundle over projective space:

\begin{equation}\label{eq:U1_bundle}
\begin{tikzcd}
      \text{U}(1) \arrow[r, hook] & \text{S}^{2N+1} \arrow[d] \\
     {} & \mathbb{P}^{N} 
\end{tikzcd}
\end{equation}The construction of well-defined forms on $\bP^{N}$ then follows from standard theory on constructing vertical and horizontal vectors/forms on fibre bundles, which we now review.

\subsection{Vertical, horizontal and basic}

In order to define functions and forms on $\bP^{N}$, we use the observation that forms that are \emph{basic} under a bundle projection can be thought of as forms living only the base of the bundle. For example, in our example where $\bP^{3}=\text{S}^{7}/\Uni 1$, well-defined forms on $\bP^{3}$ are the forms on $\text{S}^{7}$ that are basic with respect to the $\Uni 1$ action.

Let us recall how this works in general. Consider a bundle $E$ with typical fibre $F$ and base space $B$,\begin{equation}
\begin{tikzcd}
      F \arrow[r, hook,"\imath"] & E \arrow[d,"\pi"] \\
     {} & B 
\end{tikzcd}
\end{equation}where $\imath$ is the inclusion of $F$ in $E$, and $\pi$ is the projection map from the total space to the base. Now consider vectors and forms on the total space $E$. A vector field $v$ on $E$ is said to be \emph{vertical} if $\pi_{*}v=0$, where $\pi_{*}$ is the pushforward of the projection map (in coordinates, this just acts as a Jacobian on the components of $v$). Such a vector is tangent to the fibre $F$, and hence has no component lying along the base.\footnote{Note that there is no natural definition of a horizontal vector field (or a vertical form). Such a vector field should be tangent to the base and zero under some natural map between $E$ and $F$. However, there is no natural way to move a vector from $E$ to $F$ without additional data (instead, the natural map on vectors is the pushforward $\imath_{*}$ from $F$ to $E$, though even this is problematic since $\imath$ is not surjective).} Similarly, we say a form $\alpha$ on $E$ is \emph{horizontal} if $\imath_{v}\alpha=0$ for all vertical vectors $v$. Furthermore, if $\alpha$ is invariant under the Lie derivative of all vertical vectors, $\mathcal{L}_{v}\alpha=0$ for all $v$, $\alpha$ is in fact the pullback via $\pi^{*}$ of a form $\alpha_{B}$ on the base $B$:
\begin{equation}
\imath_{v}\alpha=0=\mathcal{L}_{v}\alpha\quad\forall\:\text{vertical }v\qquad\Leftrightarrow\qquad\alpha=\pi^{*}\alpha_{B}.
\end{equation}
Equivalently, both $\alpha$ and $\dd\alpha$ are horizontal. A form on the total space that is the pullback of one on the base is called \emph{basic}. The key idea is that basic forms on $E$ can be thought of as forms living on the base space $B$.

A generalisation of this is given by forms which are horizontal but have a definite charge under the vertical vectors. For example, for a $\bC^{*}$-bundle, one can consider horizontal forms $\alpha$ such that $\mathcal{L}_{v}\alpha=\nu\alpha$ where $\nu\in\bC^{*}$. Such a form is then thought of as a bundle-valued section. Indeed, the homogeneous coordinates $Z^{I}$ are precisely of this kind since they scale according to \eqref{eq:C*} under the $\bC^{*}$ action and are thus thought of as sections of $\mathcal{O}(1)$ over $\bP^{N}$.

\subsection{Derivatives}

We can use the concept of basic forms to first define forms on $\text{S}^{2N+1}$ and then $\bP^{N}$ itself. Starting on $\bC^{N+1}$ with a radial coordinate $r$ defined by \eqref{eq:radius}, the Euler vector field, which generates scaling in the radial direction, and its dual one-form read\footnote{The first of these comes from writing
\[
\dd f=\frac{\partial f}{\partial Z^{I}}\dd Z^{I}+\frac{\partial f}{\partial\bar{Z}_{I}}\dd\bar{Z}_{I},
\]
and then noting that $Z^{I}/r$ is independent of $r$. The second comes from lowering using the flat metric on $\bC^{N+1}$.}
\begin{equation}
r\partial_{r}=2\re(Z^{I}\partial_{I}),\qquad r\dd r=\re(\bar{Z}_{I}\dd Z^{I}),
\end{equation}
where we have taken the flat metric $g=\dd Z^{I}\otimes\dd\bar{Z}_{I}$ with $\imath_{\bar{\partial}_{J}}\dd Z^{I}=\delta_{J}^{I}.$ The standard complex structure on $\bC^{N+1}$ is then defined by $I_{I}{}^{J}=\ii\delta_{I}^{J}$ and $\bar{I}^{I}{}_{J}=-\ii\delta_{J}^{I}$. Using this, we can act on the Euler vector to give another vector $\xi$ and a one-form $\eta$:
\begin{equation}
\xi\equiv I^{\transpose}r\partial_{r}=-2\im(Z^{I}\partial_{I}),\qquad r^{2}\eta\equiv\imath_{r\partial_{r}}J=\im(Z^{I}\dd\bar{Z}_{I}),
\end{equation}
where $\xi$ and $\eta$ are dual in the sense that $\imath_{\xi}\eta=1$. Crucially, $\xi$ is a \emph{vertical} vector for $\Uni 1\hookrightarrow\text{S}^{2N+1}\to\bP^{N}$ and $r\partial_{r}$ is \emph{vertical} for $\bR^{+}\hookrightarrow(\bC^{N+1}-\{\boldsymbol{0}\})\to\text{S}^{2N+1}$. Since the fibres are one-dimensional in each case, all vectors tangent to the fibres are proportional to $\xi$ or $r\partial_{r}$ respectively. Thus, to check whether a form on the total space of either bundle is horizontal, the form must vanish when $\xi$ or $r\partial_{r}$ is contracted into it. Similarly, we have that $\dd r$ and $\eta$ are \emph{horizontal} for the $\Uni 1$- and $\bR^{+}$-bundles respectively, i.e.~$\imath_{\xi}\dd r=0=\imath_{r\partial_{r}}\eta$. 

It is then useful to define a projected derivative of the coordinates:
\begin{equation}
DZ^{I}\equiv\mathcal{P}^{I}{}_{J}\dd Z^{J}=\dd Z^{I}-\left(\frac{\dd r}{r}+\ii\eta\right)Z^{I},\label{eq:DZ-1}
\end{equation}
where the projector is $\mathcal{P}^{I}{}_{J}=\delta_{J}^{I}-\frac{1}{r^{2}}Z^{I}\bar{Z}_{J}$. The reason for doing this is the following. Both $\dd Z^{I}$ and $DZ^{I}$ are (complex) one-forms on $\bC^{N+1}$, but $DZ^{I}$, unlike $\dd Z^{I}$, is horizontal for both the $\Uni 1$- and $\bR^{+}$-bundles:
\begin{equation}
\imath_{r\partial_{r}}DZ^{I}=0=\imath_{\xi}DZ^{I}.
\end{equation}
It is not, however, basic. Instead, $Z^{I}$ and $DZ^{I}$ have weight $+1$ under the $\bR^{+}$ action of $r\partial_{r}$ and weight $+\ii$ under $\Uni 1$ action of $\xi$:
\begin{equation}
\begin{aligned}\bR^{+}\colon &  & \mathcal{L}_{r\partial_{r}}Z^{I} & =Z^{I}, & \mathcal{L}_{r\partial_{r}}DZ^{I} & =DZ^{I},\\
\Uni 1\colon &  & \mathcal{L}_{\xi}Z^{I} & =\ii Z^{I}, & \mathcal{L}_{\xi}DZ^{I} & =\ii DZ^{I}.
\end{aligned}
\end{equation}

\subsection{Forms on \texorpdfstring{$\text{S}^{2N+1}$}{the sphere} and \texorpdfstring{$\protect\bP^{N}$}{projective space}}

Since $Z^{I}$ and $r$ are both weight $+1$ under the $\bR^{+}$, and the Lie derivative obeys a Leibniz rule, $Z^{I}/r$ is automatically invariant under $r\partial_{r}$. Thus, $Z^{I}/r$ is a \emph{basic} function (zero-form) for the $\bR^{+}$ fibration \eqref{eq:R+_bundle} and so is a well-defined function on $\text{S}^{2N+1}$. Similarly, one can show that the one-form $DZ^{I}/r$ is invariant under $r\partial_{r}$ and so can be thought of as an honest one-form on $\text{S}^{2N+1}$. Higher-degree forms on the sphere can then be constructed from wedge products of the $DZ^{I}/r$.

From \eqref{eq:U1_bundle}, forms on $\bP^{N}$ are simply forms on $\text{S}^{2N+1}$ that are basic (horizontal and uncharged) with respect to the $\Uni 1$ action generated by $\xi$. For example, both $Z^{I}/r$ and $DZ^{I}/r$ are basic with respect to the $\bR^{+}$ but not the $\Uni 1$ since they are weight $+\ii$ under $\xi$. However, it is simple to combine them to get basic forms that are invariant under $\xi$. For example,
\begin{equation}
\frac{\bar{Z}_{I}\,DZ^{J}}{r^{2}}
\end{equation}
is basic and so is a well-defined one-form on $\bP^{N}$. Furthermore, it is actually of complex type $(1,0)$ with respect to the standard (Fubini--Study) complex structure on $\bP^{N}$ inherited from $\bC^{N+1}$ and so can be thought of as a section of $\Omega^{1,0}(\bP^{N})$. More generally, horizontal forms with non-zero charge under the combined $\bC^{*}$ action, such as $DZ^{I}$, can be thought of as bundle-valued forms, or equivalently sections of $\Omega^{p,q}(\bP^{N},\mathcal{O}(m))$. In summary:
\begin{itemize}
\item $Z^{I}/r$ and $DZ^{I}/r$ are well-defined functions and one-forms on $\text{S}^{2N+1}$ with charge $+\ii$ under the $\Uni 1$ action.
\item Well-defined functions and forms on $\bP^{N}=\text{S}^{2N+1}/\Uni 1$ are constructed by taking combinations of $Z^{I}/r$ and $DZ^{I}/r$ and their complex conjugates so that the overall $\Uni 1$ charge cancels.
\item Functions or horizontal forms with non-zero $\bC^{*}$ charge can instead be thought of as living in $\Omega^{p,q}(\bP^{N},\mathcal{O}(m))$.
\end{itemize}
We will now see how this analysis relates to the construction of an appropriate basis in which to expand the eigenmodes of the Laplacian, as discussed for scalars by Braun et al.~\cite{Braun:2008jp} and for $(p,q)$-forms by one of the present authors~\cite{Ashmore:2020ujw}. As we recalled in \eqref{eq:approx_basis}, for the scalar spectrum, one expands the eigenmodes using linear combinations of functions of the form
\begin{equation}
\frac{\left(\text{degree \ensuremath{k_{\phi}} monomials in \ensuremath{Z^{I}}}\right)\otimes\overline{\left(\text{degree \ensuremath{k_{\phi}} monomials in \ensuremath{Z^{I}}}\right)}}{(Z^{I}\bar{Z}_{I})^{k_{\phi}}},\qquad k_{\phi}\geq0,
\end{equation}
which is equivalent to expanding in the $\Uni 1$-invariant spherical harmonics on $\text{S}^{2N+1}$~\cite{Headrick:2009jz}. These are simply the well-defined functions on $\bP^{N}$ of bi-degree $(k_{\phi},k_{\phi})$ in $(Z^{I}/r,\bar{Z}_{J}/r)$ that the analysis of this appendix would point to. The basis used in \cite{Ashmore:2020ujw} for the $(p,q)$-form spectrum is a little less straightforward. For example, for $(1,0)$-forms at degree $k_\phi=4$, the basis was
\begin{equation}
\frac{\{Z^{0}\dd Z^{1}-Z^{1}\dd Z^{0},Z^{0}\dd Z^{2}-Z^{2}\dd Z^{0},\dots)\otimes\overline{\left(\text{degree 2 monomials in \ensuremath{Z^{I}}}\right)}}{(Z^{I}\bar{Z}_{I})^{2}}.
\end{equation}
This has no mention of the projected derivative $DZ^{I}$, which we seemed to need to obtain a horizontal one-form. However, from the definition of $DZ^{I}$ in \eqref{eq:DZ-1}, it is simple to check that
\begin{align}
Z^{I}DZ^{J}-Z^{J}DZ^{I} & =Z^{I}\dd Z^{J}-Z^{J}\dd Z^{I},
\end{align}
and so the projector drops out. The same is true for the higher-degree basis forms used in \cite{Ashmore:2020ujw}. The set of $\mathcal{O}(m)$-valued $(p,q)$-forms given in \eqref{eq:approx_basis} and \eqref{eq:approx_basis_pq} then follows from allowing a non-zero $\bC^{*}$ scaling, so that the basis forms are horizontal for both the $\bR^{+}$ and $\Uni 1$ actions, but not basic.

\bibliographystyle{utphys}
\bibliography{extra,inspire}

\providecommand{\href}[2]{#2}\begingroup\raggedright\begin{thebibliography}{100}

\bibitem{Braun:2005ux}
V.~Braun, Y.-H. He, B.~A. Ovrut, and T.~Pantev, ``{A Heterotic standard
  model}'', \href{http://dx.doi.org/10.1016/j.physletb.2005.05.007}{{\em Phys.
  Lett. B} {\bfseries 618} (2005)252--258},
  \href{http://arxiv.org/abs/hep-th/0501070}{{\ttfamily arXiv:hep-th/0501070}}.

\bibitem{Lukas:1998yy}
A.~Lukas, B.~A. Ovrut, K.~S. Stelle, and D.~Waldram, ``{The Universe as a
  domain wall}'', \href{http://dx.doi.org/10.1103/PhysRevD.59.086001}{{\em
  Phys. Rev. D} {\bfseries 59} (1999)086001},
  \href{http://arxiv.org/abs/hep-th/9803235}{{\ttfamily arXiv:hep-th/9803235}}.

\bibitem{Donagi:1999ez}
R.~Donagi, B.~A. Ovrut, T.~Pantev, and D.~Waldram, ``{Standard models from
  heterotic M theory}'',
  \href{http://dx.doi.org/10.4310/ATMP.2001.v5.n1.a4}{{\em Adv. Theor. Math.
  Phys.} {\bfseries 5} (2002)93--137},
  \href{http://arxiv.org/abs/hep-th/9912208}{{\ttfamily arXiv:hep-th/9912208}}.

\bibitem{Bouchard:2005ag}
V.~Bouchard and R.~Donagi, ``{An SU(5) heterotic standard model}'',
  \href{http://dx.doi.org/10.1016/j.physletb.2005.12.042}{{\em Phys. Lett. B}
  {\bfseries 633} (2006)783--791},
  \href{http://arxiv.org/abs/hep-th/0512149}{{\ttfamily arXiv:hep-th/0512149}}.

\bibitem{Blumenhagen:2006ux}
R.~Blumenhagen, S.~Moster, and T.~Weigand, ``{Heterotic GUT and standard model
  vacua from simply connected Calabi-Yau manifolds}'',
  \href{http://dx.doi.org/10.1016/j.nuclphysb.2006.06.005}{{\em Nucl. Phys. B}
  {\bfseries 751} (2006)186--221},
  \href{http://arxiv.org/abs/hep-th/0603015}{{\ttfamily arXiv:hep-th/0603015}}.

\bibitem{Lebedev:2006kn}
O.~Lebedev, H.~P. Nilles, S.~Raby, S.~Ramos-Sanchez, M.~Ratz, P.~K.~S.
  Vaudrevange, and A.~Wingerter, ``{A Mini-landscape of exact MSSM spectra in
  heterotic orbifolds}'',
  \href{http://dx.doi.org/10.1016/j.physletb.2006.12.012}{{\em Phys. Lett. B}
  {\bfseries 645} (2007)88--94},
  \href{http://arxiv.org/abs/hep-th/0611095}{{\ttfamily arXiv:hep-th/0611095}}.

\bibitem{Candelas:2007ac}
P.~Candelas, X.~de~la Ossa, Y.-H. He, and B.~Szendroi, ``{Triadophilia: A
  Special Corner in the Landscape}'',
  \href{http://dx.doi.org/10.4310/ATMP.2008.v12.n2.a6}{{\em Adv. Theor. Math.
  Phys.} {\bfseries 12} 2, (2008)429--473},
  \href{http://arxiv.org/abs/0706.3134}{{\ttfamily arXiv:0706.3134 [hep-th]}}.

\bibitem{Lebedev:2008un}
O.~Lebedev, H.~P. Nilles, S.~Ramos-Sanchez, M.~Ratz, and P.~K.~S. Vaudrevange,
  ``{Heterotic mini-landscape. (II). Completing the search for MSSM vacua in a
  Z(6) orbifold}'',
  \href{http://dx.doi.org/10.1016/j.physletb.2008.08.054}{{\em Phys. Lett. B}
  {\bfseries 668} (2008)331--335},
  \href{http://arxiv.org/abs/0807.4384}{{\ttfamily arXiv:0807.4384 [hep-th]}}.

\bibitem{Ambroso:2010pe}
M.~Ambroso and B.~A. Ovrut, ``{The Mass Spectra, Hierarchy and Cosmology of B-L
  MSSM Heterotic Compactifications}'',
  \href{http://dx.doi.org/10.1142/S0217751X11052943}{{\em Int. J. Mod. Phys. A}
  {\bfseries 26} (2011)1569--1627},
  \href{http://arxiv.org/abs/1005.5392}{{\ttfamily arXiv:1005.5392 [hep-th]}}.

\bibitem{MayorgaPena:2012ifg}
D.~K. Mayorga~Pena, H.~P. Nilles, and P.-K. Oehlmann, ``{A Zip-code for Quarks,
  Leptons and Higgs Bosons}'',
  \href{http://dx.doi.org/10.1007/JHEP12(2012)024}{{\em JHEP} {\bfseries 12}
  (2012)024}, \href{http://arxiv.org/abs/1209.6041}{{\ttfamily arXiv:1209.6041
  [hep-th]}}.

\bibitem{Anderson:2009mh}
L.~B. Anderson, J.~Gray, Y.-H. He, and A.~Lukas, ``{Exploring Positive Monad
  Bundles And A New Heterotic Standard Model}'',
  \href{http://dx.doi.org/10.1007/JHEP02(2010)054}{{\em JHEP} {\bfseries 02}
  (2010)054}, \href{http://arxiv.org/abs/0911.1569}{{\ttfamily arXiv:0911.1569
  [hep-th]}}.

\bibitem{Anderson:2011ns}
L.~B. Anderson, J.~Gray, A.~Lukas, and E.~Palti, ``{Two Hundred Heterotic
  Standard Models on Smooth Calabi-Yau Threefolds}'',
  \href{http://dx.doi.org/10.1103/PhysRevD.84.106005}{{\em Phys. Rev. D}
  {\bfseries 84} (2011)106005},
  \href{http://arxiv.org/abs/1106.4804}{{\ttfamily arXiv:1106.4804 [hep-th]}}.

\bibitem{Anderson:2013xka}
L.~B. Anderson, A.~Constantin, J.~Gray, A.~Lukas, and E.~Palti, ``{A
  Comprehensive Scan for Heterotic SU(5) GUT models}'',
  \href{http://dx.doi.org/10.1007/JHEP01(2014)047}{{\em JHEP} {\bfseries 01}
  (2014)047}, \href{http://arxiv.org/abs/1307.4787}{{\ttfamily arXiv:1307.4787
  [hep-th]}}.

\bibitem{Candelas:1985en}
P.~Candelas, G.~T. Horowitz, A.~Strominger, and E.~Witten, ``{Vacuum
  configurations for superstrings}'',
  \href{http://dx.doi.org/10.1016/0550-3213(85)90602-9}{{\em Nucl. Phys. B}
  {\bfseries 258} (1985)46--74}.

\bibitem{Greene:1986ar}
B.~R. Greene, K.~H. Kirklin, P.~J. Miron, and G.~G. Ross, ``{A Superstring
  Inspired Standard Model}'',
  \href{http://dx.doi.org/10.1016/0370-2693(86)90137-1}{{\em Phys. Lett. B}
  {\bfseries 180} (1986)69}.

\bibitem{Greene:1986bm}
B.~R. Greene, K.~H. Kirklin, P.~J. Miron, and G.~G. Ross, ``{A Three Generation
  Superstring Model. 1. Compactification and Discrete Symmetries}'',
  \href{http://dx.doi.org/10.1016/0550-3213(86)90057-X}{{\em Nucl. Phys. B}
  {\bfseries 278} (1986)667--693}.

\bibitem{Greene:1986jb}
B.~R. Greene, K.~H. Kirklin, P.~J. Miron, and G.~G. Ross, ``{A Three Generation
  Superstring Model. 2. Symmetry Breaking and the Low-Energy Theory}'',
  \href{http://dx.doi.org/10.1016/0550-3213(87)90662-6}{{\em Nucl. Phys. B}
  {\bfseries 292} (1987)606--652}.

\bibitem{Matsuoka:1986vg}
T.~Matsuoka and D.~Suematsu, ``{Realistic Models From the $E(8)$ X $E(8)$-prime
  Superstring Theory}'', \href{http://dx.doi.org/10.1143/PTP.76.886}{{\em Prog.
  Theor. Phys.} {\bfseries 76} (1986)886}.

\bibitem{Greene:1987xh}
B.~R. Greene, K.~H. Kirklin, P.~J. Miron, and G.~G. Ross, ``{27**3 Yukawa
  Couplings for a Three Generation Superstring Model}'',
  \href{http://dx.doi.org/10.1016/0370-2693(87)91151-8}{{\em Phys. Lett. B}
  {\bfseries 192} (1987)111--118}.

\bibitem{Donagi:2000zs}
R.~Donagi, B.~A. Ovrut, T.~Pantev, and D.~Waldram, ``{Standard model
  bundles}'', \href{http://dx.doi.org/10.4310/ATMP.2001.v5.n3.a5}{{\em Adv.
  Theor. Math. Phys.} {\bfseries 5} (2002)563--615},
  \href{http://arxiv.org/abs/math/0008010}{{\ttfamily arXiv:math/0008010}}.

\bibitem{Braun:2006me}
V.~Braun, Y.-H. He, and B.~A. Ovrut, ``{Yukawa couplings in heterotic standard
  models}'', \href{http://dx.doi.org/10.1088/1126-6708/2006/04/019}{{\em JHEP}
  {\bfseries 04} (2006)019},
  \href{http://arxiv.org/abs/hep-th/0601204}{{\ttfamily arXiv:hep-th/0601204}}.

\bibitem{Anderson:2010tc}
L.~B. Anderson, J.~Gray, and B.~Ovrut, ``{Yukawa Textures From Heterotic
  Stability Walls}'', \href{http://dx.doi.org/10.1007/JHEP05(2010)086}{{\em
  JHEP} {\bfseries 05} (2010)086},
  \href{http://arxiv.org/abs/1001.2317}{{\ttfamily arXiv:1001.2317 [hep-th]}}.

\bibitem{Headrick:2005ch}
M.~Headrick and T.~Wiseman, ``{Numerical Ricci-flat metrics on K3}'',
  \href{http://dx.doi.org/10.1088/0264-9381/22/23/002}{{\em Class. Quant.
  Grav.} {\bfseries 22} (2005)4931--4960},
  \href{http://arxiv.org/abs/hep-th/0506129}{{\ttfamily arXiv:hep-th/0506129}}.

\bibitem{math/0512625}
S.~K. {Donaldson}, ``{Some numerical results in complex differential
  geometry}'', \href{http://arxiv.org/abs/math/0512625}{{\ttfamily
  arXiv:math/0512625 [math.DG]}}.

\bibitem{Douglas:2006rr}
M.~R. Douglas, R.~L. Karp, S.~Lukic, and R.~Reinbacher, ``{Numerical Calabi-Yau
  metrics}'', \href{http://dx.doi.org/10.1063/1.2888403}{{\em J. Math. Phys.}
  {\bfseries 49} (2008)032302},
  \href{http://arxiv.org/abs/hep-th/0612075}{{\ttfamily arXiv:hep-th/0612075}}.

\bibitem{Douglas:2006hz}
M.~R. Douglas, R.~L. Karp, S.~Lukic, and R.~Reinbacher, ``{Numerical solution
  to the hermitian Yang-Mills equation on the Fermat quintic}'',
  \href{http://dx.doi.org/10.1088/1126-6708/2007/12/083}{{\em JHEP} {\bfseries
  12} (2007)083}, \href{http://arxiv.org/abs/hep-th/0606261}{{\ttfamily
  arXiv:hep-th/0606261}}.

\bibitem{Braun:2007sn}
V.~Braun, T.~Brelidze, M.~R. Douglas, and B.~A. Ovrut, ``{Calabi-Yau Metrics
  for Quotients and Complete Intersections}'',
  \href{http://dx.doi.org/10.1088/1126-6708/2008/05/080}{{\em JHEP} {\bfseries
  05} (2008)080}, \href{http://arxiv.org/abs/0712.3563}{{\ttfamily
  arXiv:0712.3563 [hep-th]}}.

\bibitem{Headrick:2009jz}
M.~Headrick and A.~Nassar, ``{Energy functionals for Calabi-Yau metrics}'',
  \href{http://dx.doi.org/10.4310/ATMP.2013.v17.n5.a1}{{\em Adv. Theor. Math.
  Phys.} {\bfseries 17} 5, (2013)867--902},
  \href{http://arxiv.org/abs/0908.2635}{{\ttfamily arXiv:0908.2635 [hep-th]}}.

\bibitem{Cui:2019uhy}
W.~Cui and J.~Gray, ``{Numerical Metrics, Curvature Expansions and Calabi-Yau
  Manifolds}'', \href{http://dx.doi.org/10.1007/JHEP05(2020)044}{{\em JHEP}
  {\bfseries 05} (2020)044}, \href{http://arxiv.org/abs/1912.11068}{{\ttfamily
  arXiv:1912.11068 [hep-th]}}.

\bibitem{Anderson:2010ke}
L.~B. Anderson, V.~Braun, R.~L. Karp, and B.~A. Ovrut, ``{Numerical Hermitian
  Yang-Mills Connections and Vector Bundle Stability in Heterotic Theories}'',
  \href{http://dx.doi.org/10.1007/JHEP06(2010)107}{{\em JHEP} {\bfseries 06}
  (2010)107}, \href{http://arxiv.org/abs/1004.4399}{{\ttfamily arXiv:1004.4399
  [hep-th]}}.

\bibitem{Anderson:2011ed}
L.~B. Anderson, V.~Braun, and B.~A. Ovrut, ``{Numerical Hermitian Yang-Mills
  Connections and Kahler Cone Substructure}'',
  \href{http://dx.doi.org/10.1007/JHEP01(2012)014}{{\em JHEP} {\bfseries 01}
  (2012)014}, \href{http://arxiv.org/abs/1103.3041}{{\ttfamily arXiv:1103.3041
  [hep-th]}}.

\bibitem{Anderson:2020hux}
L.~B. Anderson, M.~Gerdes, J.~Gray, S.~Krippendorf, N.~Raghuram, and F.~Ruehle,
  ``{Moduli-dependent Calabi-Yau and SU(3)-structure metrics from Machine
  Learning}'', \href{http://dx.doi.org/10.1007/JHEP05(2021)013}{{\em JHEP}
  {\bfseries 05} (2021)013}, \href{http://arxiv.org/abs/2012.04656}{{\ttfamily
  arXiv:2012.04656 [hep-th]}}.

\bibitem{Ashmore:2019wzb}
A.~Ashmore, Y.-H. He, and B.~A. Ovrut, ``{Machine Learning
  Calabi\textendash{}Yau Metrics}'',
  \href{http://dx.doi.org/10.1002/prop.202000068}{{\em Fortsch. Phys.}
  {\bfseries 68} 9, (2020)2000068},
  \href{http://arxiv.org/abs/1910.08605}{{\ttfamily arXiv:1910.08605
  [hep-th]}}.

\bibitem{Douglas:2020hpv}
M.~R. Douglas, S.~Lakshminarasimhan, and Y.~Qi, ``{Numerical Calabi-Yau metrics
  from holomorphic networks}'',
  \href{http://arxiv.org/abs/2012.04797}{{\ttfamily arXiv:2012.04797
  [hep-th]}}.

\bibitem{Jejjala:2020wcc}
V.~Jejjala, D.~K. Mayorga~Pena, and C.~Mishra, ``{Neural network approximations
  for Calabi-Yau metrics}'',
  \href{http://dx.doi.org/10.1007/JHEP08(2022)105}{{\em JHEP} {\bfseries 08}
  (2022)105}, \href{http://arxiv.org/abs/2012.15821}{{\ttfamily
  arXiv:2012.15821 [hep-th]}}.

\bibitem{Ashmore:2021ohf}
A.~Ashmore, L.~Calmon, Y.-H. He, and B.~A. Ovrut, ``{Calabi-Yau Metrics, Energy
  Functionals and Machine-Learning}'',
  \href{http://dx.doi.org/10.1142/S2810939222500034}{{\em International Journal
  of Data Science in the Mathematical Sciences} {\bfseries 1} 1, (2023)49--61},
  \href{http://arxiv.org/abs/2112.10872}{{\ttfamily arXiv:2112.10872
  [hep-th]}}.

\bibitem{Ashmore:2021rlc}
A.~Ashmore, R.~Deen, Y.-H. He, and B.~A. Ovrut, ``{Machine learning line bundle
  connections}'', \href{http://dx.doi.org/10.1016/j.physletb.2022.136972}{{\em
  Phys. Lett. B} {\bfseries 827} (2022)136972},
  \href{http://arxiv.org/abs/2110.12483}{{\ttfamily arXiv:2110.12483
  [hep-th]}}.

\bibitem{Larfors:2021pbb}
M.~Larfors, A.~Lukas, F.~Ruehle, and R.~Schneider, ``{Learning Size and Shape
  of Calabi-Yau Spaces}'', \href{http://arxiv.org/abs/2111.01436}{{\ttfamily
  arXiv:2111.01436 [hep-th]}}.

\bibitem{Larfors:2022nep}
M.~Larfors, A.~Lukas, F.~Ruehle, and R.~Schneider, ``{Numerical metrics for
  complete intersection and Kreuzer\textendash{}Skarke Calabi\textendash{}Yau
  manifolds}'', \href{http://dx.doi.org/10.1088/2632-2153/ac8e4e}{{\em Mach.
  Learn. Sci. Tech.} {\bfseries 3} 3, (2022)035014},
  \href{http://arxiv.org/abs/2205.13408}{{\ttfamily arXiv:2205.13408
  [hep-th]}}.

\bibitem{Gerdes:2022nzr}
M.~Gerdes and S.~Krippendorf, ``{CYJAX: A package for Calabi-Yau metrics with
  JAX}'', \href{http://arxiv.org/abs/2211.12520}{{\ttfamily arXiv:2211.12520
  [hep-th]}}.

\bibitem{Berglund:2022gvm}
P.~Berglund, G.~Butbaia, T.~H\"ubsch, V.~Jejjala, D.~Mayorga Pe\~na, C.~Mishra,
  and J.~Tan, ``{Machine Learned Calabi-Yau Metrics and Curvature}'',
  \href{http://arxiv.org/abs/2211.09801}{{\ttfamily arXiv:2211.09801
  [hep-th]}}.

\bibitem{Braun:2008jp}
V.~Braun, T.~Brelidze, M.~R. Douglas, and B.~A. Ovrut, ``{Eigenvalues and
  Eigenfunctions of the Scalar Laplace Operator on Calabi-Yau Manifolds}'',
  \href{http://dx.doi.org/10.1088/1126-6708/2008/07/120}{{\em JHEP} {\bfseries
  07} (2008)120}, \href{http://arxiv.org/abs/0805.3689}{{\ttfamily
  arXiv:0805.3689 [hep-th]}}.

\bibitem{Ashmore:2020ujw}
A.~Ashmore, ``{Eigenvalues and eigenforms on Calabi-Yau threefolds}'',
  \href{http://arxiv.org/abs/2011.13929}{{\ttfamily arXiv:2011.13929
  [hep-th]}}.

\bibitem{Ashmore:2021qdf}
A.~Ashmore and F.~Ruehle, ``{Moduli-dependent KK towers and the swampland
  distance conjecture on the quintic Calabi-Yau manifold}'',
  \href{http://dx.doi.org/10.1103/PhysRevD.103.106028}{{\em Phys. Rev. D}
  {\bfseries 103} 10, (2021)106028},
  \href{http://arxiv.org/abs/2103.07472}{{\ttfamily arXiv:2103.07472
  [hep-th]}}.

\bibitem{Afkhami-Jeddi:2021qkf}
N.~Afkhami-Jeddi, A.~Ashmore, and C.~Cordova, ``{Calabi-Yau CFTs and random
  matrices}'', \href{http://dx.doi.org/10.1007/JHEP02(2022)021}{{\em JHEP}
  {\bfseries 02} (2022)021}, \href{http://arxiv.org/abs/2107.11461}{{\ttfamily
  arXiv:2107.11461 [hep-th]}}.

\bibitem{Ahmed:2023cnw}
H.~Ahmed and F.~Ruehle, ``{Level Crossings, Attractor Points and Complex
  Multiplication}'', \href{http://arxiv.org/abs/2304.00027}{{\ttfamily
  arXiv:2304.00027 [hep-th]}}.

\bibitem{Strominger:1985it}
A.~Strominger and E.~Witten, ``{New Manifolds for Superstring
  Compactification}'', \href{http://dx.doi.org/10.1007/BF01216094}{{\em Commun.
  Math. Phys.} {\bfseries 101} (1985)341}.

\bibitem{Strominger:1985ks}
A.~Strominger, ``{Yukawa Couplings in Superstring Compactification}'',
  \href{http://dx.doi.org/10.1103/PhysRevLett.55.2547}{{\em Phys. Rev. Lett.}
  {\bfseries 55} (1985)2547}.

\bibitem{Mathematica}
W.~R. Inc., ``Mathematica, {V}ersion 13.2.''
\newblock \url{https://www.wolfram.com/mathematica}. Champaign, IL, 2022.

\bibitem{Anderson:2012yf}
L.~B. Anderson, J.~Gray, A.~Lukas, and E.~Palti, ``{Heterotic Line Bundle
  Standard Models}'', \href{http://dx.doi.org/10.1007/JHEP06(2012)113}{{\em
  JHEP} {\bfseries 06} (2012)113},
  \href{http://arxiv.org/abs/1202.1757}{{\ttfamily arXiv:1202.1757 [hep-th]}}.

\bibitem{GrootNibbelink:2015lme}
S.~Groot~Nibbelink, O.~Loukas, and F.~Ruehle, ``{(MS)SM-like models on smooth
  Calabi-Yau manifolds from all three heterotic string theories}'',
  \href{http://dx.doi.org/10.1002/prop.201500041}{{\em Fortsch. Phys.}
  {\bfseries 63} (2015)609--632},
  \href{http://arxiv.org/abs/1507.07559}{{\ttfamily arXiv:1507.07559
  [hep-th]}}.

\bibitem{GrootNibbelink:2015dvi}
S.~Groot~Nibbelink, O.~Loukas, F.~Ruehle, and P.~K.~S. Vaudrevange, ``{Infinite
  number of MSSMs from heterotic line bundles?}'',
  \href{http://dx.doi.org/10.1103/PhysRevD.92.046002}{{\em Phys. Rev. D}
  {\bfseries 92} 4, (2015)046002},
  \href{http://arxiv.org/abs/1506.00879}{{\ttfamily arXiv:1506.00879
  [hep-th]}}.

\bibitem{GrootNibbelin:2016ovb}
S.~Groot~Nibbelink and F.~Ruehle, ``{Line bundle embeddings for heterotic
  theories}'', \href{http://dx.doi.org/10.1007/JHEP04(2016)186}{{\em JHEP}
  {\bfseries 04} (2016)186}, \href{http://arxiv.org/abs/1601.00676}{{\ttfamily
  arXiv:1601.00676 [hep-th]}}.

\bibitem{Braun:2017feb}
A.~P. Braun, C.~R. Brodie, and A.~Lukas, ``{Heterotic Line Bundle Models on
  Elliptically Fibered Calabi-Yau Three-folds}'',
  \href{http://dx.doi.org/10.1007/JHEP04(2018)087}{{\em JHEP} {\bfseries 04}
  (2018)087}, \href{http://arxiv.org/abs/1706.07688}{{\ttfamily
  arXiv:1706.07688 [hep-th]}}.

\bibitem{hep-th/0512177}
V.~Braun, Y.-H. He, B.~A. Ovrut, and T.~Pantev, ``{The Exact MSSM spectrum from
  string theory}'', \href{http://dx.doi.org/10.1088/1126-6708/2006/05/043}{{\em
  JHEP} {\bfseries 05} (2006)043},
  \href{http://arxiv.org/abs/hep-th/0512177}{{\ttfamily arXiv:hep-th/0512177}}.

\bibitem{hep-th/0502155}
V.~Braun, Y.-H. He, B.~A. Ovrut, and T.~Pantev, ``{A Standard model from the
  E(8) x E(8) heterotic superstring}'',
  \href{http://dx.doi.org/10.1088/1126-6708/2005/06/039}{{\em JHEP} {\bfseries
  06} (2005)039}, \href{http://arxiv.org/abs/hep-th/0502155}{{\ttfamily
  arXiv:hep-th/0502155}}.

\bibitem{1112.1097}
V.~Braun, P.~Candelas, R.~Davies, and R.~Donagi, ``{The MSSM Spectrum from
  (0,2)-Deformations of the Heterotic Standard Embedding}'',
  \href{http://dx.doi.org/10.1007/JHEP05(2012)127}{{\em JHEP} {\bfseries 05}
  (2012)127}, \href{http://arxiv.org/abs/1112.1097}{{\ttfamily arXiv:1112.1097
  [hep-th]}}.

\bibitem{1007.0203}
M.~Blaszczyk, S.~Groot~Nibbelink, F.~Ruehle, M.~Trapletti, and P.~K.~S.
  Vaudrevange, ``{Heterotic MSSM on a Resolved Orbifold}'',
  \href{http://dx.doi.org/10.1007/JHEP09(2010)065}{{\em JHEP} {\bfseries 09}
  (2010)065}, \href{http://arxiv.org/abs/1007.0203}{{\ttfamily arXiv:1007.0203
  [hep-th]}}.

\bibitem{hep-th/9903052}
B.~Andreas, G.~Curio, and A.~Klemm, ``{Towards the Standard Model spectrum from
  elliptic Calabi-Yau}'',
  \href{http://dx.doi.org/10.1142/S0217751X04018087}{{\em Int. J. Mod. Phys. A}
  {\bfseries 19} (2004)1987},
  \href{http://arxiv.org/abs/hep-th/9903052}{{\ttfamily arXiv:hep-th/9903052}}.

\bibitem{Braun:2005zv}
V.~Braun, Y.-H. He, B.~A. Ovrut, and T.~Pantev, ``{Vector bundle extensions,
  sheaf cohomology, and the heterotic standard model}'',
  \href{http://dx.doi.org/10.4310/ATMP.2006.v10.n4.a3}{{\em Adv. Theor. Math.
  Phys.} {\bfseries 10} 4, (2006)525--589},
  \href{http://arxiv.org/abs/hep-th/0505041}{{\ttfamily arXiv:hep-th/0505041}}.

\bibitem{Braun:2006ae}
V.~Braun, Y.-H. He, and B.~A. Ovrut, ``{Stability of the minimal heterotic
  standard model bundle}'',
  \href{http://dx.doi.org/10.1088/1126-6708/2006/06/032}{{\em JHEP} {\bfseries
  06} (2006)032}, \href{http://arxiv.org/abs/hep-th/0602073}{{\ttfamily
  arXiv:hep-th/0602073}}.

\bibitem{Marshall:2014kea}
Z.~Marshall, B.~A. Ovrut, A.~Purves, and S.~Spinner, ``{Spontaneous $R$-Parity
  Breaking, Stop LSP Decays and the Neutrino Mass Hierarchy}'',
  \href{http://dx.doi.org/10.1016/j.physletb.2014.03.052}{{\em Phys. Lett. B}
  {\bfseries 732} (2014)325--329},
  \href{http://arxiv.org/abs/1401.7989}{{\ttfamily arXiv:1401.7989 [hep-ph]}}.

\bibitem{Marshall:2014cwa}
Z.~Marshall, B.~A. Ovrut, A.~Purves, and S.~Spinner, ``{LSP Squark Decays at
  the LHC and the Neutrino Mass Hierarchy}'',
  \href{http://dx.doi.org/10.1103/PhysRevD.90.015034}{{\em Phys. Rev. D}
  {\bfseries 90} 1, (2014)015034},
  \href{http://arxiv.org/abs/1402.5434}{{\ttfamily arXiv:1402.5434 [hep-ph]}}.

\bibitem{Ovrut:2014rba}
B.~A. Ovrut, A.~Purves, and S.~Spinner, ``{A statistical analysis of the
  minimal SUSY B\textendash{}L theory}'',
  \href{http://dx.doi.org/10.1142/S0217732315500856}{{\em Mod. Phys. Lett. A}
  {\bfseries 30} 18, (2015)1550085},
  \href{http://arxiv.org/abs/1412.6103}{{\ttfamily arXiv:1412.6103 [hep-ph]}}.

\bibitem{Ovrut:2015uea}
B.~A. Ovrut, A.~Purves, and S.~Spinner, ``{The minimal SUSY $B - L$ model: from
  the unification scale to the LHC}'',
  \href{http://dx.doi.org/10.1007/JHEP06(2015)182}{{\em JHEP} {\bfseries 06}
  (2015)182}, \href{http://arxiv.org/abs/1503.01473}{{\ttfamily
  arXiv:1503.01473 [hep-ph]}}.

\bibitem{Deen:2016vyh}
R.~Deen, B.~A. Ovrut, and A.~Purves, ``{The minimal SUSY B \ensuremath{-} L
  model: simultaneous Wilson lines and string thresholds}'',
  \href{http://dx.doi.org/10.1007/JHEP07(2016)043}{{\em JHEP} {\bfseries 07}
  (2016)043}, \href{http://arxiv.org/abs/1604.08588}{{\ttfamily
  arXiv:1604.08588 [hep-ph]}}.

\bibitem{Ovrut:2018qog}
B.~A. Ovrut, ``{Vacuum Constraints for Realistic Strongly Coupled Heterotic
  M-Theories}'', \href{http://dx.doi.org/10.3390/sym10120723}{{\em Symmetry}
  {\bfseries 10} 12, (2018)723},
  \href{http://arxiv.org/abs/1811.08892}{{\ttfamily arXiv:1811.08892
  [hep-th]}}.

\bibitem{Dumitru:2018jyb}
S.~Dumitru, B.~A. Ovrut, and A.~Purves, ``{The $R$-parity Violating Decays of
  Charginos and Neutralinos in the B-L MSSM}'',
  \href{http://dx.doi.org/10.1007/JHEP02(2019)124}{{\em JHEP} {\bfseries 02}
  (2019)124}, \href{http://arxiv.org/abs/1810.11035}{{\ttfamily
  arXiv:1810.11035 [hep-ph]}}.

\bibitem{Dumitru:2018nct}
S.~Dumitru, B.~A. Ovrut, and A.~Purves, ``{$R$-parity Violating Decays of Wino
  Chargino and Wino Neutralino LSPs and NLSPs at the LHC}'',
  \href{http://dx.doi.org/10.1007/JHEP06(2019)100}{{\em JHEP} {\bfseries 06}
  (2019)100}, \href{http://arxiv.org/abs/1811.05581}{{\ttfamily
  arXiv:1811.05581 [hep-ph]}}.

\bibitem{Dumitru:2019cgf}
S.~Dumitru, C.~Herwig, and B.~A. Ovrut, ``{$R$-parity Violating Decays of Bino
  Neutralino LSPs at the LHC}'',
  \href{http://dx.doi.org/10.1007/JHEP12(2019)042}{{\em JHEP} {\bfseries 12}
  (2019)042}, \href{http://arxiv.org/abs/1906.03174}{{\ttfamily
  arXiv:1906.03174 [hep-ph]}}.

\bibitem{Ashmore:2020ocb}
A.~Ashmore, S.~Dumitru, and B.~A. Ovrut, ``{Line Bundle Hidden Sectors for
  Strongly Coupled Heterotic Standard Models}'',
  \href{http://dx.doi.org/10.1002/prop.202100052}{{\em Fortsch. Phys.}
  {\bfseries 69} 7, (2021)2100052},
  \href{http://arxiv.org/abs/2003.05455}{{\ttfamily arXiv:2003.05455
  [hep-th]}}.

\bibitem{Ashmore:2020wwv}
A.~Ashmore, S.~Dumitru, and B.~A. Ovrut, ``{Explicit soft supersymmetry
  breaking in the heterotic M-theory B \ensuremath{-} L MSSM}'',
  \href{http://dx.doi.org/10.1007/JHEP08(2021)033}{{\em JHEP} {\bfseries 08}
  (2021)033}, \href{http://arxiv.org/abs/2012.11029}{{\ttfamily
  arXiv:2012.11029 [hep-th]}}.

\bibitem{Ashmore:2021xdm}
A.~Ashmore, S.~Dumitru, and B.~A. Ovrut, ``{Hidden Sectors from Multiple Line
  Bundles for the B\ensuremath{-}L$B-L$ MSSM}'',
  \href{http://dx.doi.org/10.1002/prop.202200071}{{\em Fortsch. Phys.}
  {\bfseries 70} 7-8, (2022)2200071},
  \href{http://arxiv.org/abs/2106.09087}{{\ttfamily arXiv:2106.09087
  [hep-th]}}.

\bibitem{Dumitru:2021jlh}
S.~Dumitru and B.~A. Ovrut, ``{Heterotic $M$-Theory Hidden Sectors with an
  Anomalous $U(1)$ Gauge Symmetry}'',
  \href{http://arxiv.org/abs/2109.13781}{{\ttfamily arXiv:2109.13781
  [hep-th]}}.

\bibitem{Dumitru:2022apw}
S.~Dumitru and B.~A. Ovrut, ``{Moduli and Hidden Matter in Heterotic M-Theory
  with an Anomalous $U(1)$ Hidden Sector}'',
  \href{http://arxiv.org/abs/2201.01624}{{\ttfamily arXiv:2201.01624
  [hep-th]}}.

\bibitem{Dumitru:2022eri}
S.~Dumitru and B.~A. Ovrut, ``{FIMP dark matter in heterotic M-theory}'',
  \href{http://dx.doi.org/10.1007/JHEP09(2022)068}{{\em JHEP} {\bfseries 09}
  (2022)068}, \href{http://arxiv.org/abs/2204.13174}{{\ttfamily
  arXiv:2204.13174 [hep-ph]}}.

\bibitem{Witten:1999eg}
E.~Witten, ``{World sheet corrections via D instantons}'',
  \href{http://dx.doi.org/10.1088/1126-6708/2000/02/030}{{\em JHEP} {\bfseries
  02} (2000)030}, \href{http://arxiv.org/abs/hep-th/9907041}{{\ttfamily
  arXiv:hep-th/9907041}}.

\bibitem{Buchbinder:2002ic}
E.~I. Buchbinder, R.~Donagi, and B.~A. Ovrut, ``{Superpotentials for vector
  bundle moduli}'', \href{http://dx.doi.org/10.1016/S0550-3213(02)01093-3}{{\em
  Nucl. Phys. B} {\bfseries 653} (2003)400--420},
  \href{http://arxiv.org/abs/hep-th/0205190}{{\ttfamily arXiv:hep-th/0205190}}.

\bibitem{Beasley:2003fx}
C.~Beasley and E.~Witten, ``{Residues and world sheet instantons}'',
  \href{http://dx.doi.org/10.1088/1126-6708/2003/10/065}{{\em JHEP} {\bfseries
  10} (2003)065}, \href{http://arxiv.org/abs/hep-th/0304115}{{\ttfamily
  arXiv:hep-th/0304115}}.

\bibitem{Basu:2003bq}
A.~Basu and S.~Sethi, ``{World sheet stability of (0,2) linear sigma models}'',
  \href{http://dx.doi.org/10.1103/PhysRevD.68.025003}{{\em Phys. Rev. D}
  {\bfseries 68} (2003)025003},
  \href{http://arxiv.org/abs/hep-th/0303066}{{\ttfamily arXiv:hep-th/0303066}}.

\bibitem{Braun:2007xh}
V.~Braun, M.~Kreuzer, B.~A. Ovrut, and E.~Scheidegger, ``{Worldsheet instantons
  and torsion curves, part A: Direct computation}'',
  \href{http://dx.doi.org/10.1088/1126-6708/2007/10/022}{{\em JHEP} {\bfseries
  10} (2007)022}, \href{http://arxiv.org/abs/hep-th/0703182}{{\ttfamily
  arXiv:hep-th/0703182}}.

\bibitem{Braun:2007tp}
V.~Braun, M.~Kreuzer, B.~A. Ovrut, and E.~Scheidegger, ``{Worldsheet
  instantons, torsion curves, and non-perturbative superpotentials}'',
  \href{http://dx.doi.org/10.1016/j.physletb.2007.03.066}{{\em Phys. Lett. B}
  {\bfseries 649} (2007)334--341},
  \href{http://arxiv.org/abs/hep-th/0703134}{{\ttfamily arXiv:hep-th/0703134}}.

\bibitem{Braun:2007vy}
V.~Braun, M.~Kreuzer, B.~A. Ovrut, and E.~Scheidegger, ``{Worldsheet Instantons
  and Torsion Curves, Part B: Mirror Symmetry}'',
  \href{http://dx.doi.org/10.1088/1126-6708/2007/10/023}{{\em JHEP} {\bfseries
  10} (2007)023}, \href{http://arxiv.org/abs/0704.0449}{{\ttfamily
  arXiv:0704.0449 [hep-th]}}.

\bibitem{Bertolini:2014dna}
M.~Bertolini and M.~R. Plesser, ``{Worldsheet instantons and (0,2) linear
  models}'', \href{http://dx.doi.org/10.1007/JHEP08(2015)081}{{\em JHEP}
  {\bfseries 08} (2015)081}, \href{http://arxiv.org/abs/1410.4541}{{\ttfamily
  arXiv:1410.4541 [hep-th]}}.

\bibitem{Buchbinder:2016rmw}
E.~I. Buchbinder and B.~A. Ovrut, ``{Non-vanishing Superpotentials in Heterotic
  String Theory and Discrete Torsion}'',
  \href{http://dx.doi.org/10.1007/JHEP01(2017)038}{{\em JHEP} {\bfseries 01}
  (2017)038}, \href{http://arxiv.org/abs/1611.01922}{{\ttfamily
  arXiv:1611.01922 [hep-th]}}.

\bibitem{Buchbinder:2019eal}
E.~I. Buchbinder, A.~Lukas, B.~A. Ovrut, and F.~Ruehle, ``{Instantons and
  Hilbert Functions}'',
  \href{http://dx.doi.org/10.1103/PhysRevD.102.026019}{{\em Phys. Rev. D}
  {\bfseries 102} 2, (2020)026019},
  \href{http://arxiv.org/abs/1912.08358}{{\ttfamily arXiv:1912.08358
  [hep-th]}}.

\bibitem{Buchbinder:2019hyb}
E.~I. Buchbinder, A.~Lukas, B.~A. Ovrut, and F.~Ruehle, ``{Heterotic Instantons
  for Monad and Extension Bundles}'',
  \href{http://dx.doi.org/10.1007/JHEP02(2020)081}{{\em JHEP} {\bfseries 02}
  (2020)081}, \href{http://arxiv.org/abs/1912.07222}{{\ttfamily
  arXiv:1912.07222 [hep-th]}}.

\bibitem{Buchbinder:2017azb}
E.~Buchbinder, A.~Lukas, B.~Ovrut, and F.~Ruehle, ``{Heterotic Instanton
  Superpotentials from Complete Intersection Calabi-Yau Manifolds}'',
  \href{http://dx.doi.org/10.1007/JHEP10(2017)032}{{\em JHEP} {\bfseries 10}
  (2017)032}, \href{http://arxiv.org/abs/1707.07214}{{\ttfamily
  arXiv:1707.07214 [hep-th]}}.

\bibitem{Buchbinder:2018hns}
E.~I. Buchbinder, L.~Lin, and B.~A. Ovrut, ``{Non-vanishing Heterotic
  Superpotentials on Elliptic Fibrations}'',
  \href{http://dx.doi.org/10.1007/JHEP09(2018)111}{{\em JHEP} {\bfseries 09}
  (2018)111}, \href{http://arxiv.org/abs/1806.04669}{{\ttfamily
  arXiv:1806.04669 [hep-th]}}.

\bibitem{Buchbinder:2002pr}
E.~I. Buchbinder, R.~Donagi, and B.~A. Ovrut, ``{Vector bundle moduli
  superpotentials in heterotic superstrings and M theory}'',
  \href{http://dx.doi.org/10.1088/1126-6708/2002/07/066}{{\em JHEP} {\bfseries
  07} (2002)066}, \href{http://arxiv.org/abs/hep-th/0206203}{{\ttfamily
  arXiv:hep-th/0206203}}.

\bibitem{Anderson:2015yzz}
L.~B. Anderson, F.~Apruzzi, X.~Gao, J.~Gray, and S.-J. Lee, ``{Instanton
  superpotentials, Calabi-Yau geometry, and fibrations}'',
  \href{http://dx.doi.org/10.1103/PhysRevD.93.086001}{{\em Phys. Rev. D}
  {\bfseries 93} 8, (2016)086001},
  \href{http://arxiv.org/abs/1511.05188}{{\ttfamily arXiv:1511.05188
  [hep-th]}}.

\bibitem{Becker:2005nb}
K.~Becker and L.-S. Tseng, ``{Heterotic flux compactifications and their
  moduli}'', \href{http://dx.doi.org/10.1016/j.nuclphysb.2006.02.013}{{\em
  Nucl. Phys. B} {\bfseries 741} (2006)162--179},
  \href{http://arxiv.org/abs/hep-th/0509131}{{\ttfamily arXiv:hep-th/0509131}}.

\bibitem{Becker:2006xp}
M.~Becker, L.-S. Tseng, and S.-T. Yau, ``{Moduli Space of Torsional
  Manifolds}'', \href{http://dx.doi.org/10.1016/j.nuclphysb.2007.07.006}{{\em
  Nucl. Phys. B} {\bfseries 786} (2007)119--134},
  \href{http://arxiv.org/abs/hep-th/0612290}{{\ttfamily arXiv:hep-th/0612290}}.

\bibitem{Anderson:2010mh}
L.~B. Anderson, J.~Gray, A.~Lukas, and B.~Ovrut, ``{Stabilizing the Complex
  Structure in Heterotic Calabi-Yau Vacua}'',
  \href{http://dx.doi.org/10.1007/JHEP02(2011)088}{{\em JHEP} {\bfseries 02}
  (2011)088}, \href{http://arxiv.org/abs/1010.0255}{{\ttfamily arXiv:1010.0255
  [hep-th]}}.

\bibitem{Melnikov:2011ez}
I.~V. Melnikov and E.~Sharpe, ``{On marginal deformations of (0,2) non-linear
  sigma models}'', \href{http://dx.doi.org/10.1016/j.physletb.2011.10.055}{{\em
  Phys. Lett. B} {\bfseries 705} (2011)529--534},
  \href{http://arxiv.org/abs/1110.1886}{{\ttfamily arXiv:1110.1886 [hep-th]}}.

\bibitem{Anderson:2011cza}
L.~B. Anderson, J.~Gray, A.~Lukas, and B.~Ovrut, ``{Stabilizing All Geometric
  Moduli in Heterotic Calabi-Yau Vacua}'',
  \href{http://dx.doi.org/10.1103/PhysRevD.83.106011}{{\em Phys. Rev. D}
  {\bfseries 83} (2011)106011},
  \href{http://arxiv.org/abs/1102.0011}{{\ttfamily arXiv:1102.0011 [hep-th]}}.

\bibitem{Anderson:2011ty}
L.~B. Anderson, J.~Gray, A.~Lukas, and B.~Ovrut, ``{The Atiyah Class and
  Complex Structure Stabilization in Heterotic Calabi-Yau Compactifications}'',
  \href{http://dx.doi.org/10.1007/JHEP10(2011)032}{{\em JHEP} {\bfseries 10}
  (2011)032}, \href{http://arxiv.org/abs/1107.5076}{{\ttfamily arXiv:1107.5076
  [hep-th]}}.

\bibitem{Anderson:2014xha}
L.~B. Anderson, J.~Gray, and E.~Sharpe, ``{Algebroids, Heterotic Moduli Spaces
  and the Strominger System}'',
  \href{http://dx.doi.org/10.1007/JHEP07(2014)037}{{\em JHEP} {\bfseries 07}
  (2014)037}, \href{http://arxiv.org/abs/1402.1532}{{\ttfamily arXiv:1402.1532
  [hep-th]}}.

\bibitem{Anderson:2013qca}
L.~B. Anderson, J.~Gray, A.~Lukas, and B.~Ovrut, ``{Vacuum Varieties,
  Holomorphic Bundles and Complex Structure Stabilization in Heterotic
  Theories}'', \href{http://dx.doi.org/10.1007/JHEP07(2013)017}{{\em JHEP}
  {\bfseries 07} (2013)017}, \href{http://arxiv.org/abs/1304.2704}{{\ttfamily
  arXiv:1304.2704 [hep-th]}}.

\bibitem{delaOssa:2014cia}
X.~de~la Ossa and E.~E. Svanes, ``{Holomorphic Bundles and the Moduli Space of
  N=1 Supersymmetric Heterotic Compactifications}'',
  \href{http://dx.doi.org/10.1007/JHEP10(2014)123}{{\em JHEP} {\bfseries 10}
  (2014)123}, \href{http://arxiv.org/abs/1402.1725}{{\ttfamily arXiv:1402.1725
  [hep-th]}}.

\bibitem{delaOssa:2015maa}
X.~de~la Ossa, E.~Hardy, and E.~E. Svanes, ``{The Heterotic Superpotential and
  Moduli}'', \href{http://dx.doi.org/10.1007/JHEP01(2016)049}{{\em JHEP}
  {\bfseries 01} (2016)049}, \href{http://arxiv.org/abs/1509.08724}{{\ttfamily
  arXiv:1509.08724 [hep-th]}}.

\bibitem{Garcia-Fernandez:2015hja}
M.~Garcia-Fernandez, R.~Rubio, and C.~Tipler, ``{Infinitesimal moduli for the
  Strominger system and Killing spinors in generalized geometry}'',
  \href{http://dx.doi.org/10.1007/s00208-016-1463-5}{{\em Math. Ann.}
  {\bfseries 369} (2017)539--595},
  \href{http://arxiv.org/abs/1503.07562}{{\ttfamily arXiv:1503.07562
  [math.DG]}}.

\bibitem{Candelas:2016usb}
P.~Candelas, X.~de~la Ossa, and J.~McOrist, ``{A Metric for Heterotic
  Moduli}'', \href{http://dx.doi.org/10.1007/s00220-017-2978-7}{{\em Commun.
  Math. Phys.} {\bfseries 356} 2, (2017)567--612},
  \href{http://arxiv.org/abs/1605.05256}{{\ttfamily arXiv:1605.05256
  [hep-th]}}.

\bibitem{Ashmore:2018ybe}
A.~Ashmore, X.~De~La~Ossa, R.~Minasian, C.~Strickland-Constable, and E.~E.
  Svanes, ``{Finite deformations from a heterotic superpotential: holomorphic
  Chern-Simons and an $L_\infty$ algebra}'',
  \href{http://dx.doi.org/10.1007/JHEP10(2018)179}{{\em JHEP} {\bfseries 10}
  (2018)179}, \href{http://arxiv.org/abs/1806.08367}{{\ttfamily
  arXiv:1806.08367 [hep-th]}}.

\bibitem{Blesneag:2018ygh}
c.~Blesneag, E.~I. Buchbinder, A.~Constantin, A.~Lukas, and E.~Palti, ``{Matter
  field K\"ahler metric in heterotic string theory from localisation}'',
  \href{http://dx.doi.org/10.1007/JHEP04(2018)139}{{\em JHEP} {\bfseries 04}
  (2018)139}, \href{http://arxiv.org/abs/1801.09645}{{\ttfamily
  arXiv:1801.09645 [hep-th]}}.

\bibitem{Ashmore:2019rkx}
A.~Ashmore, C.~Strickland-Constable, D.~Tennyson, and D.~Waldram, ``{Heterotic
  backgrounds via generalised geometry: moment maps and moduli}'',
  \href{http://dx.doi.org/10.1007/JHEP11(2020)071}{{\em JHEP} {\bfseries 11}
  (2020)071}, \href{http://arxiv.org/abs/1912.09981}{{\ttfamily
  arXiv:1912.09981 [hep-th]}}.

\bibitem{10.2748/tmj/1178228026}
R.~Kuwabara, ``{Spectrum of the Schrödinger operator on a line bundle over
  complex projective spaces}'',
  \href{http://dx.doi.org/10.2748/tmj/1178228026}{{\em Tohoku Mathematical
  Journal} {\bfseries 40} 2, (1988)199 -- 211}.

\bibitem{2302.11691}
D.~Bykov and A.~Smilga, ``{Monopole harmonics on $\mathbb{CP}^{n-1}$}'',
  \href{http://arxiv.org/abs/2302.11691}{{\ttfamily arXiv:2302.11691
  [hep-th]}}.

\bibitem{TEJEROPRIETO2006288}
C.~{Tejero Prieto}, ``Holomorphic spectral geometry of magnetic {S}chrödinger
  operators on {R}iemann surfaces'',
  \href{http://dx.doi.org/https://doi.org/10.1016/j.difgeo.2005.09.001}{{\em
  Differential Geometry and its Applications} {\bfseries 24} 3,
  (2006)288--310}.

\bibitem{Candelas:1987se}
P.~Candelas, ``{Yukawa Couplings Between (2,1) Forms}'',
  \href{http://dx.doi.org/10.1016/0550-3213(88)90351-3}{{\em Nucl. Phys. B}
  {\bfseries 298} (1988)458}.

\bibitem{Candelas:1990pi}
P.~Candelas and X.~de~la Ossa, ``{Moduli Space of {Calabi-Yau} Manifolds}'',
  \href{http://dx.doi.org/10.1016/0550-3213(91)90122-E}{{\em Nucl. Phys. B}
  {\bfseries 355} (1991)455--481}.

\bibitem{Green:1987sp}
M.~B. Green, J.~H. Schwarz, and E.~Witten, {\em {SUPERSTRING THEORY. VOL. 1:
  INTRODUCTION}}.
\newblock Cambridge Monographs on Mathematical Physics. 7, 1988.

\bibitem{Green:1987mn}
M.~B. Green, J.~H. Schwarz, and E.~Witten, {\em {SUPERSTRING THEORY. VOL. 2:
  LOOP AMPLITUDES, ANOMALIES AND PHENOMENOLOGY}}.
\newblock 7, 1988.

\bibitem{Wess:1992cp}
J.~Wess and J.~Bagger, {\em {Supersymmetry and supergravity}}.
\newblock Princeton University Press, Princeton, NJ, USA, 1992.

\bibitem{Donaldson}
S.~K. Donaldson, ``Anti self-dual {Y}ang-{M}ills connections over complex
  algebraic surfaces and stable vector bundles'',
  \href{http://dx.doi.org/10.1112/plms/s3-50.1.1}{{\em Proceedings of the
  London Mathematical Society} {\bfseries s3-50} 1, (1985)1--26}.

\bibitem{UhlenbeckYau}
K.~Uhlenbeck and S.~T. Yau, ``On the existence of hermitian-{Y}ang-{M}ills
  connections in stable vector bundles'',
  \href{http://dx.doi.org/10.1002/cpa.3160390714}{{\em Communications on Pure
  and Applied Mathematics} {\bfseries 39} S1, (1986)S257--S293}.

\bibitem{Bochner}
S.~Bochner, ``Curvature and {B}etti numbers'',
  \href{http://dx.doi.org/10.2307/1969287}{{\em Annals of Mathematics}
  {\bfseries 49} 2, (1948)379--390}.

\bibitem{Kodaira}
K.~Kodaira, ``On a differential-geometric method in the theory of analytic
  stacks'', \href{http://dx.doi.org/10.1073/pnas.39.12.1268}{{\em Proceedings
  of the National Academy of Sciences of the United States of America}
  {\bfseries 39} 12, (1953)1268--1273}.

\bibitem{Nakano}
S.~Nakano, ``{On complex analytic vector bundles.}'',
  \href{http://dx.doi.org/10.2969/jmsj/00710001}{{\em Journal of the
  Mathematical Society of Japan} {\bfseries 7} 1, (1955)1 -- 12}.

\bibitem{Demailly}
J.-P. Demailly, \href{http://dx.doi.org/10.1007/BFb0077045}{``Sur l'identite de
  bochner-kodaira-nakano en geometrie hermitienne'',} in {\em S{\'e}minaire
  d'Analyse}, P.~Lelong, P.~Dolbeault, and H.~Skoda, eds., pp.~88--97.
\newblock Springer Berlin Heidelberg, Berlin, Heidelberg, 1986.

\bibitem{Macaulay2}
D.~R. Grayson and M.~E. Stillman, ``Macaulay2, a software system for research
  in algebraic geometry.'' Available at
  \url{http://www.math.uiuc.edu/Macaulay2/}.

\bibitem{IKEDATANIGUCHI}
A.~Ikeda and Y.~Taniguchi, ``Spectra and eigenforms of the {L}aplacian on
  ${S}^n$ and {${P}^n({\bf {C}})$}'',
  \href{http://dx.doi.org/10.18910/6956}{{\em Osaka J. Math.} {\bfseries 15} 3,
  (1978)515--546}.

\bibitem{Cui:2023eqr}
W.~Cui, ``{Numerical Hermitian Yang-Mills Connection for Bundles on Quotient
  Manifold}'', \href{http://arxiv.org/abs/2302.09622}{{\ttfamily
  arXiv:2302.09622 [hep-th]}}.

\bibitem{B_rard_2016}
P.~B{\'{e}}rard and B.~Helffer, ``Courant-sharp eigenvalues for the equilateral
  torus, and for the equilateral triangle'',
  \href{http://dx.doi.org/10.1007/s11005-016-0819-9}{{\em Letters in
  Mathematical Physics} {\bfseries 106} 12, (Feb, 2016)1729--1789}.

\bibitem{Milnor64}
J.~Milnor, ``Eigenvalues of the {L}aplace operator on certain manifolds'',
  \href{http://dx.doi.org/10.1073/pnas.51.4.542}{{\em PNAS} {\bfseries 51} 4,
  (1964)542--542}.

\bibitem{1405.4944}
C.-Y. {Kao}, R.~{Lai}, and B.~{Osting}, ``{Maximization of Laplace-Beltrami
  eigenvalues on closed Riemannian surfaces}'',
  \href{http://arxiv.org/abs/1405.4944}{{\ttfamily arXiv:1405.4944 [math.DG]}}.

\bibitem{Tian}
G.~Tian, ``{On a set of polarized Kähler metrics on algebraic manifolds}'',
  \href{http://dx.doi.org/10.4310/jdg/1214445039}{{\em Journal of Differential
  Geometry} {\bfseries 32} 1, (1990)99 -- 130}.

\bibitem{1906.00392}
M.~Larfors and R.~Schneider, ``{Line bundle cohomologies on CICYs with Picard
  number two}'', \href{http://dx.doi.org/10.1002/prop.201900083}{{\em Fortsch.
  Phys.} {\bfseries 67} 12, (2019)1900083},
  \href{http://arxiv.org/abs/1906.00392}{{\ttfamily arXiv:1906.00392
  [hep-th]}}.

\bibitem{Blesneag:2021wdf}
S.~Blesneag, {\em {Holomorphic Yukawa Couplings in Heterotic String Theory}}.
\newblock PhD thesis, Oxford U., 2021.
\newblock \href{http://arxiv.org/abs/2204.01165}{{\ttfamily arXiv:2204.01165
  [hep-th]}}.

\bibitem{tomasiello_2022}
A.~Tomasiello, \href{http://dx.doi.org/10.1017/9781108635745}{{\em Geometry of
  String Theory Compactifications}}.
\newblock Cambridge University Press, 2022.

\end{thebibliography}\endgroup

\end{document}